\documentclass[11pt,letter,subeqn,fleqn]{article}
\usepackage{hyperref}
\usepackage{amsmath, amssymb,amsthm,cite}
\usepackage{graphics,color}
\usepackage{graphicx}
\usepackage{tabulary}
\usepackage{float}
\usepackage{graphicx}
\usepackage{multirow}
\usepackage{subcaption}
\usepackage{longtable}
\newcommand{\gsim}{\raisebox{-0.13cm}{~\shortstack{$<$ \\[-0.07cm]$\sim$}}~}
\title{Narrow Escape Brownian Dynamics Modeling in the Three-Dimensional Unit Sphere}

\author{ 
Vaibhava Srivastava\footnotemark[1],~~ A. F. Cheviakov \footnotemark[2]\vspace{0.5cm}\\
\small $^{\rm a}$\emph{Department of Mathematics, Iowa State University, Ames, IA 50011, USA}\vspace{0.2cm}\\
\small $^{\rm a,b}$\emph{Department of Mathematics and Statistics, University of Saskatchewan, Saskatoon, Canada}\vspace{0.2cm}\\ }

\def\div{\mathop{\hbox{\rm div}}}
\def\grad{\mathop{\hbox{\rm grad}}}

{\theoremstyle{definition}

}

\begin{document}


\footnotetext[1]{Corresponding author. Electronic mail: vaibhava@iastate.edu}
\footnotetext[2]{Alternative English spelling: Alexey Shevyakov. Electronic mail: shevyakov@math.usask.ca}

\maketitle \numberwithin{equation}{section}
\maketitle \numberwithin{remark}{section}
\numberwithin{lemma}{section}
\numberwithin{proposition}{section}

\begin{abstract}

The narrow escape problem is a first-passage problem concerned with randomly moving particles in a physical domain, being trapped by absorbing surface traps (windows), such that the measure of traps is small compared to the domain size. The expected value of time required for a particle to escape is defined as mean first passage time (MFPT), which satisfies the Poisson partial differential equation subject to a mixed Dirichlet-Neumann boundary condition. The primary objective of this work is a direct numerical simulation of multiple particles undergoing Brownian motion in a three-dimensional sphere with boundary traps, compute MFPT values by averaging Brownian escape times, and compare the results with asymptotic results obtained by solving the Poisson PDE problem. A comprehensive study of results obtained from the simulations shows that the difference between Brownian and asymptotic results for the escape times mostly not exceed $1\%$ accuracy. This comparison in some sense validates the narrow escape PDE problem itself as an approximation (averaging) of the multiple physical Brownian motion runs. This work also predicted that how many single-particle simulations are  required to match the predicted asymptotic averaged MFPT values. The next objective of this work is to study dynamics of Brownian particles near the boundary by estimating the average percentage of time spent by Brownian particle near the domain boundary for both the anisotropic and isotropic diffusion. It is shown that the Brownian particles spend more in the boundary layer than predicted by the boundary layer relative volume, with the effect being more pronounced in a narrow layer near the spherical wall. It is also shown that taking into account anisotropic diffusion yields larger times a particle spends near the boundary, and smaller escape times than those predicted by the isotropic diffusion model.

\end{abstract}

\section{Introduction}
A long time has gone since Robert Brown in 1827, observed the non-linear random motions of pollens immersed in water, yet the place of Brownian motion as a research topic among the scientific community has not changed much. Instead, it has caught interest in the 21st century due to its close connection with molecular biology and neuroscience. Rigorous mathematical models based on Brownian Dynamics acts as a template for studying the diffusive movement of various minuscule particles of differing shapes traveling in different media, namely solid, liquid, and gaseous environments. The importance of these diffusive motions can not be understated for a cell to regulate the routine processes of daily life, all cells have to transport necessary ions and molecules across semi-permeable membranes through diffusion, and therefore it would not be an overstatement to say that these diffusive processes are the drivers of life.

\begin{figure}[htbp]
\begin{subfigure}{0.45\textwidth}
 \includegraphics[width=\textwidth,height=7cm]{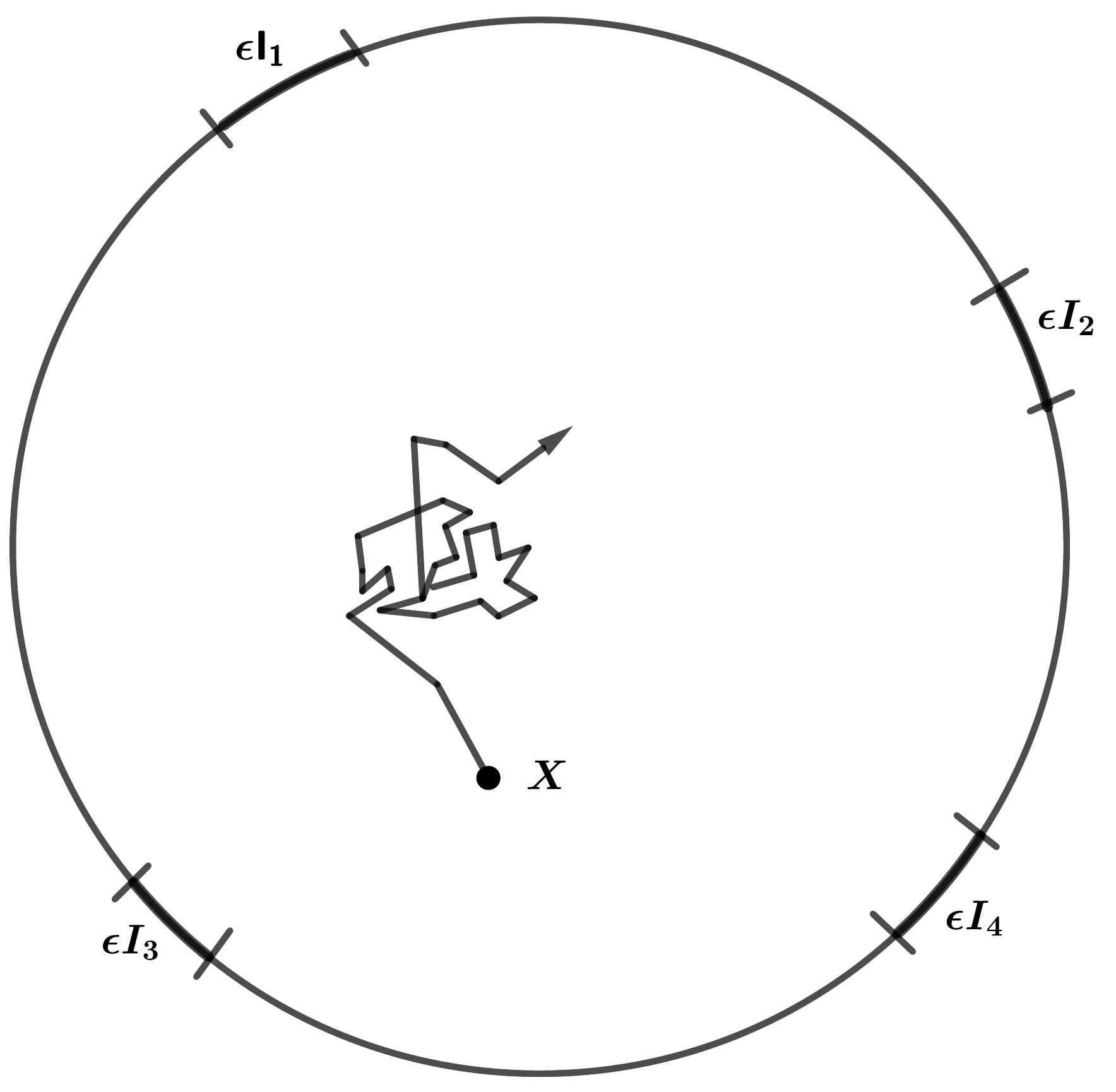}
 \caption{\label{fig:schematics_2D}}
\end{subfigure}
\hfill
\begin{subfigure}{0.45\textwidth}
\includegraphics[width=\textwidth, height=7cm]{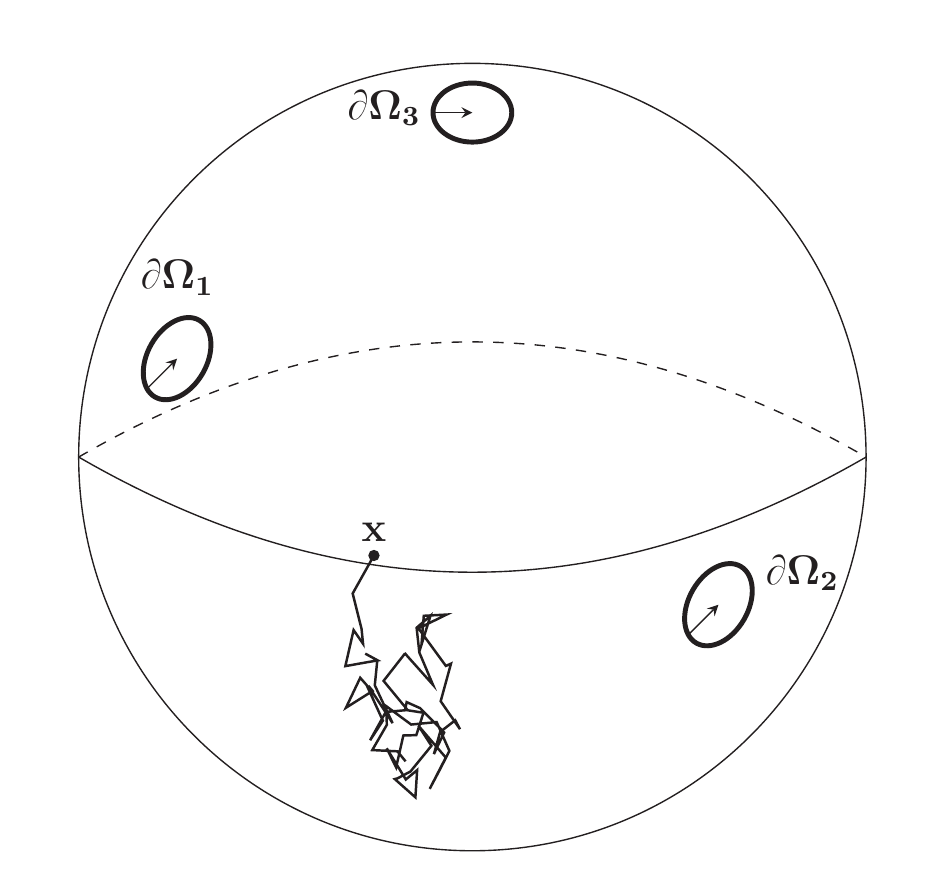}
\caption{\label{fig:schematics_3D}}
\end{subfigure}
\hfill
\caption{(\subref{fig:schematics_2D}) Schematics of the narrow escape problem in a two-dimensional spaces having surface traps of length $\{\epsilon I_j\}_{j=1}^{4}$. (\subref{fig:schematics_3D}) Illustration of the three-dimensional narrow escape problem in the unit sphere having circular absorbing windows $\{\partial \Omega_j\}_{j=1}^{3}$.}
\label{fig:schematics}
\end{figure}

Narrow escape problems arise in multiple physical settings, and form a subset of bigger class of problems known as the first-passage problem. The first-passage problems concern calculating the time required by a Brownian particle to escape a given domain through some absorbing boundary. The narrow escape problem is a first-passage problem explaining the peculiar presence of localized absorbing surface traps or windows, such that the measure of these traps is asymptotically small in some parameters (see Fig.~\ref{fig:schematics}).

Narrow escape problems have significant applicability in chemical and biological modelings, some of which include the diffusion of an ion in a biological cellular micro-domains \cite{10.2307/25449275}, receptor trafficking in a synaptic membrane \cite{HolcmanD.2004ETaS}, modeling of narrow escape kinetics to a chemical process \cite{GrigorievIgorV.2002Koet}, nanoparticle diffusion within inverse opals and related man-made materials with cavities \cite{nanoparticle}, transportation of ribonucleic acid in between cell nucleus to pores \cite{GorskiStanislawA2006Trmt}, and analyzing the narrow escape dynamics of ecological systems such as predator-prey synergies \cite{OshaninG2009Soae}.

Let $\{W(t)\}_{t \ge 0}$ represents Brownian trajectories of a particle restricted inside a two or three-dimensional bounded domain $\Omega$. Assume the boundary of the $\Omega$, $\partial\Omega=\partial\Omega_a\cup\partial\Omega_r$ is nearly everywhere reflecting denoted by $\partial \Omega_r$, except for finitely many small circular absorbing windows or regions, each centered at $x_j \in \partial \Omega$, for $j=\{1,2,\dots,N\}$ collectively marked as $\partial \Omega_a$, through which the particle can escape from the confined domain $\Omega$. A general schematic of the narrow escape problem in the case of two-dimensional (see Fig.~\ref{fig:schematics_2D}) and the three-dimensional (see Fig.~\ref{fig:schematics_3D}) domain is shown in Fig.~\ref{fig:schematics}.

The mean first passage time (MFPT) $v(x)$ defined as the expectation value of time $\tau(x)$ required for the particles having Brownian trajectories $\{W(t)\}_{t \ge 0}$ to escape through one of the boundary traps when launched from $x=W(0)$. For the asymptotically small measure of the absorbing windows $| \partial \Omega_a| = \mathcal{O}(\epsilon^{d-1})$, where $0 < \epsilon \ll 1$ is the dimensionless radius of the absorbing window, the problem acknowledging  the calculation of the MFPT $v(x)$ becomes a kind of narrow escape. Consequently, for asymptotically small choices of the trap size $\epsilon$, i.e. $\epsilon \to 0$, the question of determining the MFPT will be a singular perturbation problem.

The initial general model for all random walk-type/diffusion problems is Brownian motion, possibly with variable diffusivity. The Poisson problem \eqref{eqn:NEP_Dir_Neu} approximates such diffusion problems in the Narrow Escape context, averaging trajectories over multiple runs. A model validating the MFPT as solution of a Poisson partial differential equation (PDE) problem has been proposed and examined in the recent literature (cf.~\cite{HolcmanD.2004ETaS, RednerSidney1951-2001Agtf}), where the approximate asymptotic expressions for MFPT and average MFPT concerning narrow escape kinetics for various two and three-dimensional domain have been studied. The validity of the quality of the approximation of Brownian motion by Poisson's problem is thoroughly studied in \cite{RednerSidney1951-2001Agtf,HeinzStefan1958-2011Mm}. The Poisson PDE subject to the mixed Dirichlet-Neumann boundary conditions is given by
\begin{subequations}\label{eqn:NEP_Dir_Neu}
\begin{equation}\label{eq:NEP_pde}
	\Delta v = -\frac{1}{D}, \qquad x\in \Omega,
\end{equation}
\begin{equation}\label{eq:NEP_pde_bvp}
	v=0, \quad x\in\partial\Omega_a=\bigcup\limits_{i=1}^N\partial\Omega_{\epsilon_i},\quad j=1,2,\dots,N; \quad
	\partial_n v = 0, \quad x\in\partial\Omega_r,
\end{equation}
\end{subequations}
where $\Delta$ is the Laplacian operator, $D$ is the constant of the diffusivity for the given particle's Brownian motion, $\partial_n v$
represents the normal derivative on the surface of $\Omega$, $\partial \Omega_r$ represents the reflecting portion of the surface, $\partial \Omega_a$, the absorbing set is the collection of $N$ distinct small disjoint absorbing windows $\partial \Omega_{\epsilon_j}$ each of these centered at $x_j$, for $j=1,2,\dots,N$. For a two or three-dimensional domain $\Omega$ such that $diam(\Omega) =\mathcal{O}(1)$, the absorbing windows are identified and delineated by the help of length $|\partial \Omega_{\epsilon_j}| = \mathcal{O}(\epsilon)$ and area $|\partial \Omega_{\epsilon_j}| = \mathcal{O}(\epsilon^2)$ measures respectively, where $0<\epsilon \ll 1$. An important integral characteristic of escape times from a domain with a prescribed trap arrangement is the average mean first passage time AMFPT denoted by $\bar{v}$ and defined for the uniformly distributed initial launching positions $x \in \Omega$ concerning the Brownian walk formulate as
\begin{equation}\label{eqn:avg_mfpt}
	\bar{v}= \frac{1}{|\Omega|}\int_\Omega v(x) dx,
\end{equation}
where $|\Omega|$ is the volume of the domain $\Omega$ in three-dimensional spaces whereas represents area in the two-dimensional spaces.

Recent literature \cite{HolcmanD.2004ETaS,SingerA.2006NEPI,CheviakovAlexei2010AAAO,Gomez} contains the various asymptotic results that have been inferred concerning the narrow escape problems use the method of matched asymptotic expansion. The problem for the sphere or another domain with a single trap has been solved approximately in \cite{Brow_leak}. For the two-dimensional spaces with a smooth boundary and a small surface trap, the asymptotic expansion for MFPT is obtained in \cite{HolcmanD.2004ETaS,SingerA.2006NEPI}. Furthermore, these asymptotic results have been extended for a two-dimensional domain having a cluster of absorbing windows in \cite{PillayS2010AAAO}.

For the case of the unit sphere with $N$ well-separated asymptotically small absorbing traps or windows at the surface, Cheviakov et al.  \cite{CheviakovAlexei2010AAAO} have obtained an three-term asymptotic expansion for MFPT applying the method of matched asymptotic expansions. The asymptotic results for some non-spherical domains have studied and results are presented in \cite{Gomez}.

Exact solutions for the sphere (or other domain) Poisson problems like \eqref{eqn:NEP_Dir_Neu} are not available except for trivial cases. For understanding these Poisson problems the asymptotic solutions are much better than nothing. These asymptotic results are given by relatively simple formulas, but their precision is by default unknown. In \cite{Ashton}, it has been shown that for the Poisson problem, we have no exact, but have decently good asymptotic and numerical solutions available that is validated by numerical comparisons.

In a detailed study by Cheviakov et al. \cite{Ashton} concerning the comparison between the asymptotic results and numerical results calculated with the help of a direct finite-difference numerical solution of the Poisson PDE \eqref{eqn:NEP_Dir_Neu}, it manifests that for two-dimensional
domains like unit square and disk, and for the unit sphere that the obtain asymptotic results in  \cite{PillayS2010AAAO,CheviakovAlexei2010AAAO}
provide an accurate approximations to the AMFPT $\bar{v}$ for a wide range of trap sizes. For the case of one trap of size $\epsilon$, it is shown that the asymptotic and numerical values of the AMFPT $\bar{v}$ agree within $1\%$ for the unit disk when $\epsilon \gsim 1.25$, for unit square $\epsilon \gsim 0.2$, and for the unit sphere $\epsilon \gsim 0.8$. The above comparison produces an evident conclusion that even if the relative error between the asymptotic and numerical results is finite, irrespective of asymptotically small trap size $\epsilon$ \cite{Ashton}. This research also elucidates the benefits and the necessity of having a theory based on the extensive asymptotic analysis. This work also shows the effects of trap separability and the asymptotic formula using the numerical solution of  \eqref{eq:NEP_pde}, since the asymptotic results are obtained for the cases of well-separable absorbing surface windows, i.e., $|x_i -x_j|= \mathcal{O}(1) \quad \forall \quad i \neq j$.

Identifying optimal trap configurations that minimize the average MFPT for various geometries is considered a fundamental problem with biophysical applications. In \cite{wardellipticdomain} the authors developed and implemented a simple but highly accurate, hybrid asymptotic-numerical theory to identify optimal trap configurations in near-disk domains and the ellipse. The work manifested in \cite{BernoffAndrewJ2018NAoD} explicates the importance of the asymptotic formulae and the techniques of numerical approximation that answer a bigger problem of how the spatial configuration of planar and spherical surfaces with absorbing surface pores affects the capture rate of the target. In \cite{LawleySeanD2020RODt} presents a comprehensive study for both clustered and uniform receptor configuration on detection of external stimuli and formulated an information-theoretic framework to quantify the role of these receptor organizations.

A real-valued continuous stochastic process $W=\{W_t:t \ge 0 \}$ with a continuous sample path is considered as a Brownian motion if, for every $t \ge 0$, $W_t$ is a Gaussian process having independent time increments. For modeling the simulations of a particle undergoing Brownian motion, we use the easily implemented, and computationally favorable Wiener process \cite{HeinzStefan1958-2011Mm}. For the simulations, we have utilized the fact that the particle path is given by summing the initial position of the particle and a series of normally distributed random displacements. The \verb|MATLAB| inbuilt function \verb|rand| is used to compute the normally distributed displacements, which is proficient in returning a matrix of normally distributed random numbers with standard deviation one. The simulated Brownian motion is applied for analyzing the narrow escape kinetics in the three-dimensional unit spherical domain with several boundary traps.

The primary objective of this work is to perform a direct numerical simulation of multiple particles undergoing Brownian motion in a three-dimensional spherical domain with boundary traps, and compute MFPT values by averaging of Brownian escape times and comparing the results with the asymptotic results obtained by solving the Poisson PDE problem with mixed Dirichlet-Neumann boundary conditions \eqref{eqn:NEP_Dir_Neu}. This comparison, which does show a good agreement, in some sense validates the narrow escape PDE problem \eqref{eqn:NEP_Dir_Neu} itself as an approximation (averaging) of the multiple physical Brownian motion runs. First, we are interested in knowing how many single-particle simulations are  required to match the predicted asymptotic averaged MFPT values. Second, having multiple simulated Brownian particle trajectories at our disposal, it is of interest to study other statistics of narrow escape dynamics that include dynamics of the Brownian particle's in bulk and near the boundary by estimating the average percentage of time spent by Brownian particle near the domain boundary for both the anisotropic and isotropic diffusion. Our study numerically exhibited that Brownian particles stick to the boundary surface for both isotropic vs.~anisotropic near-boundary diffusion and travel along with it for a while before getting trapped. More precise mathematical justification can be found in \cite{LawleySeanD2019HRSD}, where the authors worked on a mathematical model of a three-dimensional anisotropic diffusion equation subjected to boundary conditions that are described by stochastic differential equations, which is solved using techniques involving the asymptotic analysis and probabilistic methods.

This paper is structured as follows. In \S \ref{sec:asymptotic_result}, we compare asymptotic results with the average Brownian escape times, which is computed with the help of a Brownian dynamics simulations performed by \verb|MATLAB| based code. In \S \ref{sec:asymptotic_result_mfpt} and \S \ref{sec:asymptotic_result_3d}, we review some asymptotic results for the narrow escape problem in the three-dimensional unit sphere with $N$ small absorbing surface windows. In \S \ref{sec:simulated_brow_motion}, we outline the idea, and scientific theories required to develop a efficient \verb|MATLAB| based code, which can model and collect three-dimensional trajectories of the Brownian particles launched from various positions specified by $(r,\phi)$, where $r$ is the radial distance and $\phi$ represents the spherical angle, up to their escape through a given trap, taking into account the boundary reflections rigorously. In \S \ref{sec:escape_condition}, we present some of the mathematical inequalities that are used for the \verb|MATLAB|-based code to ensure that Brownian particle escapes from any of the absorbing windows. In \S \ref{sec:Comparision}, we will examine and compare the asymptotic MFPT with the Brownian dynamics simulations results. In \S \ref{sec:accuracy}, we review the relative percentage difference error $\delta v(x)$, which can measure the accuracy of the results of the simulations by comparing the full simulations with the asymptotic results for the one-trap and the two-trap configurations. The study reveals that the asymptotic results are mostly within $1\%$ accuracy even when the radius of the boundary traps is $10^{-2}$. We also show that $10^{4}$ is the optimal number of iteration or runs needed for computing averaged Brownian escape times from the Brownian simulations in a unit spherical domain.

In \S \ref{sec:boundary_dynamics}, we manifest the study concerning the Brownian dynamics of a particle near boundary region titled as the $\delta$ region. Moreover, we also determine that for the unit sphere with the multiple trap configurations, the fraction of time spent by Brownian particles inside various ranges of $\delta$ region is constant and does not depend upon the launching positions. Also, \S \ref{sec:boundary_dynamics} presents a comprehensive study of results obtained from the Brownian dynamics simulations in the unit sphere having several boundary traps. It is shown that the Brownian particles spend more in the boundary layer than predicted by the boundary layer relative volume, with the effect being more pronounced in a narrow layer near the spherical wall.

In \S \ref{sec:var_vs_constant_diffisuion}, we extend our work and use our working \verb|MATLAB| code to present the study of the Brownian dynamics simulation for computing MFPT and boundary dynamics of a particle for both the anisotropic and isotropic diffusion. In \S \ref{sec:var_vs_constant_diffisuion_escape_time}, we compare the asymptotic results involving computation of MFPT with anisotropic diffusion and isotropic diffusion. In \S \ref{sec:var_vs_constant_diffisuion_boundary_time}, we  exhibits the comparison of Brownian boundary dynamics for both the anisotropic diffusion and isotropic diffusion. The Brownian dynamic simulation is modified to take into account anisotropic variable discussion \cite{Lin,booklindif} that is a better physical approximation in close proximity to the domain wall. Effects of this discussion variability and anisotropy on the escape times (MFPT) and time spent near the boundary are studied. It is found that accounting for the modified near-wall discussion leads to smaller escape times than predicted by the PDE model \eqref{eqn:NEP_Dir_Neu}, and a larger time spent by particles near the boundary.

The last and final \S \ref{sec:conclusion_disc} includes a review of the work, obtained results, and some open problems.

All numerical computations have been performed with a \verb|MATLAB|-based code \cite{GithubCode} developed in the current work.

\section{Comparison of Asymptotic and Brownian Simulation Result for the Three-Dimensional Unit Sphere}\label{sec:asymptotic_result}
\subsection{Asymptotic Formulas and Results for the Mean First-Passage Times}\label{sec:asymptotic_result_mfpt}
The current section exhibits some asymptotic results associated with the MFPT for the three-dimensional unit sphere with $N$ well-separated locally circular absorbing surface windows, which has been discussed and derived using the method of matched asymptotic expansions (MMAE) by Cheviakov et al. \cite{CheviakovAlexei2010AAAO}.

If $y(x)$ represents the solution of any singularly perturbed boundary layer problem, an asymptotic expansion for $y(x)$ can be approximated using the MMAE. To employ this method, one first identifies the region of rapid change in the domain. Next, one tries to find a solution in a region that is away from the region of rapid change known as the outer solution, say $y_{outer}(x)$, which has valid representation in the outer region. Later, one tries to construct another solution in a region inside the region of rapid change known as the inner solution, say $y_{inner}(X)$ defined in the inner region in terms of inner-variable or stretched variable $X$. The uniform or composite solution $y(x)$ is obtained by matching both the inner and outer solutions in some overlapped regions. The matching condition was first proposed by Ludwig Prandtl's in 1905.

Let us consider a domain $\Omega \subset \mathbb{R}^3$, having a trap on the surface, having it's center at $x_j$. The stretch variable $y = \epsilon^{-1} (x-x_j)$ is used to describe an asymptotic expansion for the inner solution near the trap center $x_j$, i.e., $x-x_j = \mathcal{O}(\epsilon)$
\begin{align*}
v(x) = u(y) = u_0 + u_1 + u_2+ \dots
\end{align*}
In \cite{CheviakovAlexei2010AAAO}, it has been explained that the terms in the asymptotic expansion of the inner solution are proportional to $\epsilon^{p} \log \epsilon$, where $p \in \mathbb{Z}$, i.e.
\begin{align*}
u_0 = \mathcal{O}(\epsilon^{-1}), \quad u_1 = \mathcal{O}(\log \epsilon), \quad u_2 = \mathcal{O}\Big(\epsilon^0 \Big), \quad \dots.
\end{align*}
Furthermore, the asymptotic expansion of the outer solution has a valid representation in a region away from the trap, $|x-x_j| \gg \epsilon$, and it is given by
\begin{align}\label{eq:outer_sol}
v(x) =  v_0 + v_1 + v_2+ \dots.
\end{align}
The terms in the asymptotic expansion of the outer solutions are systematically obtained by substituting the inner $u(y)$ and outer $v(x)$ in the Poisson's PDE \eqref{eq:NEP_pde} and boundary conditions \eqref{eq:NEP_pde_bvp}, and further matching both expressions in some overlapped region where $x \to x_j$ and $y \to \infty$.

To calculate the correction and higher-order terms of the outer solution, define Neumann Green's function $G(x;x_j)$ for $\Omega \subset \mathbb{R}^3$, which has a singularity at $x_j$, which lies on the surface of $\Omega$. For specific details, see \cite{CheviakovAlexei2010AAAO}. The Green function $G(x;x_j)$ satisfies the PDE along with the boundary conditions
\begin{equation}\label{eq:neuman_green_func}
\begin{array}{ll}
\Delta G=\dfrac{1}{|\Omega|},\quad x\in \Omega; \\[3ex]
\partial_n G=0,\quad x\in \partial \Omega \setminus\{\xi\}; \\[3ex]
\quad \displaystyle \int_{\Omega}G \hspace{0.03in} dx=0,
\end{array}
\end{equation}
where $\partial_n$ represents the normal derivative to $\partial \Omega$. $\mathcal{H}_m$ is the mean curvature of the boundary at $\xi \in \partial \Omega$ and $R(\xi;\xi)$ represents the bounded regular part of the $G(x;\xi)$. The above PDE admits a unique solution for the domain $\Omega \subset \mathbb{R}^3$ with a smooth boundary for $x \to \xi$, which is defined as
\begin{equation}\label{eq:green_func_general_3d}
G(x;\xi)=\dfrac{1}{2\pi|x-\xi|} - \dfrac{\mathcal{H}_m}{4 \pi} \log |x-\xi|+R(\xi;\xi).
\end{equation}

For $\epsilon \to 0$, the MFPT, $v(x)$ in the outer region $|x-x_j| \gg \epsilon$, by making use of the Neumann-Green's function and the average MFPT \eqref{avg_mfpt_nep} for a domain $\Omega$ with $N$ well-separated absorbing windows on the boundary is given by
\begin{align}\label{eq:kappa_j}
v(x) = \bar{v} + \sum_{i=1}^{N} \kappa_j G(x;x_j) + \mathcal{O}(\epsilon),
\end{align}
where for each $j=\{1,\dots,N\}$, $\kappa_j$ is the constant that depends upon the trap size $\epsilon$, trap arrangements, domain shape, and boundary is derived in Ref. \cite{CheviakovAlexei2010AAAO,Ashton}.

\subsubsection{A Case of the Three-Dimensional Unit Sphere}\label{sec:asymptotic_result_3d}

This section sketches some of the asymptotic formulae for the NEP in the three-dimensional unit sphere. Consider $\Omega$ being the unit sphere centered at the origin having $N$ locally, well-separated absorbing surface windows. Each absorbing window has a center at $x_j$, which lies on the surface of the unit sphere and has the radius $\epsilon a_j$, where $j \in  \{ 1,2,\dots,N\}$.

All these traps are distinguished by a physical quantity known as electrostatic capacitance $c_j$, which can be obtained by solving a local PDE problem which is a  tangent plane approximation to the surface of $\Omega$ at $x_j$
\begin{equation}\label{eq:tangen_plane_approx}
\begin{array}{ll}
u_{\xi_1 \xi_1}+u_{\xi_2 \xi_2} + u_{\xi_3 \xi_3} = 0, \quad \xi_1 \ge 0, \quad -\infty < \xi_2,\xi_3 < \infty, \\[3ex]
u=1, \quad \text{on} \quad \xi_1 =0, \quad \xi_2^2 + \xi_3^2 < a_j^2; \\[3ex]
u_{\xi_1} = 0, \quad \text{on} \quad  \xi_1=0, \quad \xi_2^2 + \xi_3^2 > a_j^2, \\[3ex]
u \sim \dfrac{c_j}{|\xi|}, \quad \text{as} \quad |\xi| \to \infty,
\end{array}
\end{equation}
where $\xi= (\xi_1,\xi_2,\xi_3).$
The above \eqref{eq:tangen_plane_approx} does not have an analytic solution for an absorbing window of arbitrary shape.

\begin{figure}[htbp]
  \centering
  \includegraphics[width=10cm,height=7cm]
  {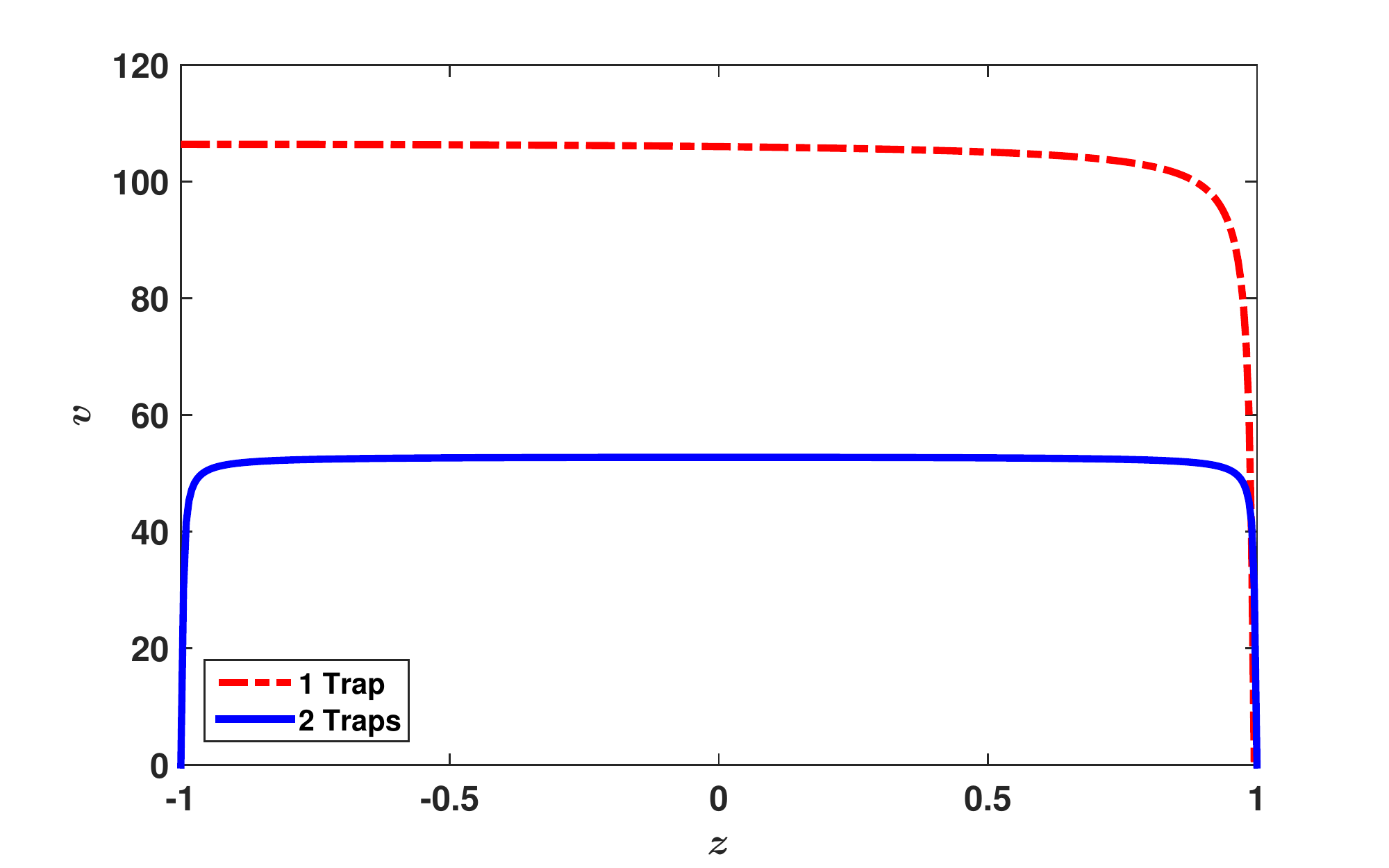}
\caption{ MFPT for the unit sphere with several traps arrangements of sizes $\epsilon=0.01$ when a Brownian particle is launched along the $Z$-axis. }
\label{fig:alltrap_mfpt}
\end{figure}

In \cite{CheviakovAlexei2010AAAO}, the surface Neumann-Green's function for the three-dimensional unit sphere with $N$ small well separated absorbing windows on the boundary of the sphere each centered at $x_j$ with $j =1,2,\dots,N$ is derived and defined as
\begin{equation}\label{eq:green_func}
G_s (x;\xi)= \dfrac{1}{2 \pi |x- \xi|} + \dfrac{1}{8 \pi} (|x|^2 +1) + \dfrac{1}{4 \pi} \log \Big( \dfrac{2}{1 - |x| \cos \gamma+ |x-\xi|} \Big)-\dfrac{7}{10 \pi},
\end{equation}
where the angle between the vector $x \in \Omega$ and $\xi \in \partial \Omega$ is labeled as $\gamma$ and $|x| \cos \gamma$ equals $\langle x ,\xi \rangle$ for Euclidean inner product $\langle,\rangle$.

The self-interaction term corresponding to the above surface Neumann-Green's function $\eqref{eq:green_func_general_3d}$ is given by
\begin{equation}\label{eq:self_intera_term}
R(\xi;\xi) = - \dfrac{9}{20 \pi}.
\end{equation}
To define the MFPT and average MFPT, $\kappa_j$ \eqref{eq:kappa_j}, is expressed in terms of the average capacitance
\begin{align*}
\bar{c} = \dfrac{c_1+ \dots + c_n}{N},
\end{align*}
consequently, the $\kappa_j$ is given by
\begin{align*}
\kappa_j = \dfrac{c_j}{2} \Big[ 2 \log 2 - \dfrac{3}{2} + \log a_j \Big].
\end{align*}

For $\epsilon \to 0$, the detailed analysis in \cite{CheviakovAlexei2010AAAO} has shown that the MFPT $v(x)$ for the unit sphere with $N$ well separated locally traps defined in the outer region is defined as
\begin{equation}\label{eq:mfpt_nep}
v(x)=\bar{v}-\dfrac{|\Omega|}{ DN\bar{c}}\sum_{j=1}^{N}c_jG_s(x;x_j)+\mathcal{O}(\epsilon\log \epsilon).
\end{equation}

Moreover, for $\epsilon \to 0$, the average MFPT $\bar{v}$ equals
\begin{equation}\label{avg_mfpt_nep}
\bar{v}=\dfrac{|\Omega| }{2\pi \epsilon DN\bar{c}}\bigg[ 1+\epsilon \log\left(\dfrac{2}{\epsilon}\right)\dfrac{\sum_{j=1}^{N}c_j^2}{2N\bar{c}}+\dfrac{2\pi\epsilon}{ N\bar{c}}p_c(x_1,\dots,x_N)- \dfrac{\epsilon}{ N\bar{c}}\sum_{j=1}^{N}c_j\kappa_j+\mathcal{O}(\epsilon^2\log\epsilon)\bigg].
\end{equation}
In expression for average MFPT \eqref{avg_mfpt_nep}, the terms of order $\mathcal{O}(\epsilon)$, depends upon the energy-like functions
\begin{align*}
p_c(x_1,\dots,x_N)=\mathcal{C}^{T}\mathcal{G}_s \mathcal{C},
\end{align*}
which is defined in terms of the capacitance vector $C$
\begin{align*}
\mathcal{C} = (c_1,\dots,c_N)^T,
\end{align*}
and the Green's matrix $\mathcal{G}_s$
\begin{equation}\label{eq:green_matrix}
\mathcal{G}_s \equiv\begin{pmatrix}R &G_{s12}&\dots&G_{s1N}\\G_{s21}&R&\dots&G_{s2N}\\\vdots&\vdots&\ddots&\vdots\\G_{sN1}&\dots&G_{sN,N-1}&R\end{pmatrix} , \quad
R\equiv R(x_j;x_j)=-\dfrac{9}{ 20\pi}, \quad G_{sij}\equiv G_s(x_i;x_j).
\end{equation}

The formula concerning the average MFPT is explicitly shown for the unit sphere with $N$ well-separated circular traps with common radius $a_j=1$ in \cite{Ashton}. The PDE problem \eqref{eq:tangen_plane_approx}, corresponding to the circular traps is classified as the electrostatic problem and its solution is the electrostatic capacitance, which is given by
\begin{align*}
c_j = \dfrac{2 a_j}{\pi},
\end{align*}
for each $j=1,\dots,N$. Correspondingly, the average MFPT $\bar{v}$ is
\begin{equation}\label{eq:avg_mfpt_cir_traps}
\bar{v}\sim \dfrac{|\Omega|}{ 4\epsilon DN}\bigg[1+\dfrac{\epsilon}{ \pi}\log\left(\dfrac{2}{\epsilon}\right)+\frac{\epsilon}{\pi}\bigg(-\frac{9N}{5}+2(N-2)\log 2+\dfrac{3}{2}+\dfrac{4}{N}\mathcal{H}(x_1,\dots,x_N)\bigg)\bigg].
\end{equation}
In this expression the interactive energy term $\mathcal{H}(x_1,\dots,x_N)$ is
\begin{equation}\label{eq:interact_cir_trap}
\mathcal{H}(x_1,\dots,x_N)=\sum_{i=1}^{N}\sum_{j=i+1}^{N}h(x_i;x_j),
\end{equation}
whereas the pairwise interaction energy is given by
\begin{equation}\label{eq:pairwiseinteract_cir_trap}
h(x_i;x_j)=\dfrac{1}{|x_i-x_j|}-\dfrac{1}{2}\log|x_i-x_j|-\dfrac{1}{2}\log(2+|x_i-x_j|).
\end{equation}

In \cite{Ashton}, it is also revealed that the average MFPT for a case of the unit sphere with two kinds of traps, specified by two kinds of radii $a_j=1$ for $j=1,\dots,N$ and $a_j=\alpha$ for $j=N+1,\dots,2N$ is given by
\begin{equation}\label{eq:avg_mfpt_diff_trap_size}
\bar{v} \sim \dfrac{|\Omega|}{4 \epsilon DN (1+\alpha)} \Big[ 1+ \dfrac{\epsilon}{\pi} \log \Big( \dfrac{2}{\epsilon}\Big) \Big( \dfrac{1+\alpha^2}{1+\alpha} \Big) +\dfrac{\epsilon}{\pi} \Big( S+ \dfrac{4}{N(1+\alpha)} \tilde{\mathcal{H}}(x_1,\dots,x_N) \Big) \Big].
\end{equation}

\begin{figure}[H]
\begin{subfigure}{0.45\textwidth}
 \includegraphics[width=\textwidth,height=6cm]{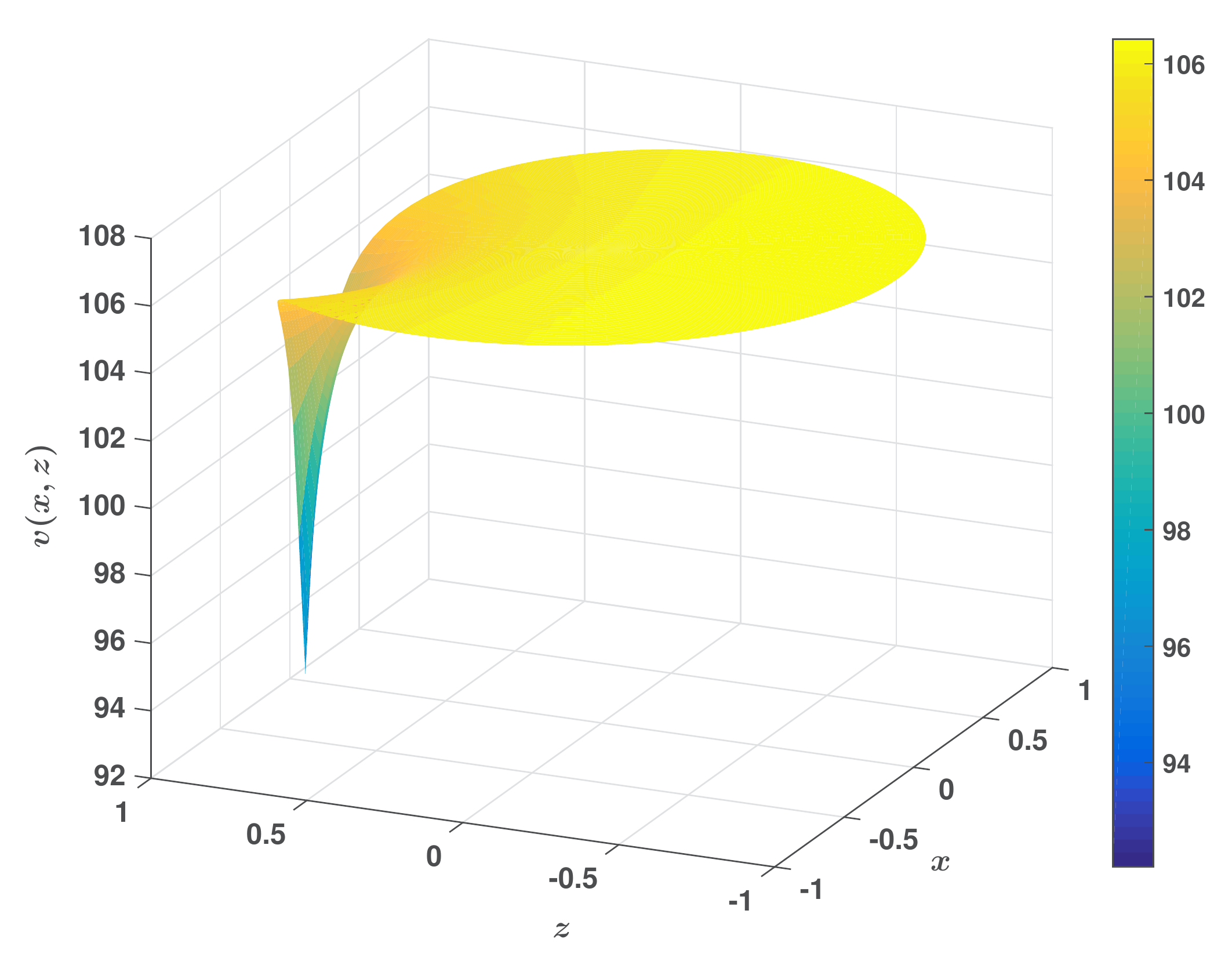}
 \caption{\label{fig:asym_res_1trap_surface}}
\end{subfigure}
\hfill
\begin{subfigure}{0.45\textwidth}
\includegraphics[width=\textwidth,height=6cm]{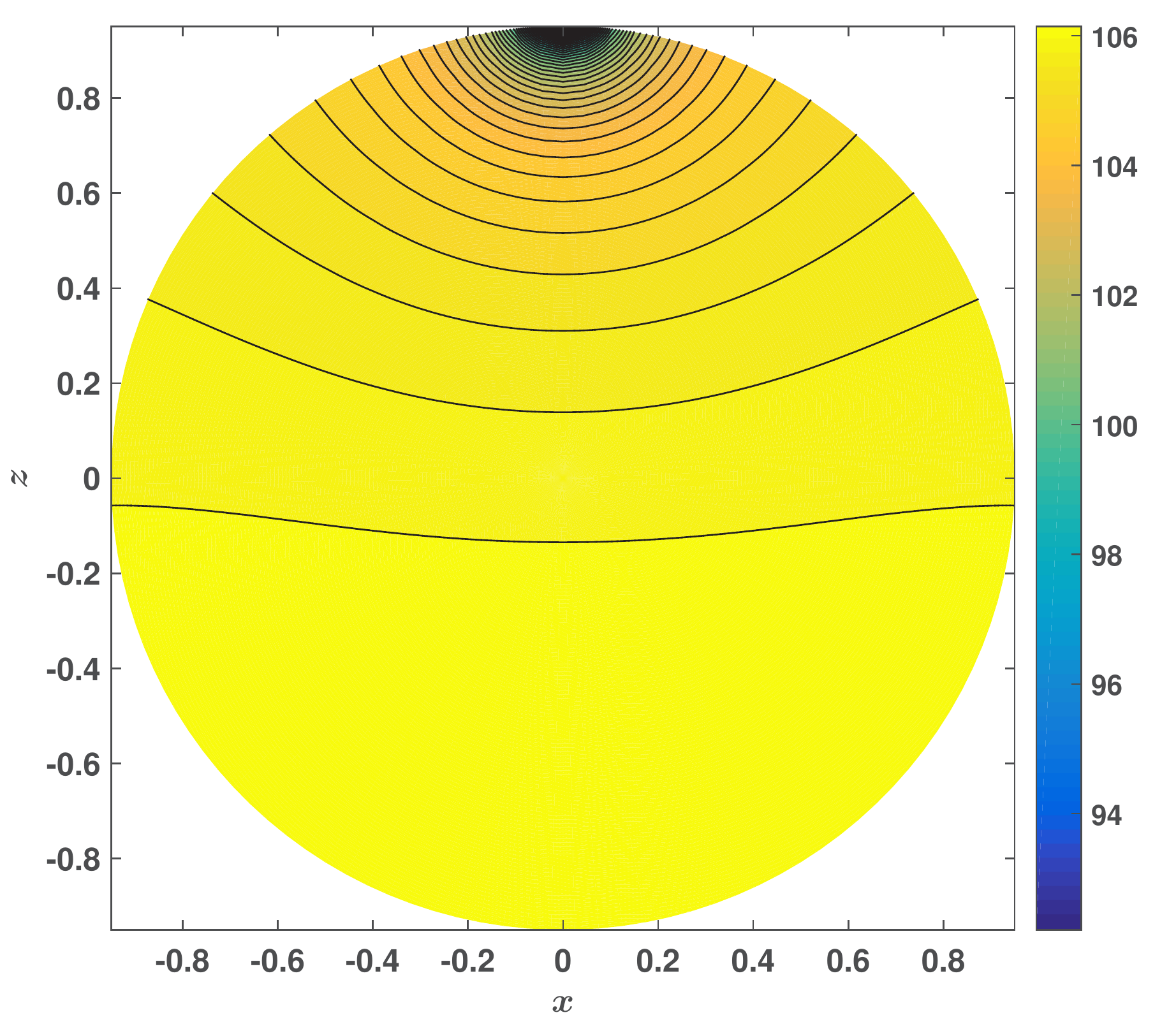}
\caption{\label{fig:asym_res_1trap_contour}}
\end{subfigure}
\hfill
\caption{Surface (\subref{fig:asym_res_1trap_surface}) and contour (\subref{fig:asym_res_1trap_contour}) plots of the asymptotic MFPT $v(x)$ for one boundary trap having length $\epsilon=0.01$, and located at $(0,0,1)$.}
\label{fig:asym_res_1trap}
\end{figure}

\begin{figure}[H]
\begin{subfigure}{0.45\textwidth}
 \includegraphics[width=\textwidth,height=6cm]{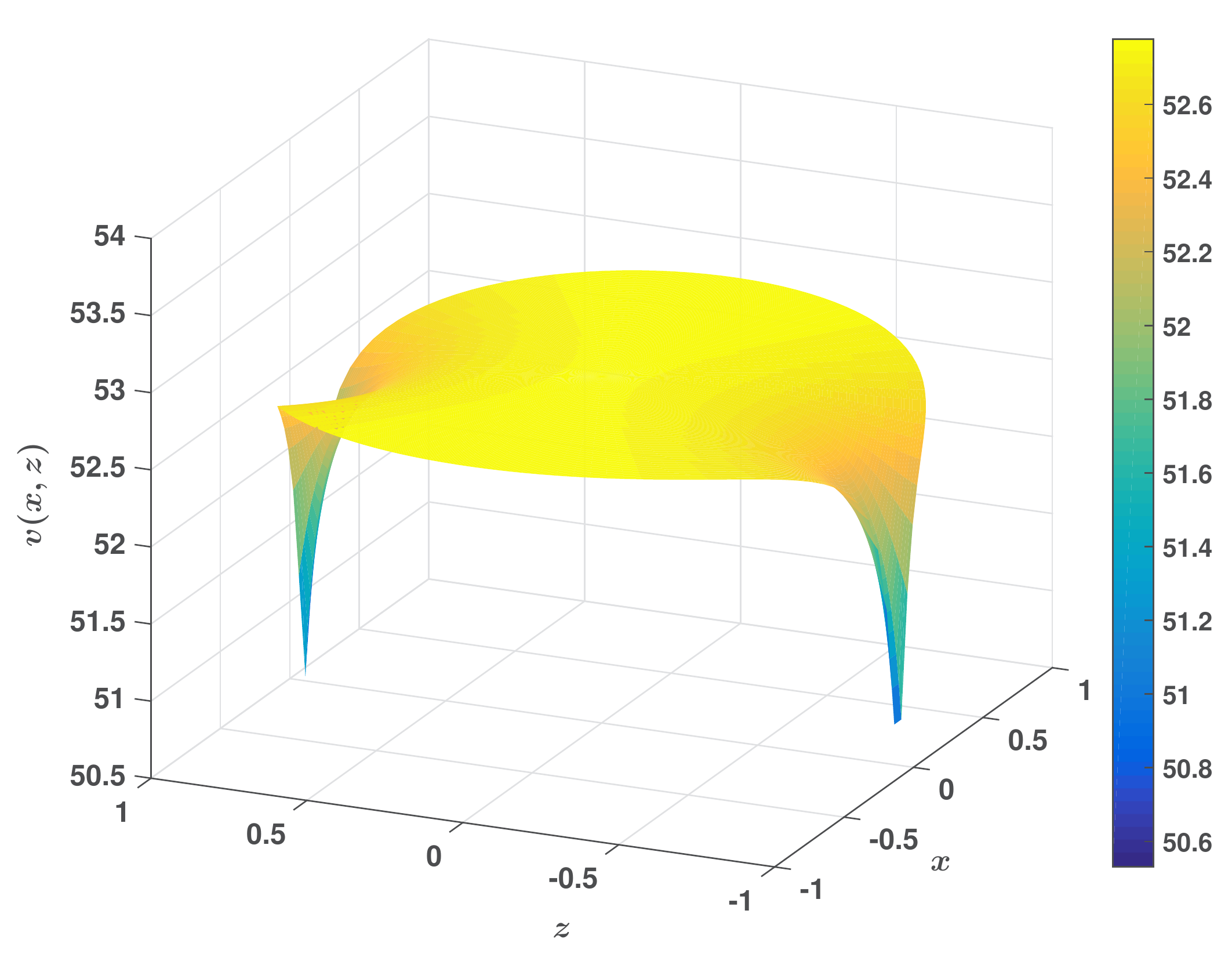}
 \caption{\label{fig:asym_res_2trap_surface}}
\end{subfigure}
\hfill
\begin{subfigure}{0.45\textwidth}
\includegraphics[width=\textwidth,height=6cm]{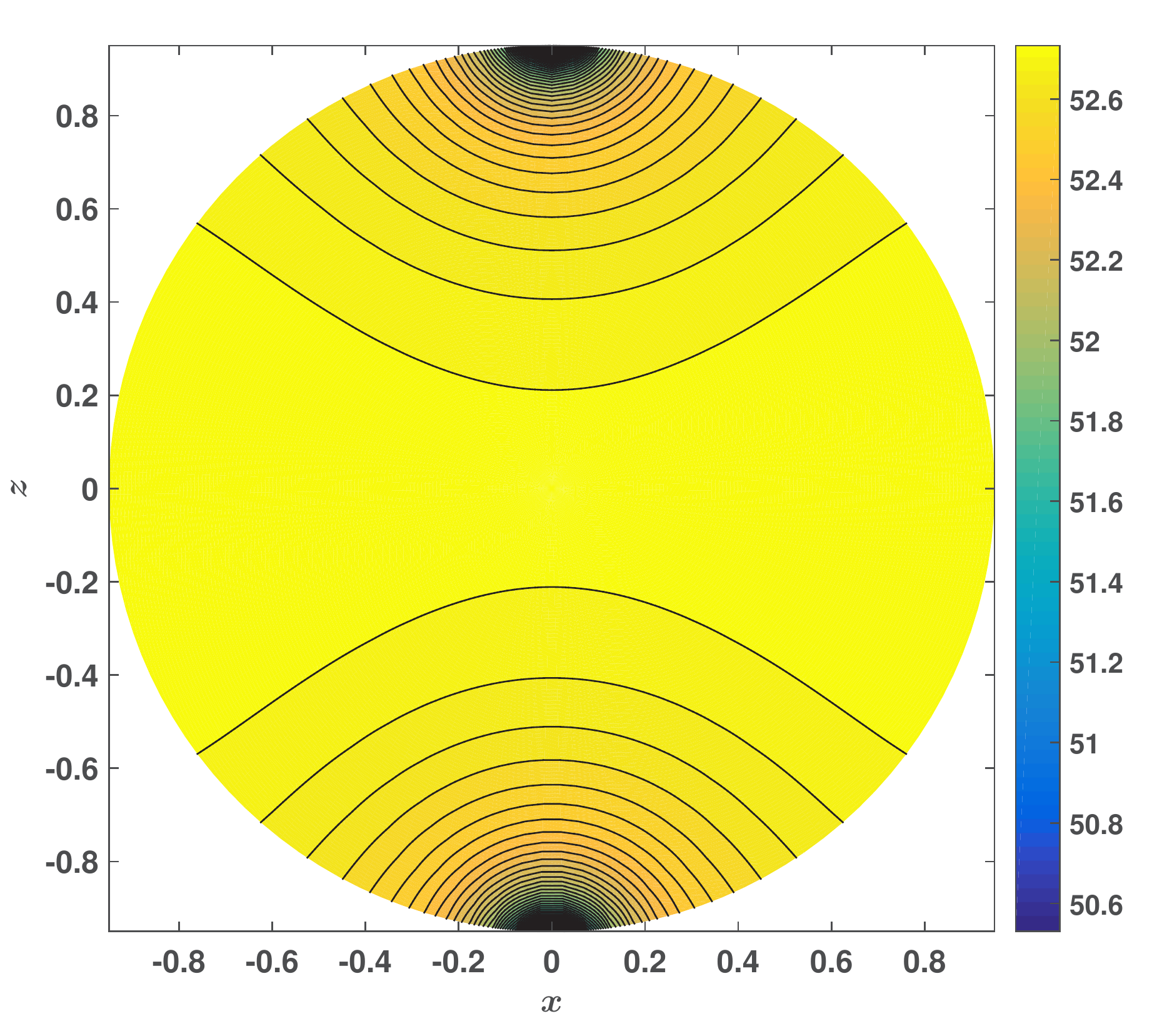}
\caption{\label{fig:asym_res_2trap_contour}}
\end{subfigure}
\hfill
\caption{Surface (\subref{fig:asym_res_2trap_surface}) and contour (\subref{fig:asym_res_2trap_contour}) plots of the asymptotic MFPT $v(x)$ for two boundary traps having common length $\epsilon=0.01$, and located at $(0,0,1)$ and $(0,0,-1)$.}\label{fig:asym_res_2trap}
\end{figure}

In this particular case, the elements of Green's matrix
$\mathcal{G}_s$ is given by \eqref{eq:green_matrix},
and the capacitance vector $\mathcal{C}$ becomes
\begin{align*}
\mathcal{C} = \dfrac{2}{\pi} (1,\dots,1,\alpha,\dots,\alpha)^T.
\end{align*}
Further, the constant $S$ in \eqref{eq:avg_mfpt_diff_trap_size} is
\begin{align*}
S= - \dfrac{9}{5} N (1+\alpha) +2 \log 2 \Big( (N-2)(1+\alpha) + \dfrac{4 \alpha}{1+\alpha} \Big) + \dfrac{3}{2} \Big( \dfrac{1+\alpha^2}{1+\alpha} \Big) -\dfrac{\alpha^2}{1+\alpha} \log \alpha,
\end{align*}
whereas the interaction energy term is
\begin{equation}\label{eq:intera_diff_trap_size}
\tilde{\mathcal{H}}(x_1,\dots,x_N) = \sum_{i=1}^{N}\sum_{j=i+1}^{N} h(x_i;x_j)+ \alpha \sum_{i=1}^{N}\sum_{j=N+1}^{2N} h(x_i;x_j)+ \alpha^2 \sum_{i=N+1}^{2N}\sum_{j=i+1}^{2N} h(x_i;x_j),
\end{equation}
here $h(x_i;x_j)$ is defined by \eqref{eq:pairwiseinteract_cir_trap}.

For a better understanding of the asymptotic MFPT $v(x)$ \eqref{eq:mfpt_nep}, we have plotted the MFPT $v(x)$ for the one-trap and the two-trap configuration, where the boundary traps have the radius $10^{-2}$ (See Fig.~\ref{fig:asym_res_1trap} and \ref{fig:asym_res_2trap}).

\subsection{Simulated Brownian Motion}\label{sec:simulated_brow_motion}

\begin{figure}[htbp]
\begin{subfigure}{0.45\textwidth}
 \includegraphics[width=\textwidth,height=7cm]{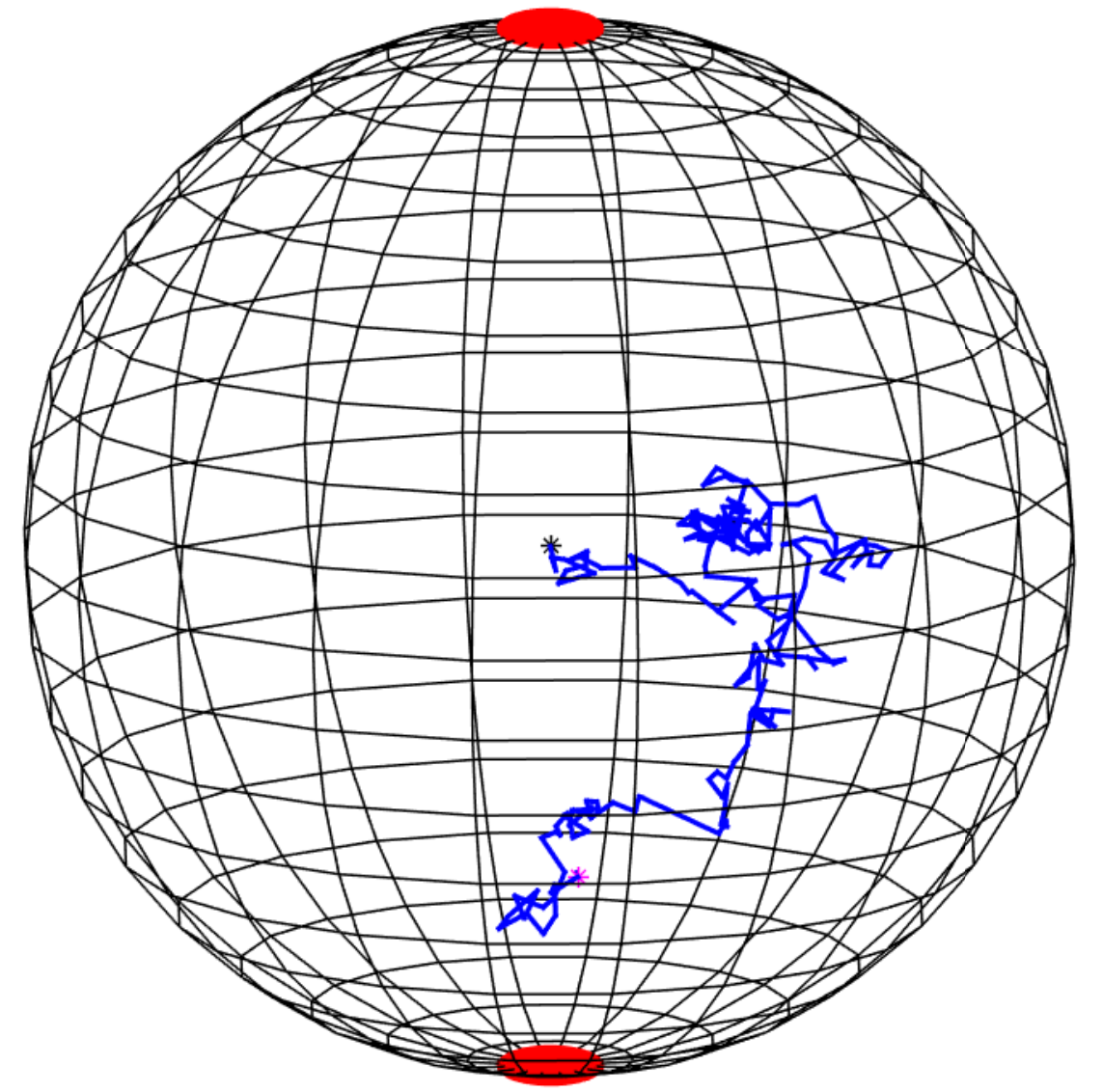}
 \caption{\label{fig:code_schema_1}}
\end{subfigure}
\hfill
\begin{subfigure}{0.45\textwidth}
\includegraphics[width=\textwidth,height=7cm]{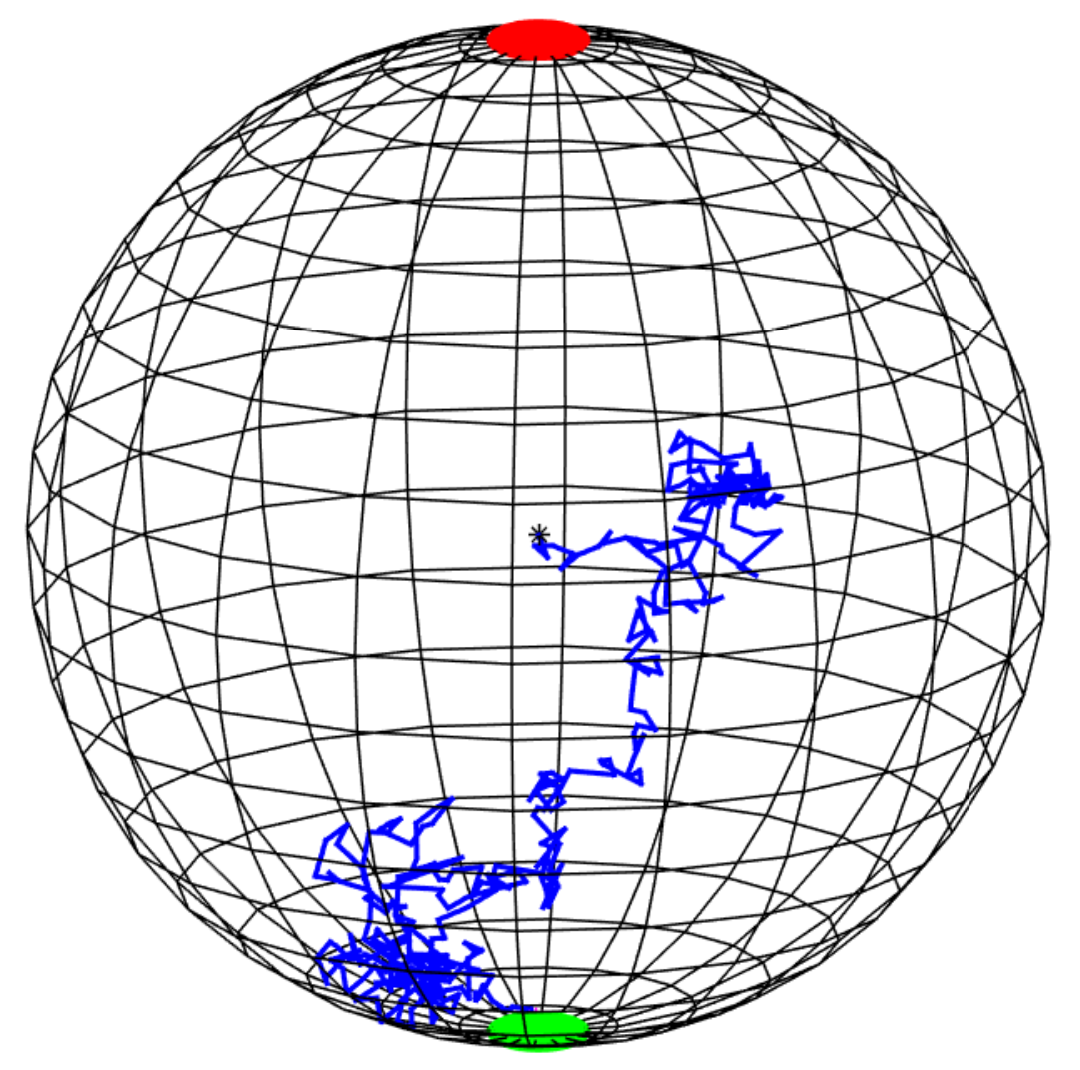}
\caption{\label{fig:code_schema_2}}
\end{subfigure}
\hfill
\caption{(\subref{fig:code_schema_1}, \subref{fig:code_schema_2}) Schematics of simulation of a Brownian particle for the narrow escape problem in the three dimensional unit sphere having absorbing windows denoted by red, leading to an escape through a absorbing window denoted by green.}
\label{fig:code_schema}
\end{figure}

This section mulls over the scientific and analytical ideas utilized to develop an efficient \verb|MATLAB| based code that is competent in simulating the Brownian motion, retaining the account of trajectories of a Brownian particles up to their escape through a trap arrangement, and taking into account boundary reflections rigorously (see Fig.~\ref{fig:code_schema}). The trap configuration is merely another terminology for locations of multiple traps on the surface of the domain, which satisfies the boundary conditions of \eqref{eq:NEP_pde_bvp}.
In this work, we successfully develop a \verb|MATLAB| based code to perform a direct numerical simulation of various particles, undergoing Brownian motion in the three-dimensional spherical domain with the boundary traps, to compute MFPT by averaging of Brownian escape times. We also predict how many single-particle simulations are needed to match the foretold asymptotic averaged MFPT values closely. The code is also employed to examine the boundary dynamics of the Brownian particle's trajectories in $\delta$ regions (See \S.~\ref{sec:boundary_dynamics}).

For the modeling of the Brownian dynamics simulations, we use the Weiner process model \cite{HeinzStefan1958-2011Mm}. The model concerning the Weiner process is interpreted in terms of a linear stochastic first-order difference equation.
\begin{align*}\label{eq:weiner_SDE}
    x_{n} = x_{n-1} + r \epsilon_{n-1}, \quad n \ge 0
\end{align*}
where $x_n$ represents the position of a particle undergoing Brownian motion, and $r$ is a dimensionless parameter that controls the strength of randomness, and  $\epsilon_{n-1}$ is a normally distributed random variable, i.e. $\epsilon_{n-1} \sim \mathcal{N}(0,1)$.
The normally distributed displacements are determined using the \verb|MATLAB| inbuilt function \verb|rand|, which is proficient in returning the matrix of normally distributed random numbers with standard-deviation $1$. The simulated Brownian motion helps in analyzing the narrow escape kinetics in the three-dimensional unit spherical domain with several boundary trap configurations.

The mean square displacements of any particle undergoing Brownian motion is given by
\begin{equation}\label{eq:weiner_process}
 \Delta x_n=\sqrt{3D\tau}\, \gamma_{n-1}; \quad \gamma_{n} \sim \mathcal{N}(0,1).
\end{equation}
where $D$ is the constant of the diffusivity. It is evident from the \eqref{eq:weiner_process} that the re-scaling of the Brownian process is proportional to the $D$. For specific details, see \cite{HeinzStefan1958-2011Mm}.

\medskip\noindent\textbf{Boundary reflections:} In order to that ensure the code works, we have to make sure that the particle is reflected at the boundary. We are interested in studying the confinement of Brownian particles inside the unit sphere. Therefore, to restrain the particle inside the unit sphere, we impose some constraints. A trivial constraint to ensure the confinement is whether the norm of the particle's position $x_{n-1}$ is strictly less than one, i.e., $|x_{n-1}|<1$. Let $\lambda$ symbolizes the displacement, and if $|x_{n-1}+\lambda| \ge 1$, then we find the intersection point say $u=(u_x,u_y,u_z)$ on the surface where the particle hits and reflects back. The displaced path of the particle can be explicitly expressed as
\begin{equation}\label{eq:boudary_reflection_intersection}
    |u| = 1, \quad u= x_{n-1}+ \alpha \dfrac{\lambda}{|\lambda|}.
\end{equation}
The boundary intersection point $u$ can be obtained by expressing $\alpha$ in terms of known $x_{n-1}$ (initial position) and $\lambda$ (displacement of the particle). The coefficient $\alpha$ is found to be
\begin{equation*}
    \alpha = \sqrt{1+\kappa^2-|x_{n-1}|^2}-\kappa
\end{equation*}
where,
\begin{equation*}
    \kappa = \dfrac{ \langle x_{n-1}, \lambda \rangle}{|\lambda|}.
\end{equation*}

\begin{figure}
    \centering
    \includegraphics[width=3in,height=3in]{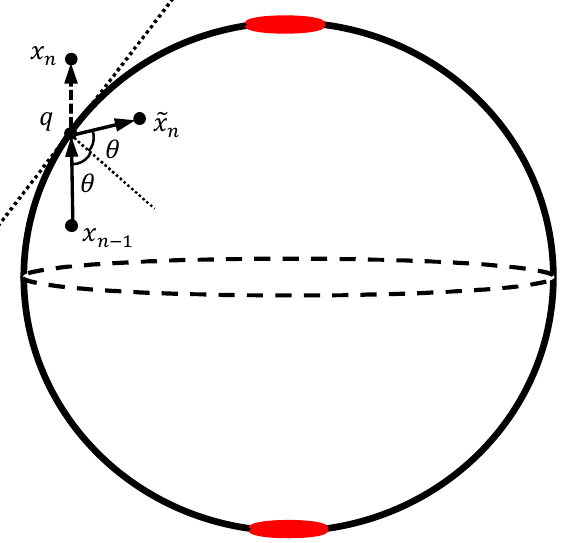}
    \caption{Schematics of the Brownian particle's boundary reflection dynamics in a three-dimensional unit sphere having circular absorbing window (denoted by color red).}
    \label{fig:boundary_reflection}
\end{figure}
The equation \eqref{eq:boudary_reflection_intersection} demonstrates that we can express the path of the particle after deflection in terms of the initial position, and displacement of the particle. Mathematically, this can be done by rotation and scaling of the displacement vector $\lambda$, such that
\begin{equation}
    |x_n-u| = |\lambda - u|, \quad \theta'=\theta,
\end{equation}
where $\theta'$ is the angle between $x_n$ and $u$, and $\theta$ is the angle between $x_{n-1}$ and $u$. Moreover, we can also express the deflected particle's position $x_{n}$ in terms of its preceding position $x_{n-1}$, and displacement $\lambda$, using the rotation matrix $\mathcal{R}(\Tilde{\theta})$, with angle of rotation $\Tilde{\theta}=180^{\circ}$, and axis of rotation $u=(u_x,u_y,u_z)$, i.e.,
\begin{equation*}
    x_n = u - \mathcal{R}(\Tilde{\theta}) |\lambda - u| \dfrac{\lambda}{|\lambda|},
\end{equation*}

where $\mathcal{R} (\Tilde{\theta})$ at $180^{\circ}$ is given by
\begin{align*}
\mathcal{R} (180^{\circ})  = \begin{pmatrix} 2u_{x}^{2}-1 & 2u_{x}u_{y} & 2u_{x}u_{z} \\
2u_{y}u_{x} & 2u_{y}^{2}-1 & 2u_{y}u_{z}\\
2u_{z}u_{x} & 2u_{z}u_{y} & 2u_{z}^2-1
\end{pmatrix}.
\end{align*}

The general rotation matrix $\mathcal{R}(\Tilde{\theta})$ about an arbitary axis is represented by
\begin{align*}
\mathcal{R} (\Tilde{\theta}) =
\begin{pmatrix}
cos\Tilde{\theta}+ u_{x}^2(1-cos\Tilde{\theta}) & u_{x}u_{y}(1-cos\Tilde{\theta})-u_{z}sin\Tilde{\theta} & u_{x}u_{z}(1-cos\Tilde{\theta}) + u_{y}sin\Tilde{\theta} \\
u_{x}u_{y}(1-cos\Tilde{\theta})+ u_{z} sin\Tilde{\theta} & \cos \Tilde{\theta} + u_{y}^2(1-cos\Tilde{\theta}) & u_{y}u_{z}(1-cos\Tilde{\theta})- u_{x}sin\Tilde{\theta}\\
u_{z}u_{x}(1-cos\Tilde{\theta}) - u_{y}sin\Tilde{\theta} & u_{z}u_{y}(1-cos\Tilde{\theta}) + u_{x}sin\Tilde{\theta} & cos\Tilde{\theta}+ u_{z}^2(1-cos\Tilde{\theta})
\end{pmatrix}.
\end{align*}

\medskip\noindent\textbf{Parameters used in the code:}
To simulate the Brownian simulation, we have chosen the time step to be $6 \times 10^{-6}$. The \verb|MATLAB| code is optimized with parallel computing and vector operations, which allows the code to increase its computational speed and efficiency. The trap radii $\epsilon$ for each trap configuration (See Fig.~\ref{fig:trap_config}) is chosen as $10^{-2}$, which is $1\%$ of the domain's radius. By re-scaling the PDE \eqref{eq:NEP_pde}, we have selected $D$ as $1$. We also determine that $10^4$ is the optimal number of iteration or runs chosen for computing averaged Brownian escape times from the direct numerical simulations in the unit spherical domain with boundary traps.

\medskip\noindent\textbf{Specifications of the machine to run the Brownian dynamics simulations:}
The simulations are run on the machine with operating system Red Hat Linux $7.6$, with the processor Intel Xeon(R) E5-2687W (3.10GHz) $\times$ 16 and memory $128$ GiB.

\subsection{Escape Conditions for Various Trap Configurations}\label{sec:escape_condition}

To ensure the proper functioning of the code, we impose some conditions that will guarantee whenever the particle hits the boundary, it needs to follow the laws of reflection. In this work, the trap configuration is referred to as the location of the boundary traps. This section will summarize all the mathematical inequalities that are used for the proper functioning of the code.

\begin{figure}[htbp]
\begin{subfigure}{0.45\textwidth}
 \includegraphics[width=\textwidth,height=6cm]{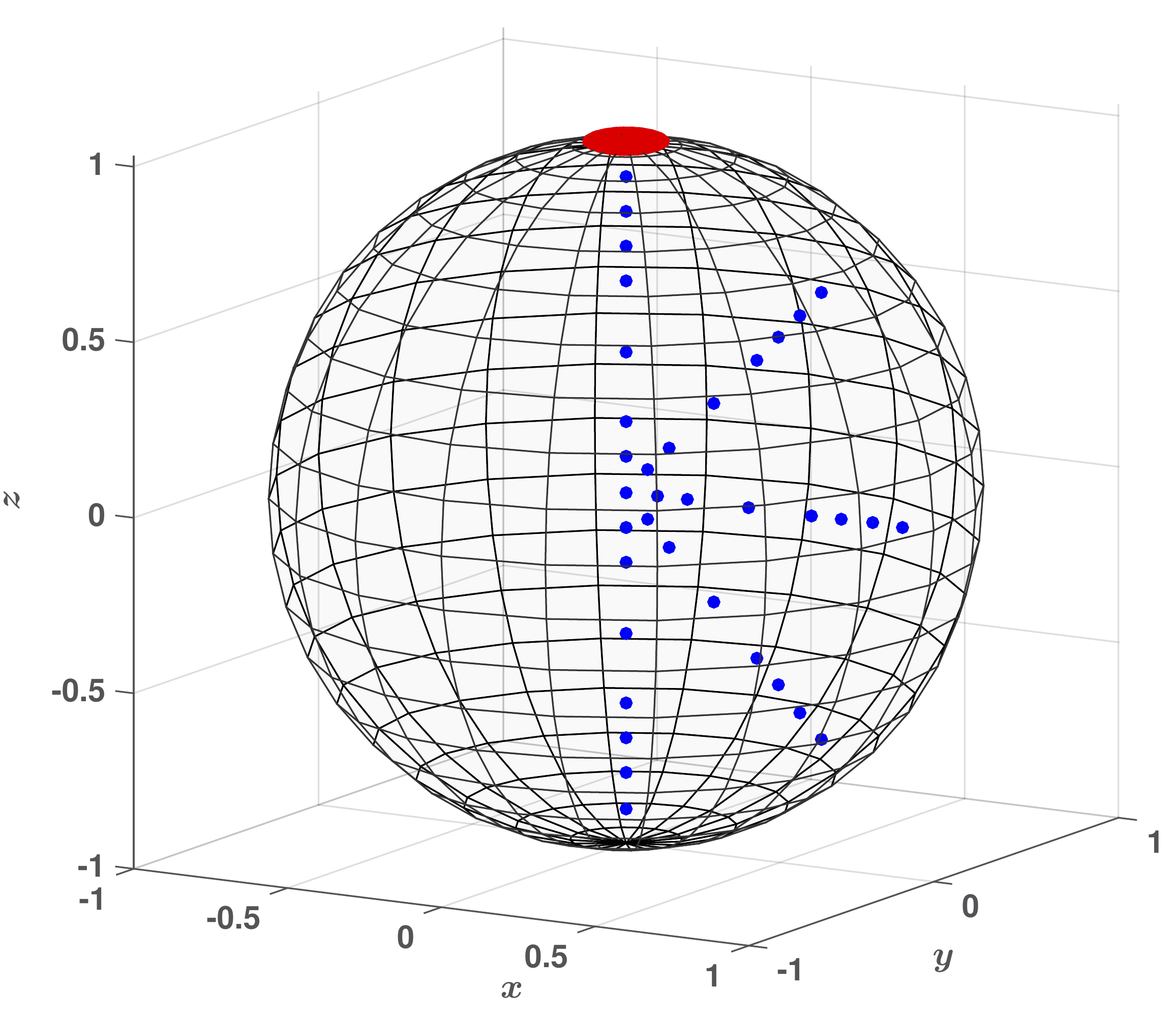}
 \caption{\label{fig:1trap_config}}
\end{subfigure}
\hfill
\begin{subfigure}{0.45\textwidth}
\includegraphics[width=\textwidth,height=6cm]{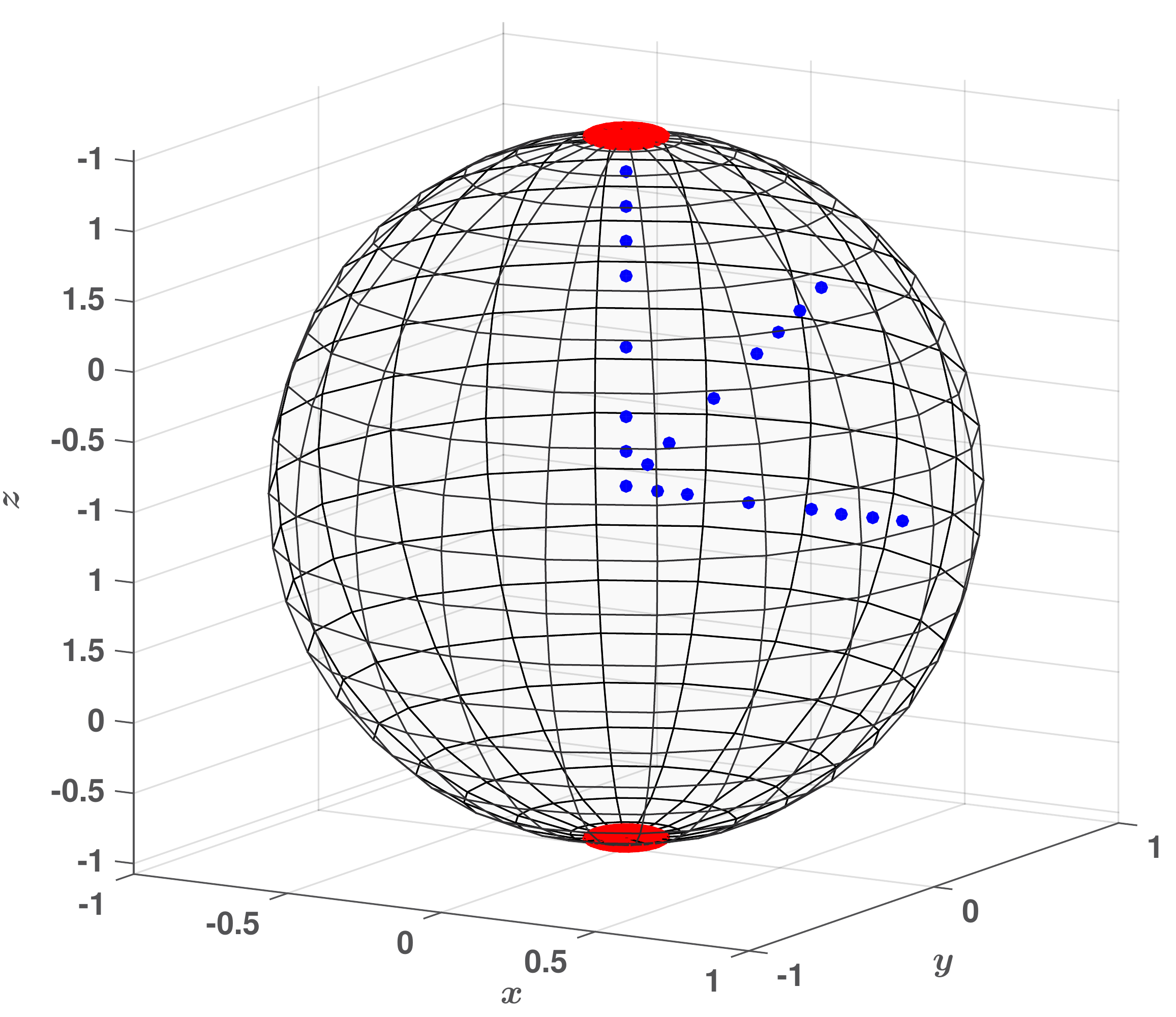}
\caption{\label{fig:2trap_config}}
\end{subfigure}
\hfill
\caption{(\subref{fig:1trap_config}, \subref{fig:2trap_config}) Sketch of the one-trap and two-trap configurations, the traps are represented by red, whereas the set of blue dots represent the sufficient number of launching coordinates concerned with our study.}
\label{fig:trap_config}
\end{figure}

Let us consider that $x = (x_1,x_2,x_3)$ give the Brownian particle's position. The escape condition of a particle with the position $x$ is assured by calculating the angular distance of the particle from the axis passing through the center of the trap and correlating it with the trap radius $\epsilon$ using the conditional operators \verb|and| $\&$ and \verb|or| $||$. This work fundamentally deals with the one-trap and the two-trap configurations, for which we formulated the \verb|MATLAB| based code and obtained precise and accurate approximations concerning the asymptotic MFPT $v(x)$ and boundary dynamics. For the one-trap configuration, the trap is located at the north pole $(0,0,1)$, with a radius of $\epsilon=0.01$, which is $1\%$ of the radius of the unit sphere (see Fig.~\ref{fig:1trap_config}).

On using some elementary trigonometric relations, we infer that a Brownian particle with position $x$ is said to be escaped from the unit sphere through the trap if the escape condition is given by
\begin{equation}\label{eq:one_trap_condition}
\Big[ \cos^{-1} (x_3) < \epsilon \quad \& \quad |x|=1 \Big].
\end{equation}

On extending the similar idea and techniques, we estimated the escape conditions for the two-trap configuration. For the two-trap configuration, the traps are positioned at the north pole $(0,0,1)$, and the south pole $(0,0,-1)$ with the common radius of $\epsilon=0.01$ (see Fig.~\ref{fig:2trap_config}). The escape condition for the two-trap configuration is given by
\begin{equation}\label{eq:two_trap_condition}
\Big[ \Big( \cos^{-1} (x_3) > \pi - \epsilon \quad || \quad \cos^{-1} (x_3) < \epsilon \Big) \quad \& \quad |x|=1 \Big].
\end{equation}

\subsection{Comparison of the Asymptotic MFPT with the Averaged Brownian Escape Times for Various Traps Configurations}\label{sec:Comparision}

In this section, we will examine and compare the asymptotic MFPT with the full direct numerical Brownian dynamics simulation results computed by the \verb|MATLAB| based code, which can effectively model and collect three-dimensional trajectories of Brownian particles launched from various points specified by $(r,\phi)$, for a radial distance $0\le r \le 1$ and a spherical angle $\phi$, up to their escape through a given trap, taking into account boundary reflections rigorously. The asymptotic results for the unit sphere having $N$ boundary traps are discussed in \S.~\ref{sec:asymptotic_result}. This work primarily considers the narrow escape kinematics concerning the trap configurations mentioned in Fig.~\ref{fig:trap_config}.

Averaged Brownian numerically simulated escape times $v^B_{N}$, when a particle is launched from $x$ is computed by averaging escape times $\{v_{i}\}_{i=1}^{N}$ of Brownian particles launched from the same starting point $x$
\begin{equation}\label{eq:average_brow_escape_times}
	v^B_N(x)=\dfrac{1}{N}\sum_{i=1}^N v_i.
\end{equation}

The study also shows that $N=10^{4}$ is the optimal number of iteration or runs needed for computing averaged Brownian escape times by numerical simulations in the unit spherical domain with boundary traps.

\subsubsection{One Trap}\label{sec:comp_1_trap}

For one boundary trap located at the north pole $(0,0,1)$ in the unit sphere, we ran the \verb|MATLAB| based code for the optimal number of runs, which is $10^4$. The result received from the Brownian dynamics simulations, which calculates the averaged Brownian escape times, displays precise and accurate agreement between the asymptotic result.

The simulations are initially run along the $Z$-axis when Brownian particles are launched from the various coordinates ranging over the $Z$-axis. Moreover, after getting sufficient agreement between the simulations results and asymptotic results, we launch the Brownian particles from the $XZ$-plane for various choices of the radial distance $r$ and the spherical angle $\phi$. For our work, we are interested in launching the Brownian particles from the coordinate
\begin{equation}\label{eq:lauch_coords}
\begin{array}{ll}
 x= \Big(r \cdot \sin(\phi),0,r \cdot \cos(\phi)\Big) \\[3ex]
 r=\Big[0,0.1,0.2,0.4,0.6,0.7,0.8,0.9\Big] \quad \& \quad\phi=\Big[0,\dfrac{\pi}{4},\dfrac{\pi}{2},\dfrac{3 \pi}{4},\pi,\dfrac{5 \pi}{4},\dfrac{3 \pi}{2},\dfrac{7 \pi}{4}\Big].
\end{array}
\end{equation}

The trap symmetry for the one-trap configuration assures that for the above set of Brownian launches, it suffices to launch the particles for the spherical angle $\Big[0,\dfrac{\pi}{4},\dfrac{\pi}{2},\dfrac{3 \pi}{4},\pi\Big]$. The schematics of these launching coordinates are shown in Fig.~\ref{fig:1trap_config}, where the trap is symbolized by the red, and blue dots denote the sufficient number of launches required for our study.

\begin{figure}[htbp]
\begin{subfigure}{0.40\textwidth}
 \includegraphics[width=7.8cm,height=7cm]{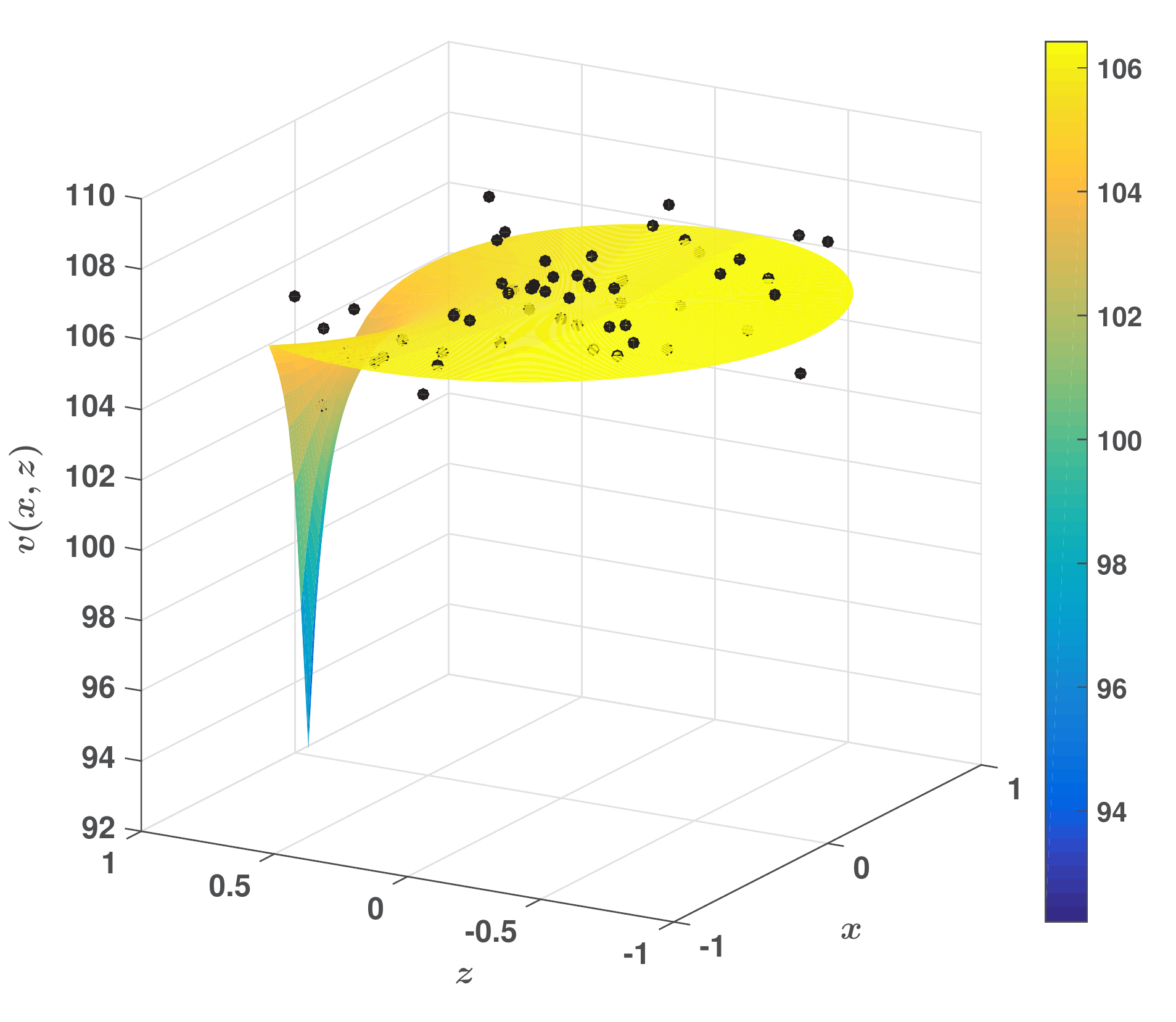}
 \caption{\label{fig:asym_res_1trap_main}}
\end{subfigure}
\hfill
\begin{subfigure}{0.49\textwidth}
\includegraphics[width=8.5cm,height=7cm]{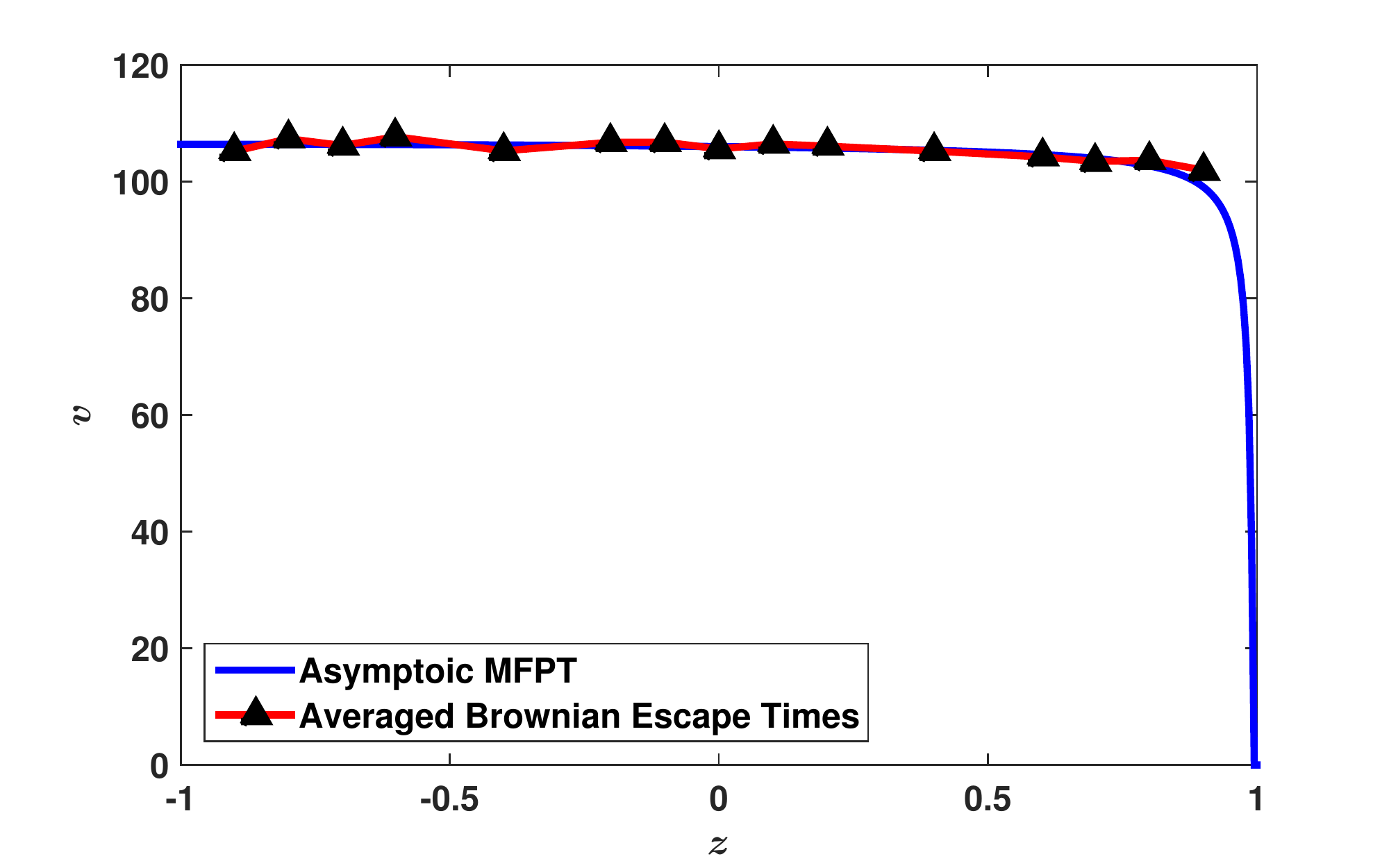}
\caption{\label{fig:asym_res_1trap_zaxis}}
\end{subfigure}
\hfill
\caption{(\subref{fig:asym_res_1trap_main}) Scatter plot of the averaged Brownian escape times on the surface plot of the asymptotic MFPT for various launching coordinates in $XZ$-plane for the one-trap configuration. (\subref{fig:asym_res_1trap_zaxis}) Averaged Brownian escape times vs MFPT, when the particles are launched from $Z$-axis.}
\label{fig:1_trap_comp}
\end{figure}

As in Fig.~\ref{fig:asym_res_1trap}, we have plotted the surface and contour curves of the asymptotic MFPT $v(x)$ for the unit sphere with one absorbing window at $(0,0,1)$ with radius $0.01$. We essentially require to inscribe a \verb|MATLAB| code that efficiently performs a direct numerical simulation of multiple particles undergoing Brownian motion in the three-dimensional spherical domain with boundary traps, and computes MFPT values by averaging of Brownian escape times and comparing the results with the asymptotic results calculated by solving the Poisson PDE problem with mixed Dirichlet-Neumann boundary conditions \eqref{eqn:NEP_Dir_Neu}. We have launched the Brownian particles from various points on the $Z$-axis to validate the code's accuracy and effectiveness, and the computed averaged Brownian escape times are plotted against the asymptotic MFPT in the Fig.~\ref{fig:asym_res_1trap_zaxis}.

The close agreement between the asymptotic and direct simulation results motivates us to extend the work and launch the particles from various launching coordinates, each uniquely identified by the radial distance $r$ and the spherical angle $\phi$. The averaged Brownian escape times obtained by direct numerical simulation when initiated from all such points are plotted over the surface curve of the asymptotic MFPT in Fig.~\ref{fig:1_trap_comp}. Here, in Fig.~\ref{fig:1_trap_comp} every dark-colored black dot denotes the averaged Brownian escape times concerned with a launching coordinate that lies above the surface whereas light-colored black dots lies below or on the surface.

\subsubsection{Two Traps}\label{sec:comp_2_trap}

The good agreement in the asymptotic and simulations result motivates us to extend the code for the two-trap configuration. For the two-trap configuration, we have two traps located at the north pole $(0,0,1)$ and the south pole $(0,0,-1)$ in the unit sphere, for which we run the Brownian dynamics simulations for the optimal number of time. In this case, the study shows that the optimal number of runs required for good agreement is found to be $10^4$.

\begin{figure}[htbp]
\begin{subfigure}{0.40\textwidth}
 \includegraphics[width=7.8cm,height=7cm]{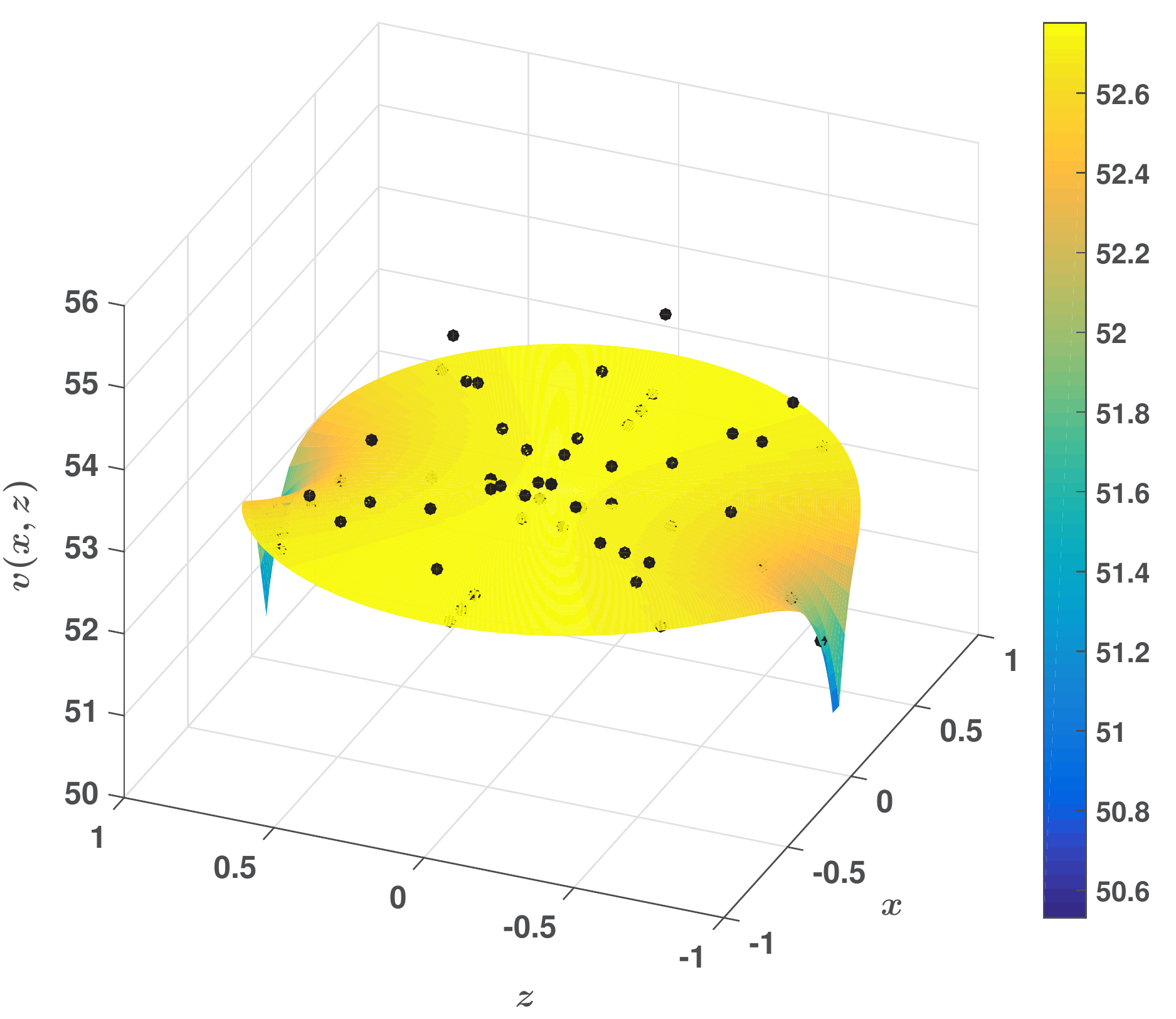}
 \caption{\label{fig:asym_res_2trap_main}}
\end{subfigure}
\hfill
\begin{subfigure}{0.49\textwidth}
\includegraphics[width=8.5cm,height=7cm]{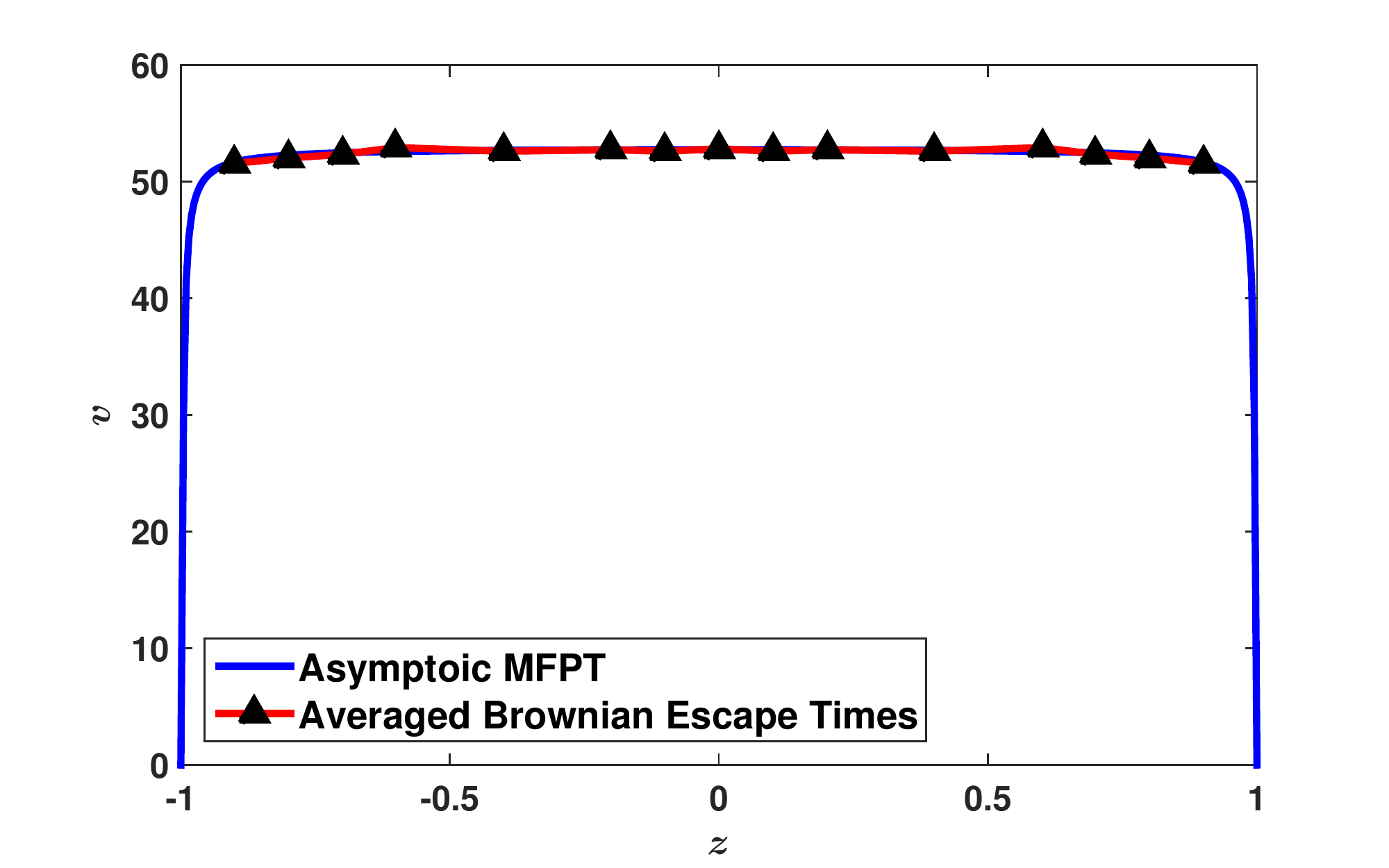}
\caption{\label{fig:asym_res_2trap_zaxis}}
\end{subfigure}
\hfill
\caption{(\subref{fig:asym_res_2trap_main}) Scatter plot of the averaged Brownian escape times on the surface plot of the asymptotic MFPT for various launching coordinates in $XZ$-plane for the two-trap configuration. (\subref{fig:asym_res_2trap_zaxis}) Averaged Brownian escape times vs MFPT, when the particles are launched from $Z$-axis.}
\label{fig:2_trap_comp}
\end{figure}

We simulate the Brownian particles from the $Z$-axis for preliminary analysis and found the relative error is within $0.7\%$. We are interested in launching the Brownian particle from the $XZ$-plane for various choices of the radial distance $r$ and the spherical angle $\phi$. We are interested in launching the Brownian particles from the coordinates  \eqref{eq:lauch_coords} for analysis regarding the two-trap configuration. The trap symmetry is apparent from the trap location, which ensures that it will be sufficient to launch the particles for the spherical angle $\Big[0,\dfrac{\pi}{4},\dfrac{\pi}{2} \Big]$. The schematics of these launching coordinates are manifested in Fig.~\ref{fig:2trap_config}, where the trap is denoted by the red and blue dots that depict the sufficient number of launches needed for our study.

In Fig.~\ref{fig:asym_res_2trap}, we have plotted the surface and contour curves of the asymptotic MFPT $v(x)$ for the unit sphere having two absorbing windows positioned respectively at $(0,0,1)$ and $(0,0,-1)$, and share a common radius $0.01$. The results obtained from running Brownian dynamics simulations are plotted against the asymptotic MFPT $v(x)$ \eqref{eq:mfpt_nep} in Fig.~\ref{fig:asym_res_1trap_zaxis} when the Brownian particle is launched from the $Z$-axis. Moreover, when a Brownian particle is launched from the $XZ$-plane, each coordinate is uniquely identified by the radial distance $r$ and the spherical angle $\phi$. The averaged Brownian escape times computed from numerical simulations over all such points are plotted over the surface curve of the asymptotic MFPT in Fig.~\ref{fig:2_trap_comp}. Here, in the Fig.~\ref{fig:2_trap_comp} each dark-colored black dot represents the averaged Brownian escape times, which lies above the surface, whereas light-colored black dots lies below or on the surface.

\subsection{Accuracy of the Averaged Brownian Escape Times Computed Using Simulations}\label{sec:accuracy}

In the first passage problems like the narrow escape problem, the asymptotic expansion concerning the MFPT \eqref{eq:mfpt_nep} described within the certain valid ranges of the limiting parameter renders a tool for analyzing and comparing the results with the Brownian dynamics simulations. This section will review the relative percentage difference error $\delta v(x)$, which can measure the accuracy of the results computed using the simulations by comparing it with the asymptotic results for the one-trap and two-trap configuration. Let $v$ be the asymptotic MFPT for the unit sphere with $N$ boundary traps given by  \eqref{eq:mfpt_nep}, and let the $v^{B}_{N}$ describes the averaged Brownian escape times given by \eqref{eq:average_brow_escape_times}. Define relative percentage error $\delta v(x)$
\begin{equation}\label{eq:relative_perc_difference}
	\delta v(x) = \dfrac{|v(x) - v^B_N(x)|} {v(x)} 	\times 100\%,
\end{equation}
when the Brownian particle is launched form the position $x$.

\begin{center}
\begin{longtable}{|c|c|c|}
\caption{A comprehensive list illustrating the relative error difference $\delta v(x)$ between the asymptotic results $v(x)$ versus averaged Brownian escape times $v^{B}_{N}$ obtained from Brownian dynamics simulations, when a Brownian particle is launched in the unit sphere with the one-trap configuration.}
\label{table:mfpt_1trap}
\\
\hline \multicolumn{1}{|c|}{\textbf{Launching Coordinates}} & \multicolumn{1}{c|}{\textbf{Averaged Brownian Escape Times}} & \multicolumn{1}{c|}{\textbf{$\delta v$}} \\ \hline
\endfirsthead

\multicolumn{3}{c}%
{{\bfseries \tablename\ \thetable{} -- Continued}} \\
\hline \multicolumn{1}{|c|}{\textbf{Launching Coordinates}} & \multicolumn{1}{c|}{\textbf{Averaged Brownian Escape Times}} & \multicolumn{1}{c|}{\textbf{$\delta v$}} \\ \hline
\endhead

\hline \multicolumn{3}{|r|}{{Continued on next page}} \\ \hline
\endfoot

\hline
\endlastfoot
$(0,0,0)$ & $105.6267$ & $0.3745$  \\
$(0,0,0.1)$ & $106.4793$ & $0.5348$  \\
$(0,0,0.2)$ & $106.1091$ & $0.3149$  \\
$(0,0,0.4)$ & $105.2495$ & $0.1261$  \\
$(0,0,0.6)$ & $104.2623$ & $0.3784$  \\
$(0,0,0.7)$ & $103.4924$ & $0.4739$  \\
$(0,0,0.8)$ & $103.6674$ & $0.9282$  \\
$(0,0,0.9)$ & $102.0075$ & $2.9119$  \\ \hline
$(0.0707,0,0.0707)$ & $106.5675$ & $0.5833$  \\
$(0.1414,0,0.1414)$ & $106.8009$ & $0.8802$  \\
$(0.2828,0,0.2828)$ & $104.9331$ & $0.726$  \\
$(0.4243,0,0.4243)$ & $104.9480$ & $0.5725$  \\
$(0.495,0,0.495)$ & $106.3650$ & $0.8208$  \\
$(0.5657,0,0.5657)$ & $105.8812$ & $0.3958$  \\
$(0.6364,0,0.6364)$ & $106.8603$ & $1.34$  \\ \hline
$(0.1,0,0)$ & $106.6579$ & $0.5958$  \\
$(0.2,0,0)$ & $106.9766$ & $0.8896$  \\
$(0.4,0,0)$ & $105.8399$ & $0.2051$  \\
$(0.6,0,0)$ & $106.9358$ & $0.8016$  \\
$(0.7,0,0)$ & $107.3109$ & $1.1437$  \\
$(0.8,0,0)$ & $106.0947$ & $0.0116$  \\
$(0.9,0,0)$ & $105.5033$ & $0.5746$  \\
\hline
$(0.0707,0,-0.0707)$ & $106.4983$ & $0.3843$  \\
$(0.1414,0,-0.1414)$ & $105.9568$ & $0.1815$  \\
$(0.2828,0,-0.2828)$ & $105.7599$ & $0.455$  \\
$(0.4243,0,-0.4243)$ & $106.9267$ & $0.583$  \\
$(0.495,0,-0.495)$ & $106.3211$ & $0.006$  \\
$(0.5657,0,-0.5657)$ & $107.4646$ & $1.054$  \\
$(0.6364,0,-0.6364)$ & $107.2201$ & $0.8164$  \\
\hline
$(0,0,-0.1)$ & $106.7835$ & $0.6305$  \\
$(0,0,-0.2)$ & $106.7765$ & $0.5533$  \\
$(0,0,-0.4)$ & $105.2995$ & $0.9409$  \\
$(0,0,-0.6)$ & $107.7019$ & $1.2518$  \\
$(0,0,-0.7)$ & $106.2090$ & $0.1734$  \\
$(0,0,-0.8)$ & $107.3674$ & $0.9004$  \\
$(0,0,-0.9)$ & $105.2420$ & $1.105$  \\
\end{longtable}
\end{center}

\begin{figure}[htbp]
  \centering
  \includegraphics[width=10cm,height=7cm]
  {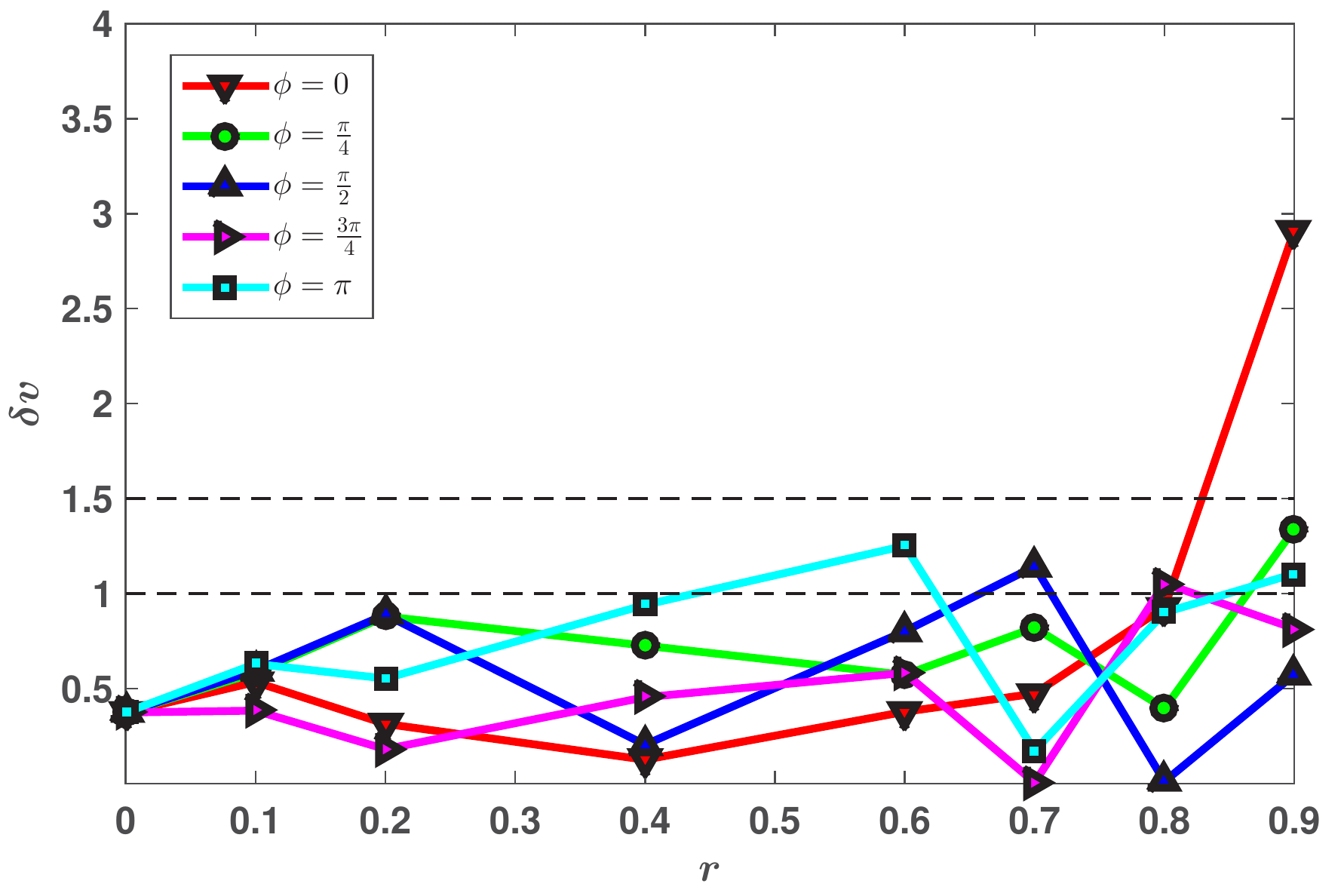}
\caption{Line plot for the relative percentage difference $\delta v(x)$ for various choices of the radial distance $r$ and the spherical angle $\phi$ for the one-trap configuration.}
\label{fig:accu_1}
\end{figure}

The Table.~\ref{table:mfpt_1trap} presents a comprehensive list showing the comparison between the asymptotic MFPT formula \eqref{eq:mfpt_nep} and the averaged Brownian numerically simulated escape times \eqref{eq:average_brow_escape_times} for the unit sphere having one boundary trap at the north pole. The calculation illustrates that the simulation shows a close approximation of the mean first passage time MFPT. The study exhibits that the relative percentage difference error $\delta v(x)$ between the asymptotic results are mostly within $1\%$ accuracy, even when the radius of the boundary trap is $10^{-2}$ (See Fig.~\ref{fig:accu_1}).

\begin{figure}[H]
  \begin{minipage}{.450\textwidth}
\begin{subfigure}[b]{.45\textwidth}
  \centering
  \subcaptionbox{\label{fig:accu_1_surface}}
	{\includegraphics[width=9cm,height=8.1cm]{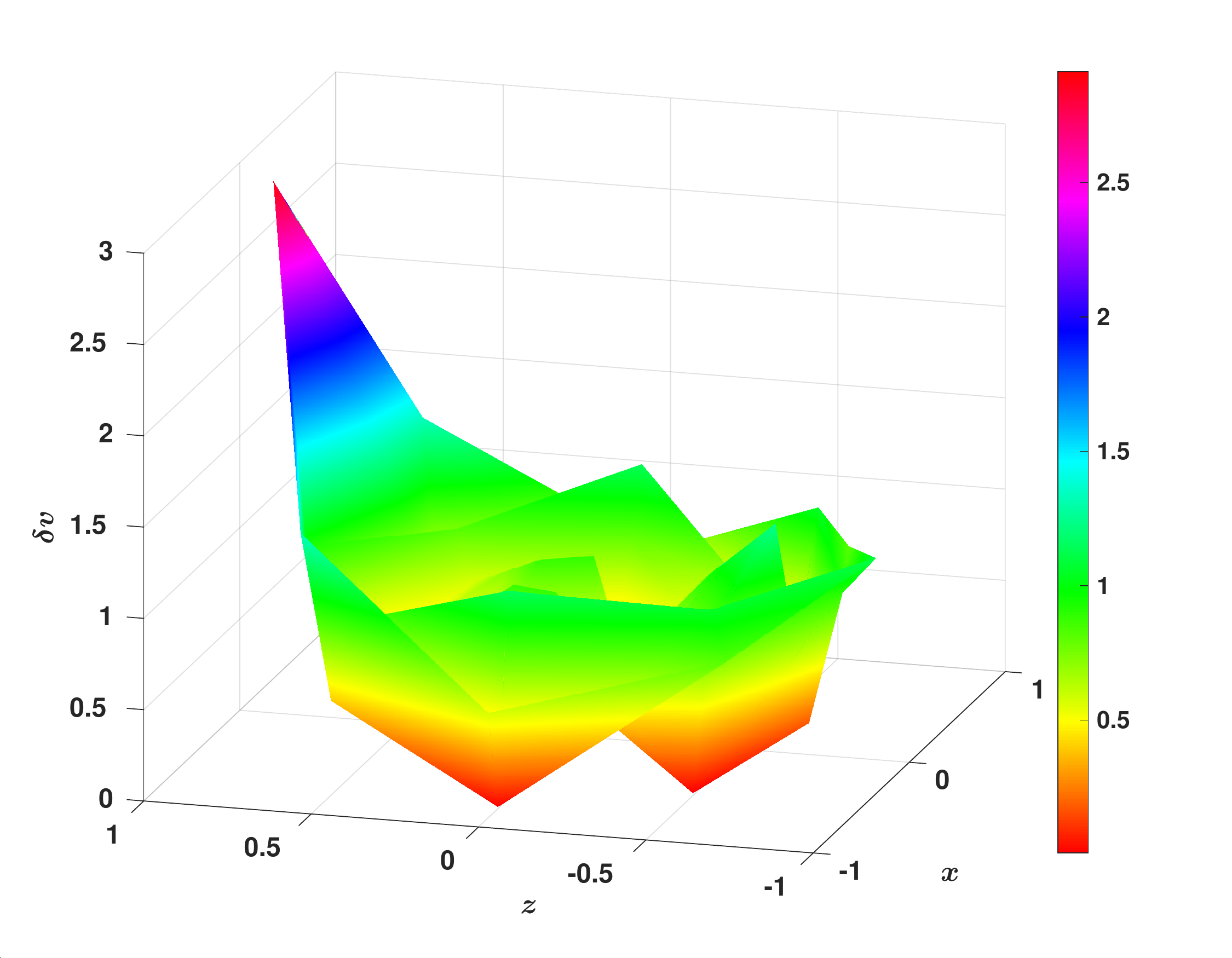}}
\end{subfigure}
\end{minipage}\hspace{1in}
\begin{minipage}{.45\textwidth}
\begin{subfigure}[b]{.45\textwidth}
\centering
\subcaptionbox{\label{fig:accu_1_top}}		{\includegraphics[width=4cm,height=3.8cm]{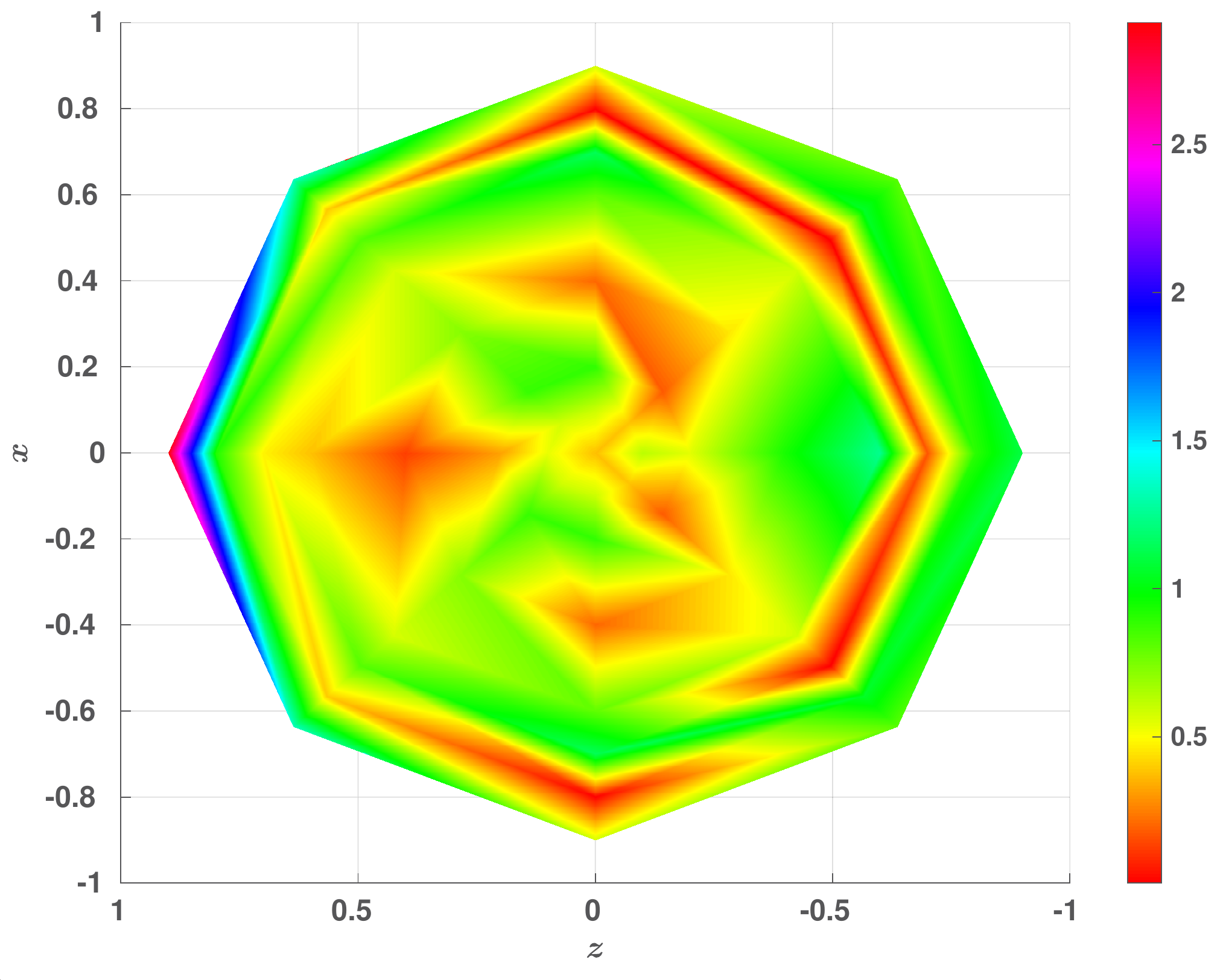}}
\end{subfigure}
\newline
\begin{subfigure}[b]{.5\textwidth}
\centering
\subcaptionbox{\label{fig:accu_1_contour}}	{\includegraphics[width=4cm,height=3.8cm]{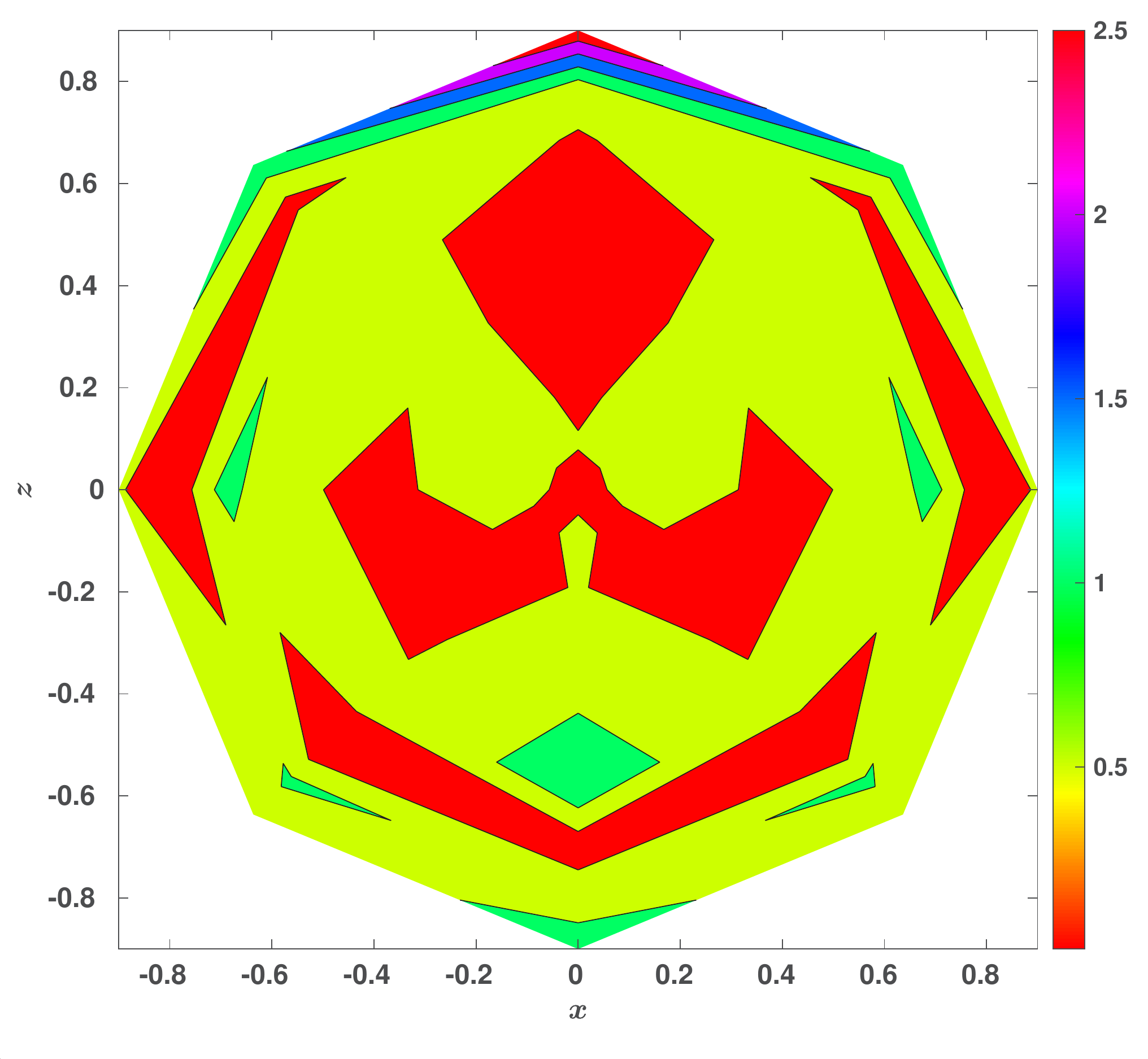}}
\end{subfigure}
\hfill
\end{minipage}
\hfill
\caption{ (\subref{fig:accu_2_surface}) Surface plot perspective view, (\subref{fig:accu_2_top}) top view, and (\subref{fig:accu_2_contour}) contour
plot for the relative percentage difference $\delta v(x)$ for a radial distance $r$ and a spherical angle $\phi$ in $XZ$-plane for the unit sphere with the one-trap configuration. }
\label{fig:accu_12}
\end{figure}

\begin{figure}[H]
  \centering
  \includegraphics[width=10cm,height=7cm]
  {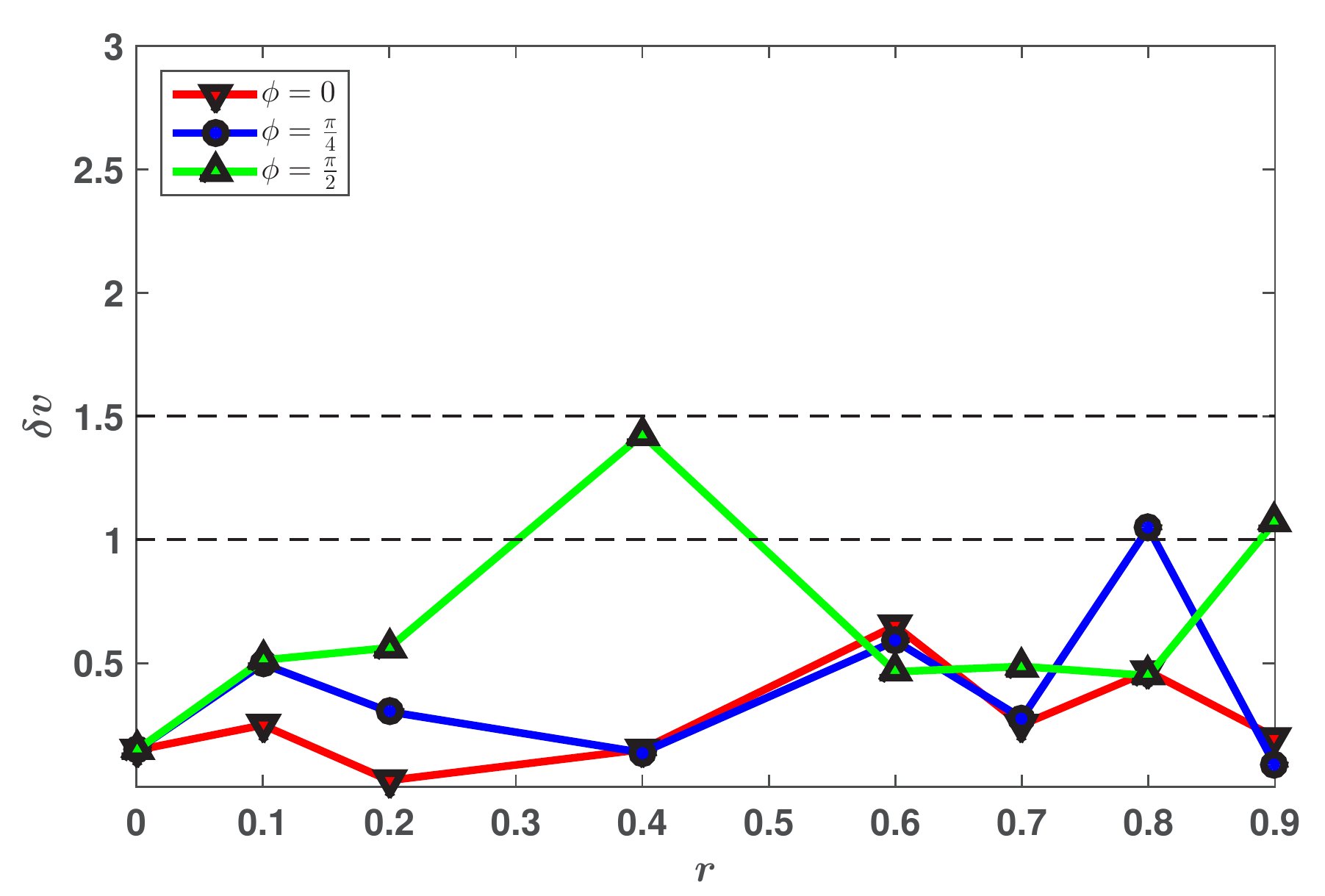}
\caption{Line plot for the relative percentage difference $\delta v(x)$ for the various radial distance $r$ and the spherical angle $\phi$ for the two-trap configuration.}
\label{fig:accu_21}
\end{figure}

It is aptly said that the $3$-D plots allow us to comprehend and visualize the result from different angles of perspectives and provides us more depth in the information. In the \S.~\ref{sec:comp_1_trap}, we have presented the set of sufficient Brownian launching coordinates needed for our study. Fig.~\ref{fig:accu_12} shows that the simulations provide an accurate approximation of the asymptotic MFPT for the unit sphere with a boundary trap at the north pole.

\begin{figure}[H]
  \begin{minipage}{.450\textwidth}
\begin{subfigure}[b]{.45\textwidth}
  \centering
  \subcaptionbox{\label{fig:accu_2_surface}}
	{\includegraphics[width=9cm,height=8.1cm]{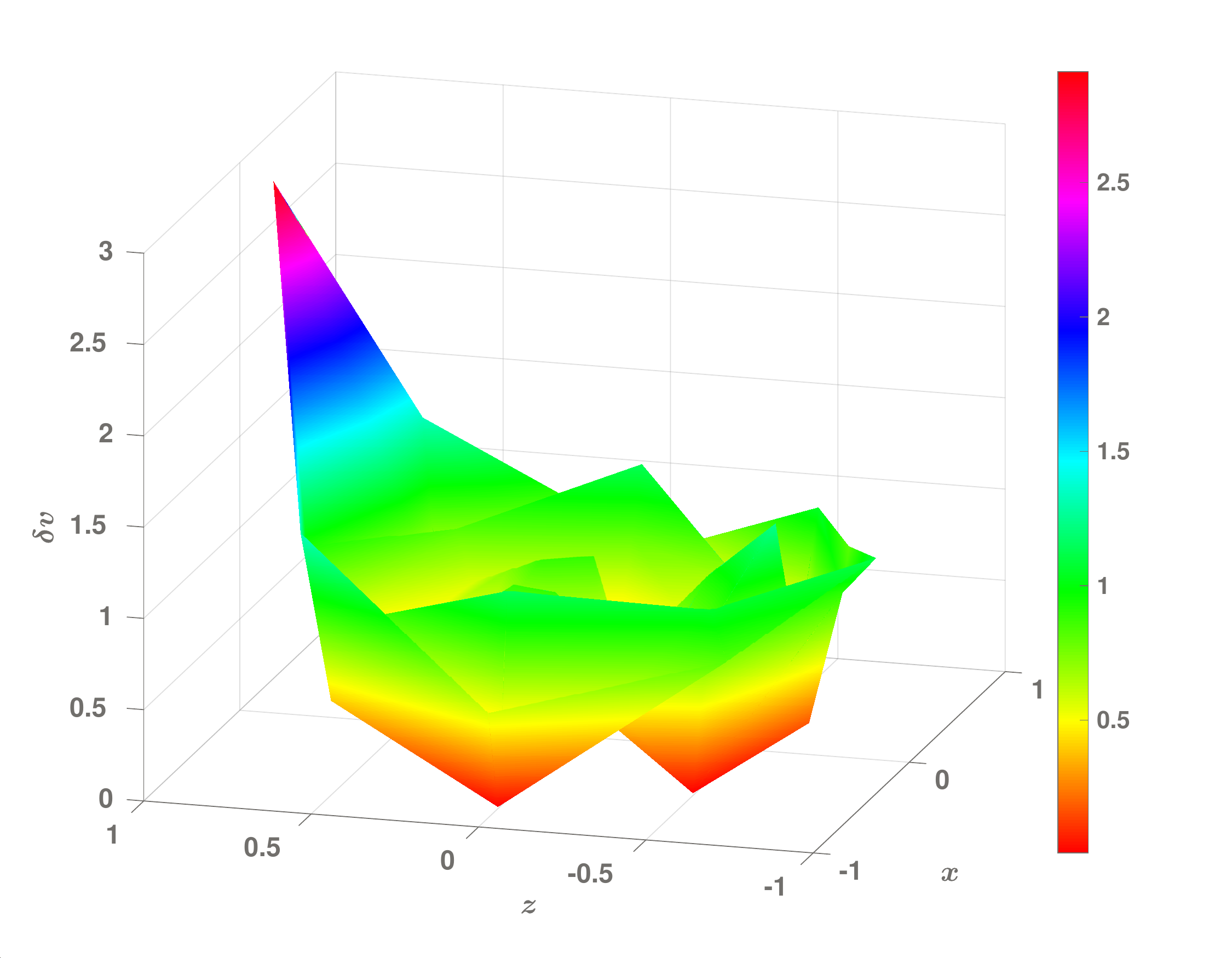}}
\end{subfigure}
\end{minipage}\hspace{1in}
\begin{minipage}{.45\textwidth}
\begin{subfigure}[b]{.45\textwidth}
\centering
\subcaptionbox{\label{fig:accu_2_top}}		{\includegraphics[width=4cm,height=3.8cm]{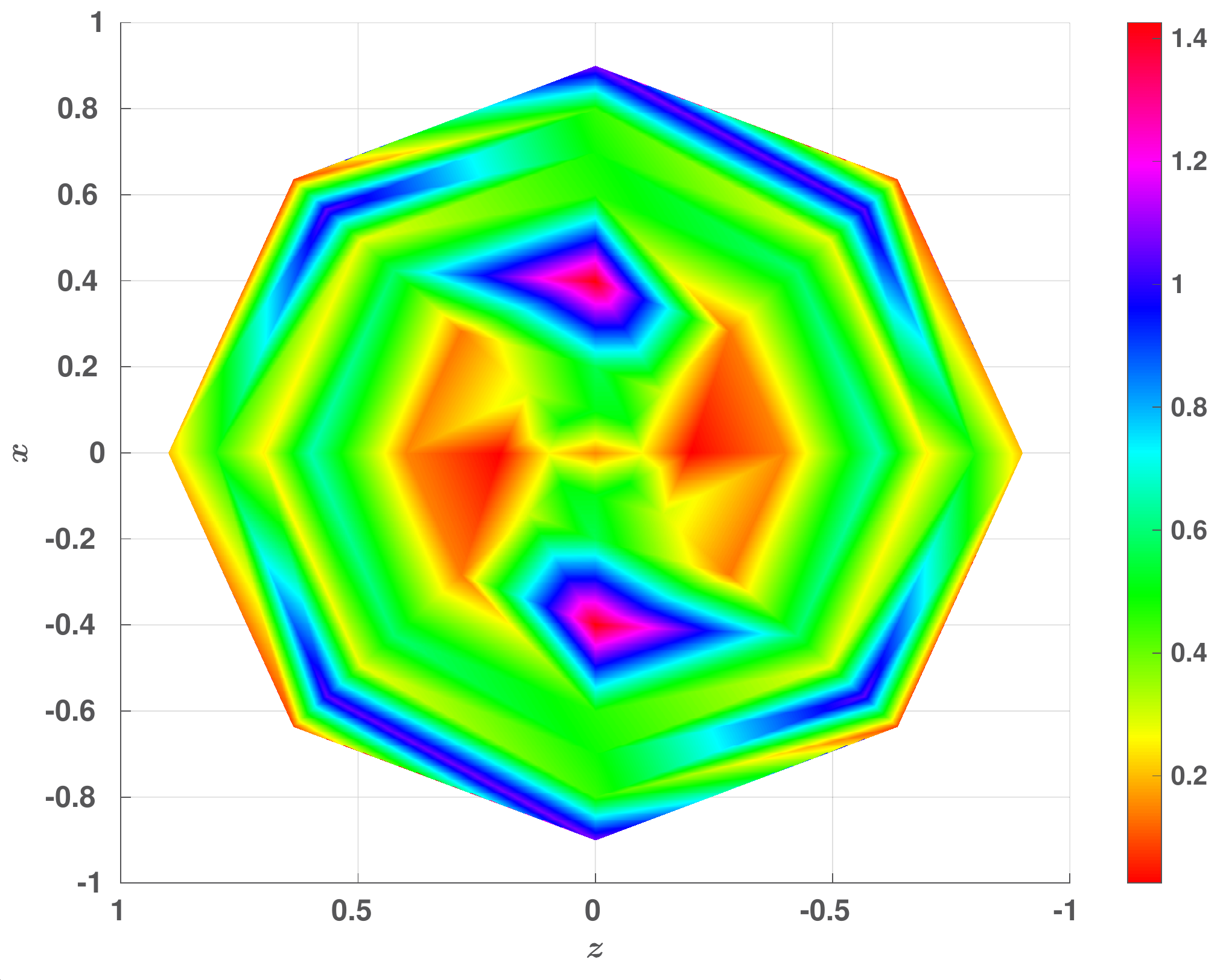}}
\end{subfigure}
\newline
\begin{subfigure}[b]{.5\textwidth}
\centering
\subcaptionbox{\label{fig:accu_2_contour}}	{\includegraphics[width=4cm,height=3.8cm]{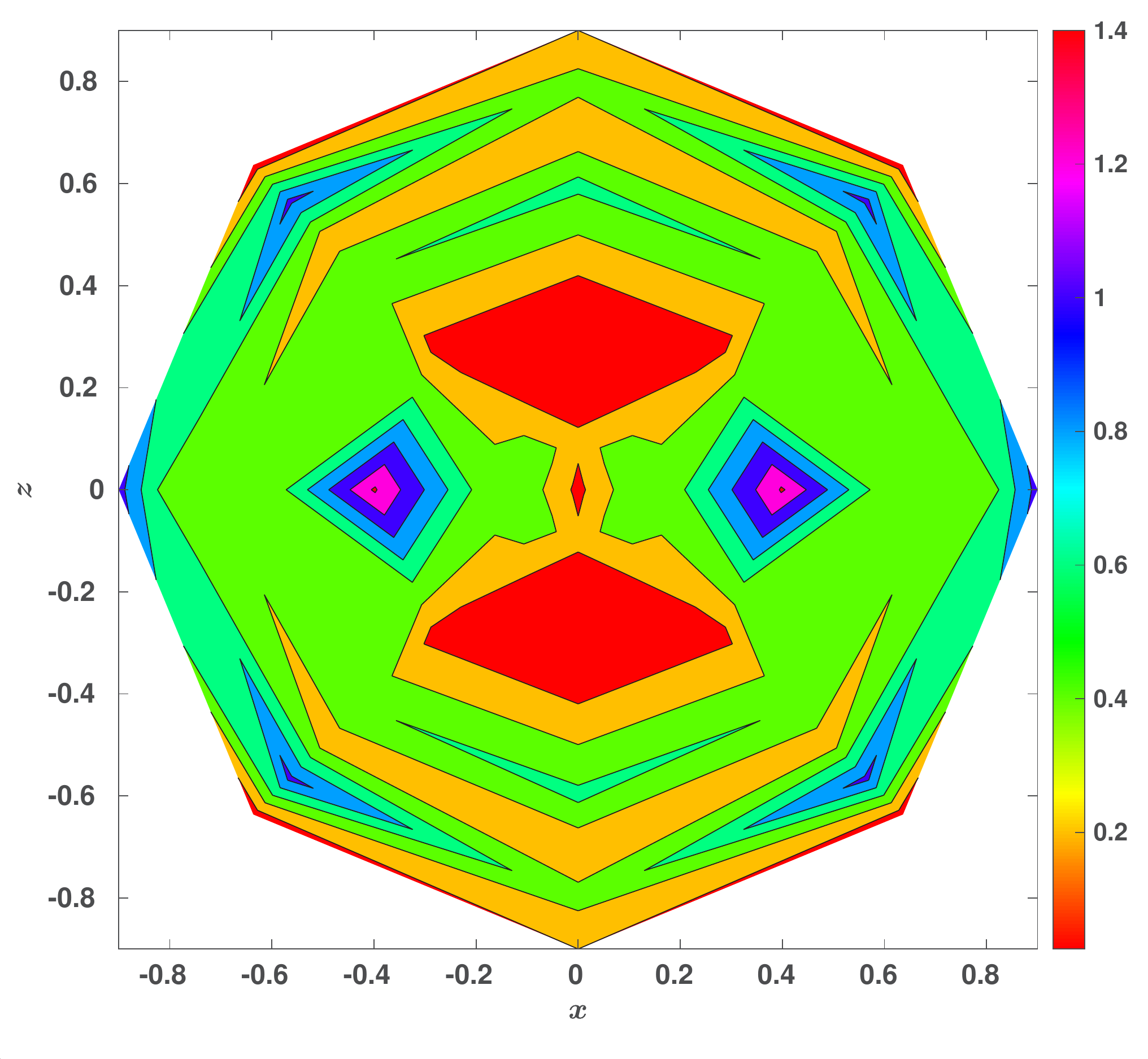}}
\end{subfigure}
\hfill
\end{minipage}
\hfill
\caption{ (\subref{fig:accu_2_surface}) Surface plot perspective view, (\subref{fig:accu_2_top}) top view, and (\subref{fig:accu_2_contour}) contour
plot for the relative percentage difference $\delta v(x)$ for a radial distance $r$ and a spherical angle $\phi$ in $XZ$-plane for the unit sphere with the two-trap configuration. }
\label{fig:accu_22}
\end{figure}

For the two-trap configuration, the good agreement between the asymptotic and simulation results is illustrated in the Table.~\ref{table:mfpt_2trap} for the unit sphere having two boundary traps located sequentially at the north pole and the south pole. The study reveals that the relative percentage difference error $\delta v(x)$ between the results computed from the simulations is within the $0.7\%$ when the Brownian particles are launched from the $Z$-axis, whereas $\delta v(x)$ is mostly within $1\%$ when the particle is launched from various positions specified by \eqref{eq:lauch_coords} even when the radius of the boundary traps are $10^{-2}$ (See Fig.~\ref{fig:accu_22}). The surface plots with both perspective and top view, and contour plot (See Fig.~\ref{fig:accu_22}) are added, which help us discern the obtained results.

\begin{center}
\begin{longtable}{|c|c|c|}
\caption{A comprehensive list illustrating the relative error difference $\delta v(x)$ between the asymptotic result $v(x)$ versus averaged Brownian escape times $v^{B}_{N}$ obtained from Brownian dynamics simulations, when a Brownian particle is launched in the unit sphere with the two-trap configuration.}
\label{table:mfpt_2trap}
\\
\hline \multicolumn{1}{|c|}{\textbf{Launching Coordinates}} & \multicolumn{1}{c|}{\textbf{Averaged Brownian Escape Times}} & \multicolumn{1}{c|}{\textbf{$\delta v$}} \\ \hline
\endfirsthead

\multicolumn{3}{c}%
{{\bfseries \tablename\ \thetable{} -- Continued}} \\
\hline \multicolumn{1}{|c|}{\textbf{Launching Coordinates}} & \multicolumn{1}{c|}{\textbf{Averaged Brownian Escape Times}} & \multicolumn{1}{c|}{\textbf{$\delta v$}} \\ \hline
\endhead

\hline \multicolumn{3}{|r|}{{Continued on next page}} \\ \hline
\endfoot

\hline
\endlastfoot
$(0,0,0)$ & $52.8241$ & $0.1473$  \\
$(0,0,0.1)$ & $52.6118$ & $0.2495$  \\
$(0,0,0.2)$ & $52.7475$ & $0.0254$  \\
$(0,0,0.4)$ & $52.6111$ & $0.1509$  \\
$(0,0,0.6)$ & $52.9296$ & $0.6519$  \\
$(0,0,0.7)$ & $52.3500$ & $0.2502$  \\
$(0,0,0.8)$ & $52.0311$ & $0.4668$  \\
$(0,0,0.9)$ & $51.5775$ & $0.2003$  \\ \hline
$(0.0707,0,0.0707)$ & $52.4829$ & $0.4975$  \\
$(0.1414,0,0.1414)$ & $52.9024$ & $0.3037$  \\
$(0.2828,0,0.2828)$ & $52.8041$ & $0.1381$  \\
$(0.4243,0,0.4243)$ & $53.0296$ & $0.5901$  \\
$(0.495,0,0.495)$ & $52.8598$ & $0.2779$  \\
$(0.5657,0,0.5657)$ & $53.2618$ & $1.0478$  \\
$(0.6364,0,0.6364)$ & $52.6601$ & $0.0884$  \\ \hline
$(0.1,0,0)$ & $53.0176$ & $0.5126$  \\
$(0.2,0,0)$ & $53.0463$ & $0.5627$  \\
$(0.4,0,0)$ & $53.5098$ & $1.4258$  \\
$(0.6,0,0)$ & $52.5211$ & $0.4658$  \\
$(0.7,0,0)$ & $52.5139$ & $0.487$  \\
$(0.8,0,0)$ & $52.5367$ & $0.4498$  \\
$(0.9,0,0)$ & $53.3428$ & $1.0738$  \\
\end{longtable}
\end{center}

\section{Dynamics of Brownian Particle Near the Boundary}\label{sec:boundary_dynamics}

\begin{figure}[htbp]
\begin{subfigure}{0.45\textwidth}
 \includegraphics[width=\textwidth,height=7cm]{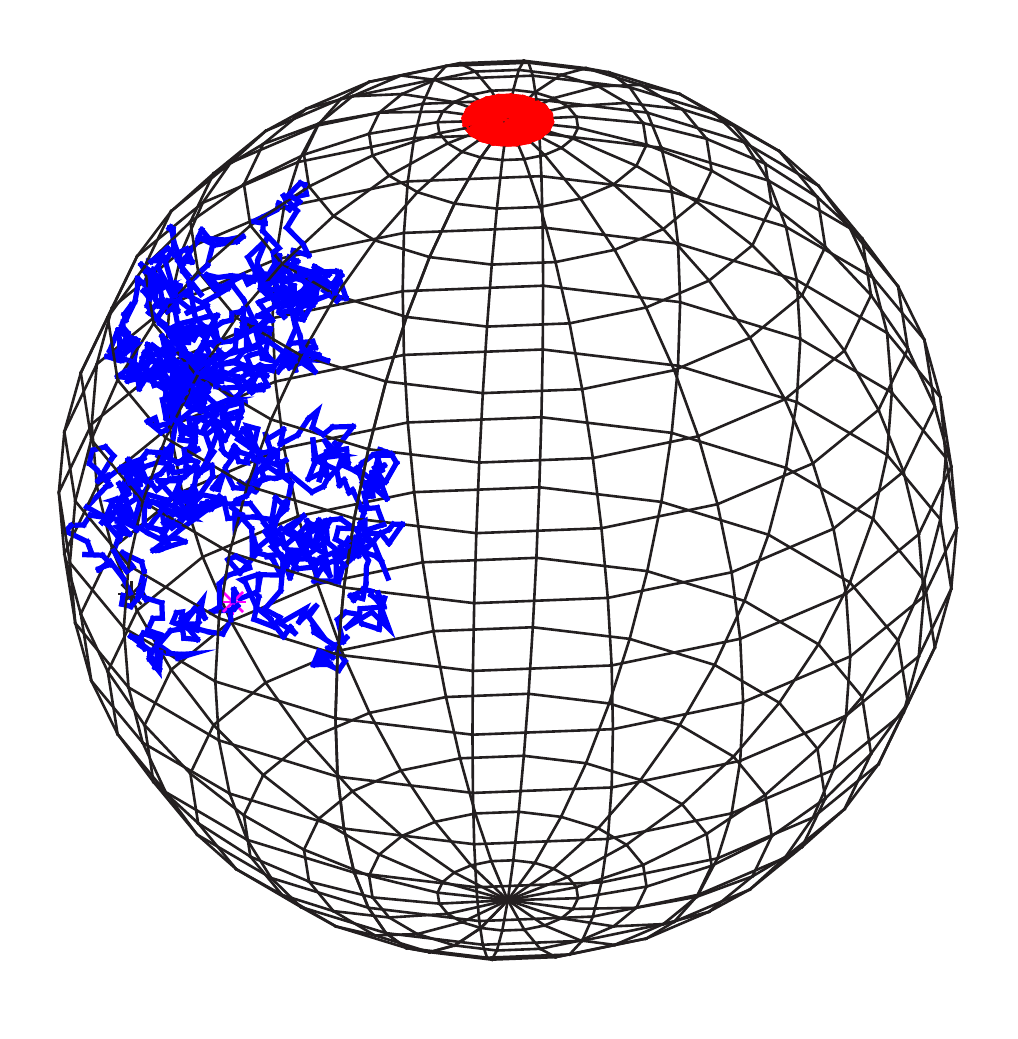}
 \caption{\label{fig:bdd_gen_dyna}}
\end{subfigure}
\hfill
\begin{subfigure}{0.45\textwidth}
\centering
\includegraphics[width=0.5\textwidth,height=7cm]{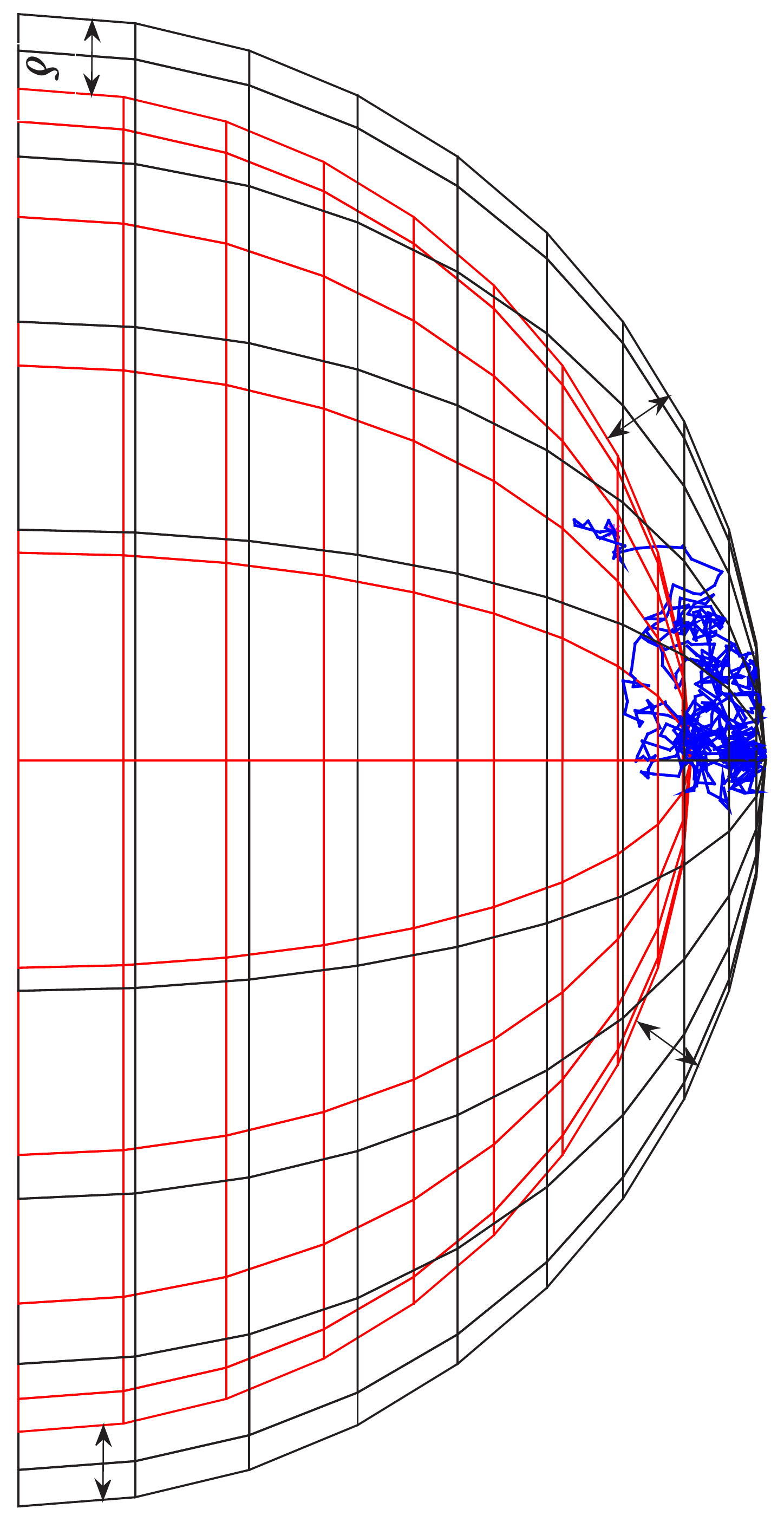}
\caption{\label{fig:bdd_gen_delta}}
\end{subfigure}
\hfill
\caption{(\subref{fig:bdd_gen_dyna}) Schematics of the boundary dynamics of a Brownian particle for the narrow escape problem in the three dimensional unit sphere having absorbing windows denoted by red. (\subref{fig:bdd_gen_delta}) Sketch of the trajectories of a Brownian particle near the boundary and illustration of $\delta$ region sized region $\delta_{\text{region}}$.}
\label{fig:bdd_gen_dyna1}
\end{figure}

It is apparent from the \S.~\ref{sec:Comparision} that the averaged Brownian escape times computed using the direct numerical simulation of Brownian particle for the narrow escape problem in the unit sphere with multiple trap configurations offers an accurate approximation when compared with the asymptotic results \eqref{eq:mfpt_nep}. A good agreement in the results prompts us to study one of the significant and widely applicable problems known as the boundary dynamics problems. These problems concern study of Brownian particle in a certain region near the boundary, where the particle obeys the laws of reflection at the boundary.

In this section we will study the dynamics of a Brownian particle near the boundary for the unit sphere having multiple trap configurations, which has applications in various physical and biological settings. For instance, inside a biological cell membrane, ions like $Na^{+}$, $K^{+}$ cannot escape through the confined cellular membranes due to the high relative dielectric constants difference between the lipid cell membranes and intracellular salt solutions \cite{Brow_boundary}.

We have analyzed the boundary dynamics by running the full discrete numerical computer simulations. The detailed mathematical model concerning the Brownian particles near the boundary interfaces are explained in \cite{Brow_boundary}. A schematics of the simulations of Brownian particles illustrating a sketch of the trajectories near the boundary for the one-trap configuration are displayed in Fig.~\ref{fig:bdd_gen_dyna1}.

It is evident that the MFPT PDE problem given by \eqref{eq:NEP_pde} does not retain any information about the $3$-D trajectories of Brownian particles. With the help of a working \verb|MATLAB| code capable of modeling and accumulating the $3$-D trajectories of Brownian particle, and following the physical laws of reflection everywhere on the boundary except the traps, we will try to understand the dynamics of the Brownian particle near the boundary. This section will attempt to answer how much dimensionless seconds a Brownian particle spends on an average in a delta-sized region near the boundary (See Fig.~\ref{fig:bdd_gen_delta}).

Define delta-sized region $\delta_{\text{region}}$
\begin{equation*}
 \delta_{\text{region}} = \{ x \in \Omega \quad | \quad \delta \leq |x| \leq 1\}.
\end{equation*}
We are interested in computing boundary time $\tau$, which is defined as the relative time a Brownian particle spends inside the $\delta_{\text{region}}$
\begin{equation}\label{eq:tau:delta}
	\tau = \left\langle \dfrac{T_{\delta} }{T} \right\rangle,
\end{equation}
where $T$ represents the total time spent by the Brownian particle before escaping through one of the absorbing windows, and $T_\delta$ describes the total time spent by the Brownian particle in $\delta_{\text{region}}$ region. Our study involves an in-depth analysis of the boundary time $\tau$ when expressed as a function of launching coordinates $\tau (\phi,r)$ for both the one-trap and the two-trap configurations.

\subsection{One Trap}

For the one-trap configuration in the unit sphere when the trap is located at the north pole $(0,0,1)$, we are interested in calculating the boundary time $\tau$, when the Brownian particles are launched from various positions \eqref{eq:lauch_coords}, each position uniquely specified by a spherical angle $\phi$ and a radial distance $r$, for $10^4$ time, which is the optimal number of runs required for a good agreement of the asymptotic results with the direct numerical simulation results. For each run, the trajectories of Brownian particles are collected and stored in a matrix, which is used for calculating the $\tau(\phi,r)$. It was a subject of interest to determine how many dimensionless seconds particle spends inside the $\delta_{\text{region}}$ for the ranges of $\delta$. For this work, $\delta$'s are chosen to be $0.1$ and $0.01$.

\begin{center}
\begin{longtable}{|c|c|c|}
\caption{ A comprehensive list illustrating the boundary time $\tau$, when a Brownian particle is launched in the unit sphere with the one-trap configuration.}
\label{table:boundary_dynamics_12trap}
\\
\hline
\multirow{2}{*}{\textbf{Launching Coordinates}} &
\multicolumn{2}{c|}{\textbf{Boundary Time}}  \\
& $\delta=0.1$ & $\delta=0.01$\\ \hline

\endfirsthead

\multicolumn{2}{c}%
{{\bfseries \tablename\ \thetable{} -- Continued}} \\
\hline \multirow{2}{*}{\textbf{Launching Coordinates}} & \multicolumn{2}{c|}{\textbf{Boundary Time}} \\
& $\delta=0.1$ & $\delta=0.01$\\ \hline
\endhead

\hline \multicolumn{3}{|r|}{{Continued on next page}} \\ \hline

\endfoot

\hline

\hline Average $\tau$ & 0.2829
 & 0.0401
  \\ \hline

  \endlastfoot
$(0,0,0)$ & 0.2804 & 0.0397\\
$(0,0,0.1)$ & 0.2808 &  0.0398\\
$(0,0,0.2)$ & 0.2802 & 0.0397\\
$(0,0,0.4)$ & 0.2809 & 0.0398\\
$(0,0,0.6)$ & 0.2816 & 0.0399 \\
$(0,0,0.7)$ & 0.2824 & 0.0400 \\
$(0,0,0.8)$ & 0.2861 & 0.0407 \\
$(0,0,0.9)$ & 0.3040 & 0.0438 \\  \hline
$(0.0707,0,0.0707)$ & 0.2803 & 0.0397 \\
$(0.1414,0,0.1414)$ & 0.2806 & 0.0397 \\
$(0.2828,0,0.2828)$ & 0.2805 & 0.0397  \\
$(0.4243,0,0.4243)$ & 0.2812 & 0.0398 \\
$(0.495,0,0.495)$ & 0.2817 & 0.0399 \\
$(0.5657,0,0.5657)$ & 0.2823 & 0.0400  \\
$(0.6364,0,0.6364)$ & 0.2826 & 0.0400  \\  \hline
$(0.1,0,0)$ & 0.2803 & 0.0397   \\
$(0.2,0,0)$ & 0.2804 & 0.0397 \\
$(0.4,0,0)$ & 0.2810 & 0.0398  \\
$(0.6,0,0)$ & 0.2809 & 0.0398 \\
$(0.7,0,0)$ & 0.2810 & 0.0398 \\
$(0.8,0,0)$ & 0.2817 & 0.0399  \\
$(0.9,0,0)$ & 0.2817 & 0.0399  \\
\hline
$(0.0707,0,-0.0707)$ & 0.2807 & 0.0398 \\
$(0.1414,0,-0.1414)$ & 0.2808 & 0.0398 \\
$(0.2828,0,-0.2828)$ & 0.2809 & 0.0398 \\
$(0.4243,0,-0.4243)$ & 0.2809 & 0.0398 \\
$(0.495,0,-0.495)$ & 0.2812 & 0.0398  \\
$(0.5657,0,-0.5657)$ & 0.2814 & 0.0399  \\
$(0.6364,0,-0.6364)$ & 0.2816 & 0.0399  \\ \hline
$(0,0,-0.1)$ & 0.2806 & 0.0398  \\
$(0,0,-0.2)$ & 0.2803 & 0.0397  \\
$(0,0,-0.4)$ & 0.2808 & 0.0398   \\
$(0,0,-0.6)$ & 0.2809 & 0.0398 \\
$(0,0,-0.7)$ & 0.2812 & 0.0398 \\
$(0,0,-0.8)$ & 0.2815 & 0.0399  \\
$(0,0,-0.9)$ & 0.2816 & 0.0399 \\ \hline
\end{longtable}
\end{center}

\begin{figure}[htbp]
\begin{subfigure}[b]{0.45\textwidth}
 \includegraphics[width=\textwidth,height=7cm]{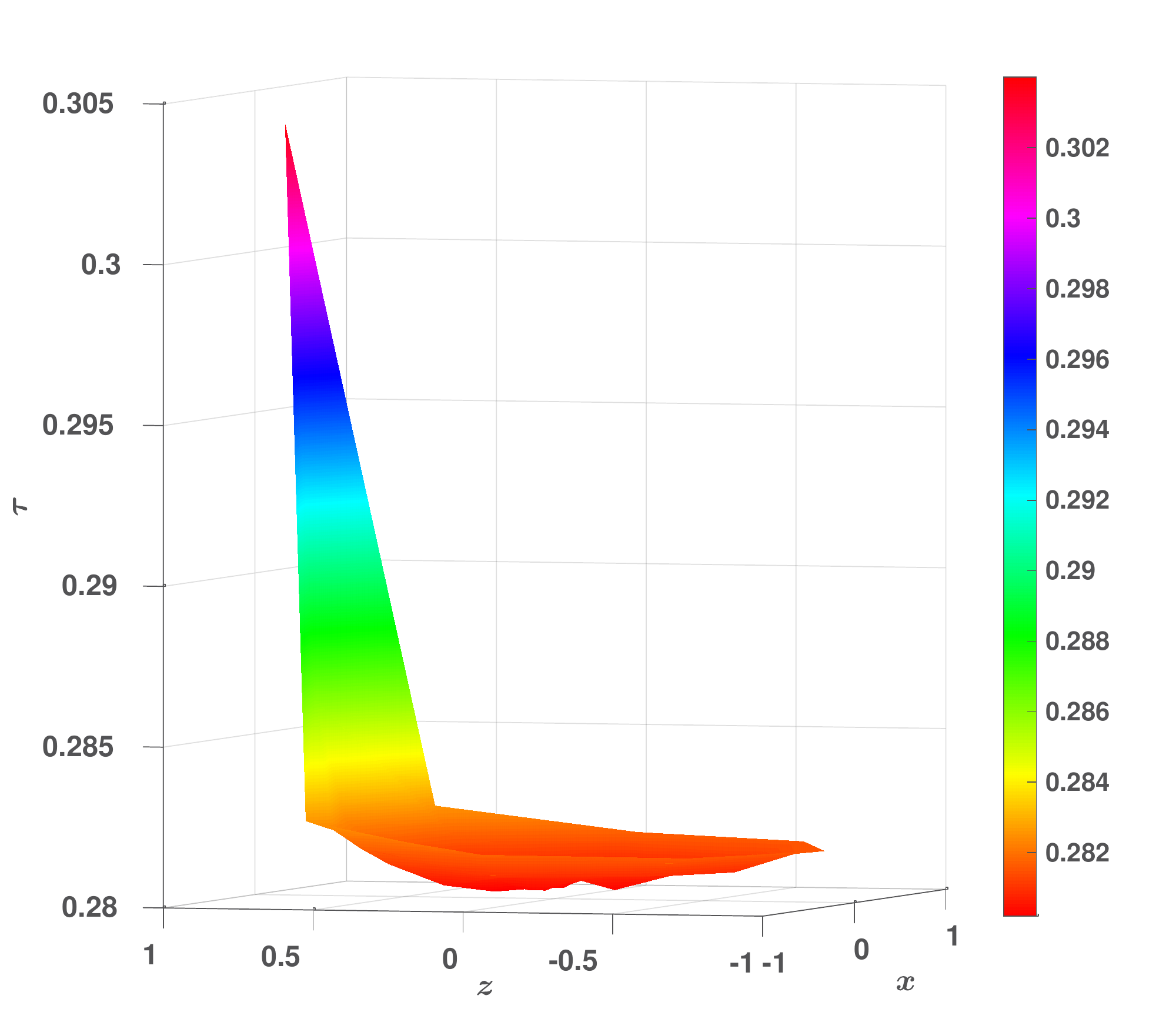}
 \caption{\label{fig:bdd_1_del_1_surface}}
\end{subfigure}
\hfill
\begin{subfigure}[b]{0.45\textwidth}
\includegraphics[width=\textwidth,height=7cm]{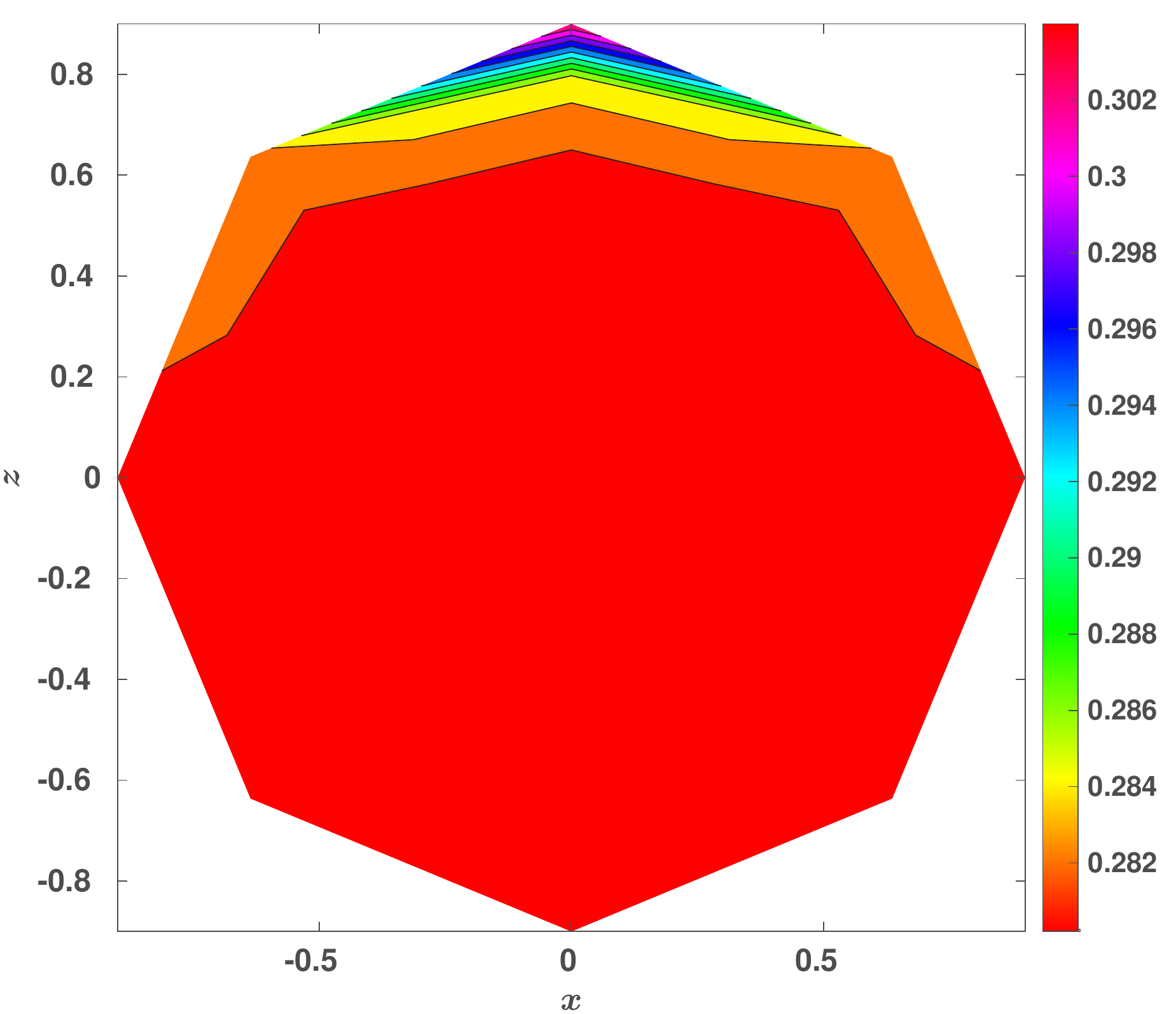}
\caption{\label{fig:bdd_1_del_1_contour}}
\end{subfigure}
\hfill
\caption{ (\subref{fig:bdd_1_del_1_surface}) Surface plot perspective view and (\subref{fig:bdd_1_del_1_contour}) contour plot of the relative boundary time $\tau(\phi, r)$, a Brownian particle spends in $\delta_{\text{region}}$ when $\delta=0.1$ in the unit sphere with the one-trap configuration.}
\label{fig:bdd_1_del_1}
\end{figure}

\begin{figure}[H]
\begin{subfigure}[b]{0.45\textwidth}
 \includegraphics[width=\textwidth,height=6cm]{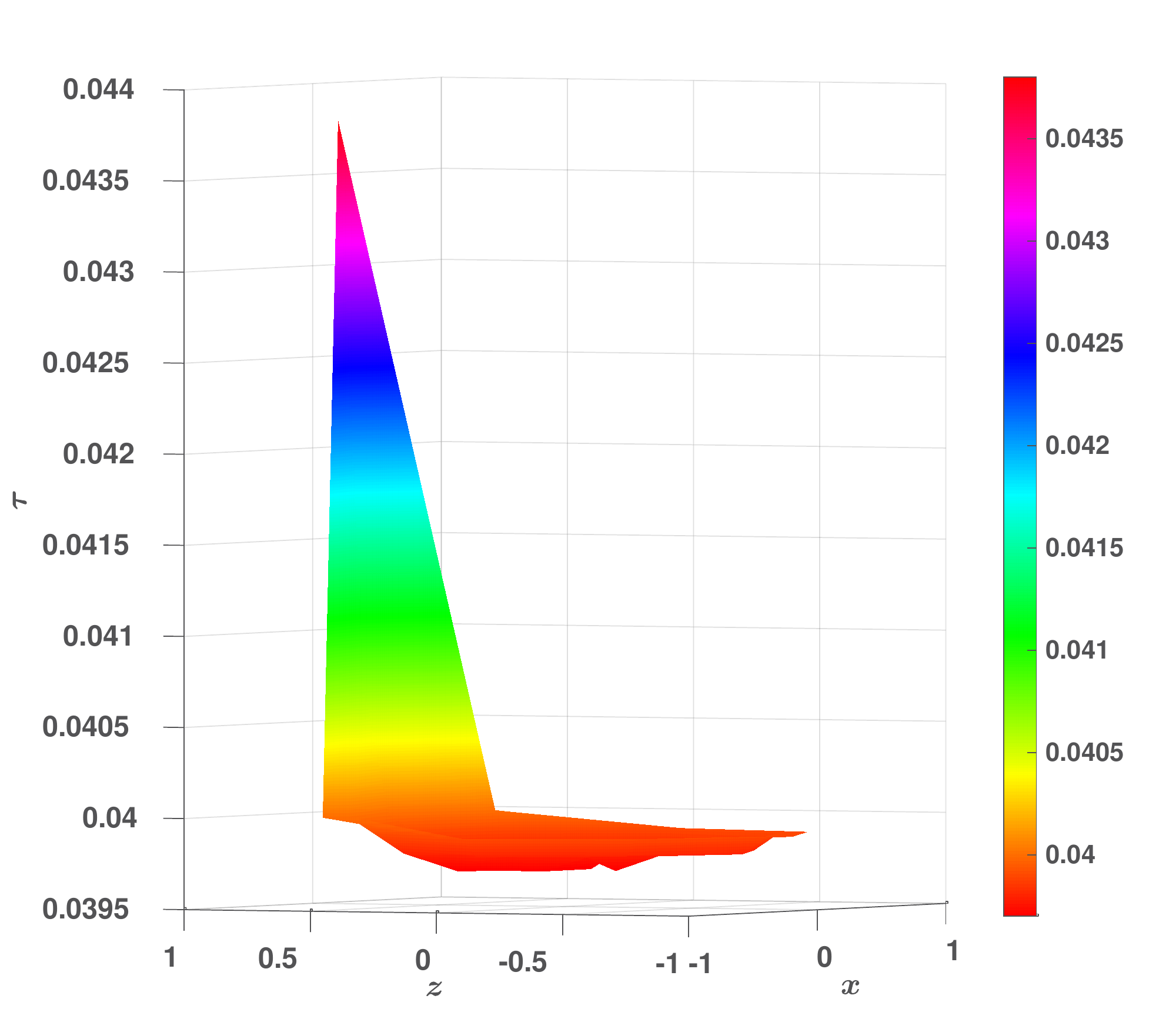}
 \caption{\label{fig:bdd_1_del_2_surface}}
\end{subfigure}
\hfill
\begin{subfigure}[b]{0.45\textwidth}
\includegraphics[width=\textwidth,height=6cm]{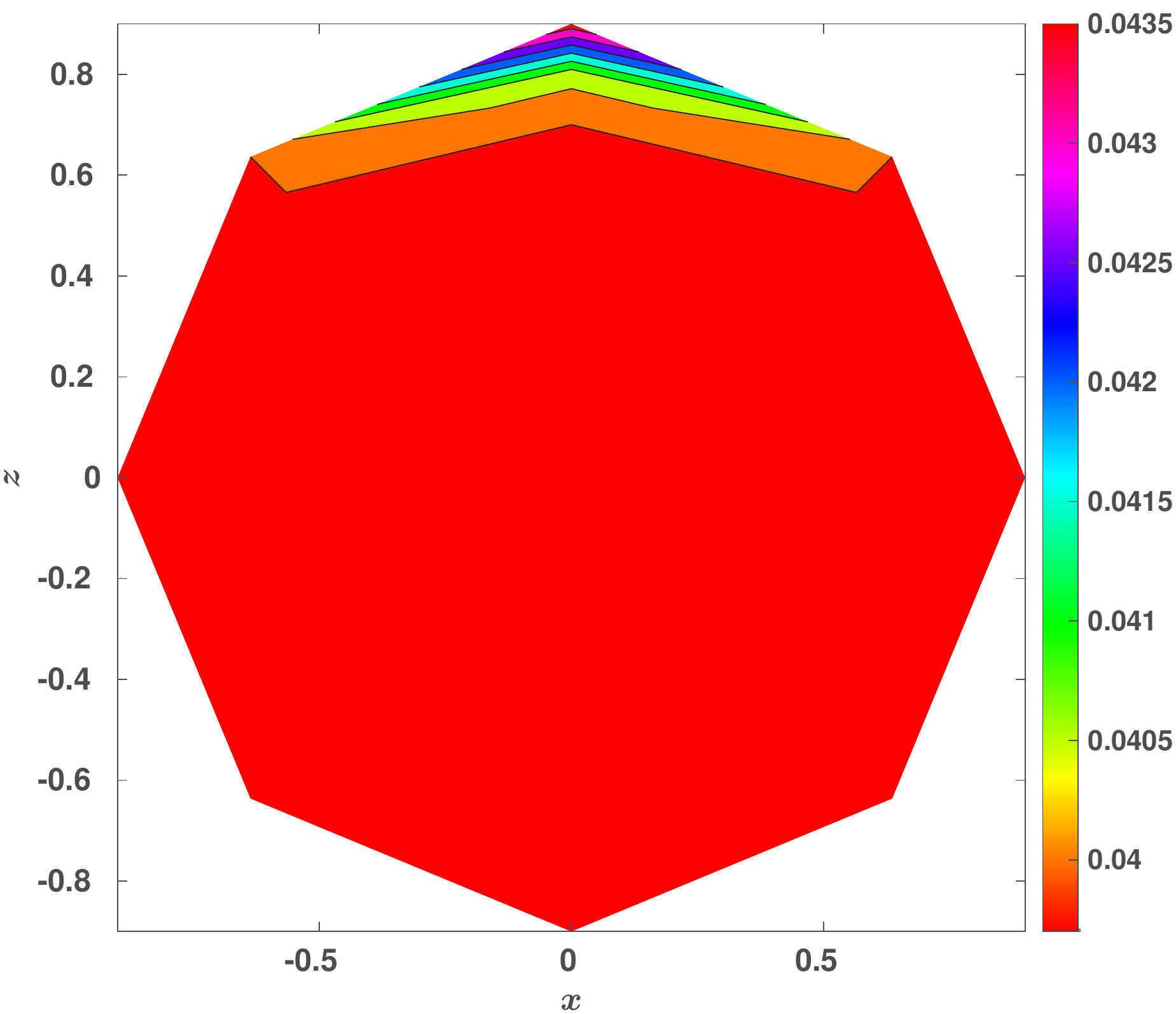}
\caption{\label{fig:bdd_1_del_2_contour}}
\end{subfigure}
\hfill
\caption{ (\subref{fig:bdd_1_del_2_surface}) Surface plot perspective view and (\subref{fig:bdd_1_del_2_contour}) contour plot of the relative boundary time $\tau(\phi, r)$, a Brownian particle spends in $\delta_{\text{region}}$ when $\delta=0.01$ in the unit sphere with the one-trap configuration. }
\label{fig:bdd_1_del_2}
\end{figure}

For the one-trap configuration, the boundary time $\tau$, when expressed as a function of launching coordinates, $\tau (\phi, r)$ is constant $\sim 28.29 \%$  throughout the domain, inside $\delta_{\text{region}}$ for $\delta=0.1$, which encompassed the $\sim 27.1 \%$ of the total volume of the domain. The study reveals that the $\tau(\phi,r)$ is constant $\sim 4.01 \%$ for $\delta_{\text{region}}$ when $\delta=0.01$, which is $\sim 2.97 \%$ of the total volume of the sphere with one absorbing window at the north pole.

\subsection{Two Traps}

For the two-trap configuration in the unit sphere when the traps are placed at the north pole $(0,0,1)$, and the south pole $(0,0,-1)$, we are interested in calculating the boundary time $\tau$, when expressed as the function of the spherical angle $\phi$ and the radial distance $r$. For this work, we have launched the Brownian particles from various positions \eqref{eq:lauch_coords}, each uniquely defined by a $\phi$ and a $r$, for $10^4$ times, which is the optimal number of runs required for a good agreement of the asymptotic results with the direct numerical simulation results. For the two-trap configuration, we have chosen the $\delta$ to be $0.1$ and $0.01$, for which we have calculated the boundary time $\tau(\phi,r)$, when a particle is inside the $\delta_{\text{region}}$.

\begin{figure}[H]
\begin{subfigure}[b]{0.45\textwidth}
 \includegraphics[width=\textwidth,height=7cm]{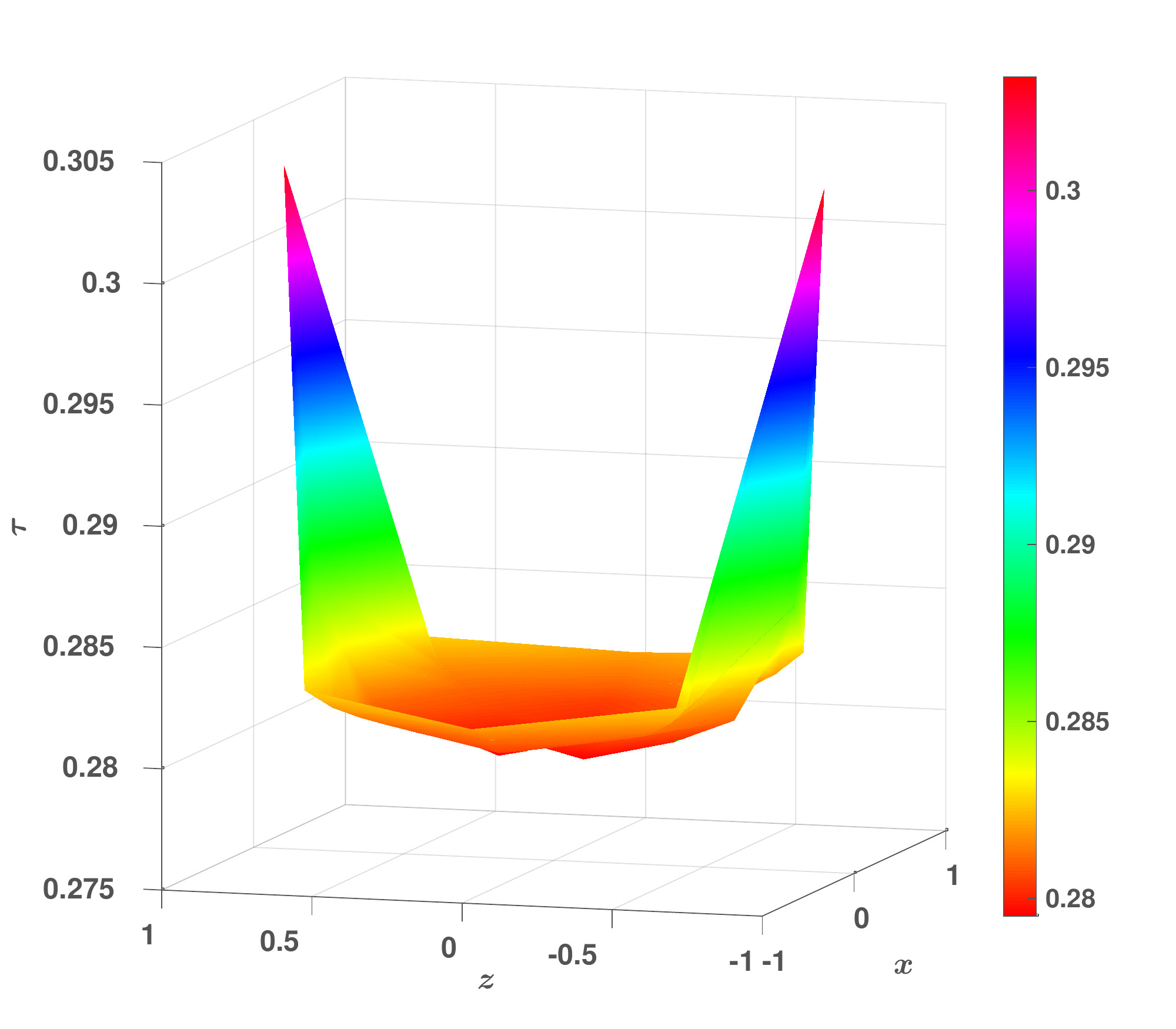}
 \caption{\label{fig:bdd_2_del_1_surface}}
\end{subfigure}
\hfill
\begin{subfigure}[b]{0.45\textwidth}
\includegraphics[width=\textwidth,height=7cm]{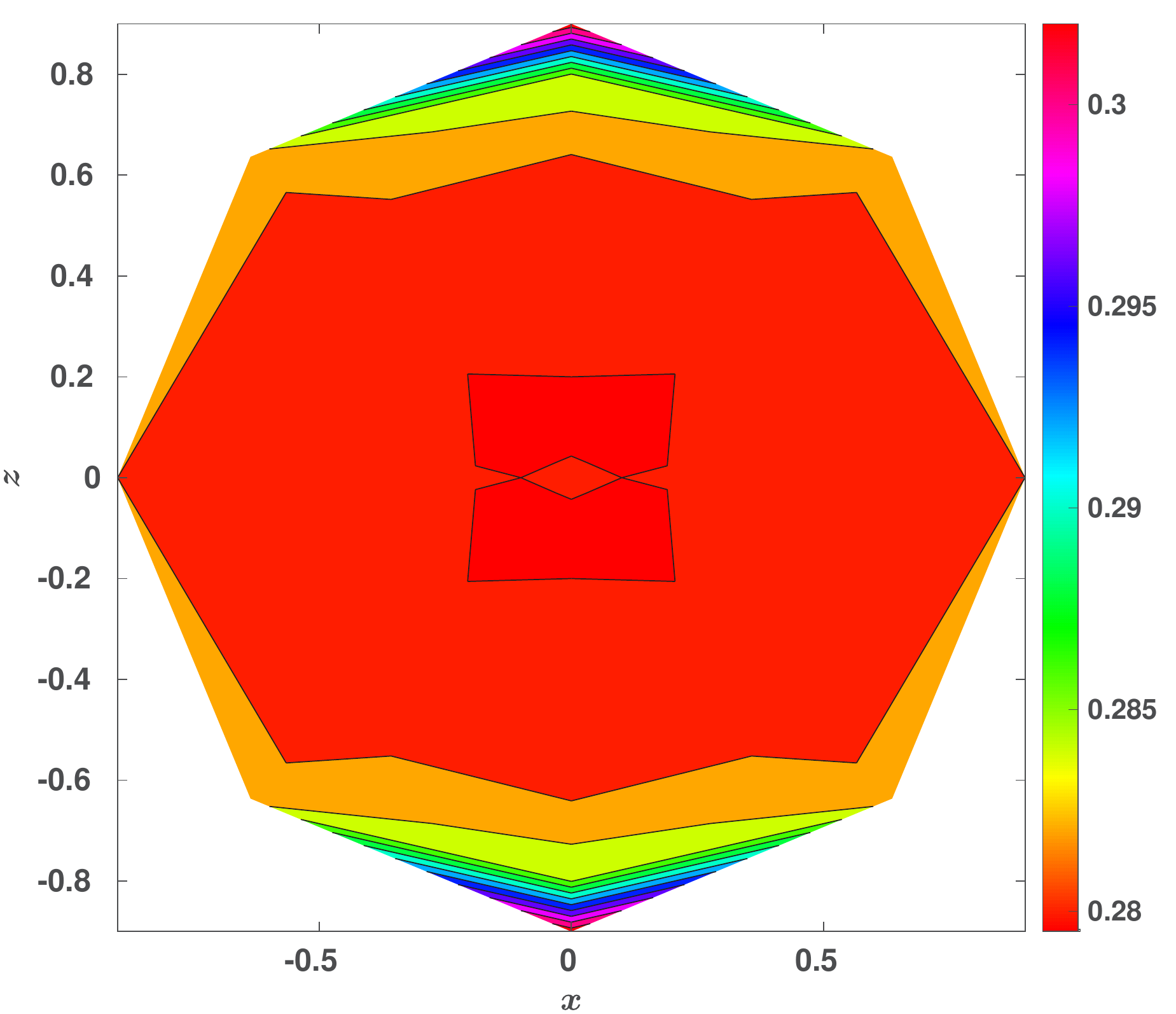}
\caption{\label{fig:bdd_2_del_1_contour}}
\end{subfigure}
\hfill
\caption{ (\subref{fig:bdd_2_del_1_surface}) Surface plot perspective view and (\subref{fig:bdd_2_del_1_contour})
contour plot of the relative boundary time $\tau(\phi, r)$, a Brownian particle spends in $\delta_{\text{region}}$ when $\delta=0.1$ in the unit sphere with the two-trap configuration. }
\label{fig:bdd_2_del_1}
\end{figure}

\begin{center}
\begin{longtable}{|c|c|c|}
\caption{A comprehensive list illustrating the boundary time $\tau$, when a Brownian particle is launched in the unit sphere with the two-trap configuration.}
\label{table:boundary_dynamics_1trap}
\\
\hline
\multirow{2}{*}{\textbf{Launching Coordinates}} &
\multicolumn{2}{c|}{\textbf{Boundary Time}}  \\
& $\delta=0.1$ & $\delta=0.01$ \\ \hline

\endfirsthead

\multicolumn{2}{c}%
{{\bfseries \tablename\ \thetable{} -- Continued}} \\
\hline \multirow{2}{*}{\textbf{Launching Coordinates}} & \multicolumn{2}{c|}{\textbf{Boundary Time}} \\
& $\delta=0.1$ & $\delta=0.01$ \\ \hline
\endhead

\hline \multicolumn{3}{|r|}{{Continued on next page}} \\ \hline
\endfoot

\hline

\hline Average $\tau$ & 0.2821
 & 0.0400
  \\ \hline

  \endlastfoot

$(0,0,0)$ & 0.2803 & 0.0397   \\
$(0,0,0.1)$ & 0.2796 &  0.0396\\
$(0,0,0.2)$ & 0.2800 & 0.0397 \\
$(0,0,0.4)$ & 0.2801 & 0.0397 \\
$(0,0,0.6)$ & 0.2811 & 0.0399  \\
$(0,0,0.7)$ & 0.2833 & 0.0402  \\
$(0,0,0.8)$ & 0.2859 & 0.0407 \\
$(0,0,0.9)$ & 0.3032 & 0.0439  \\  \hline
$(0.0707,0,0.0707)$ & 0.2796 & 0.0396   \\
$(0.1414,0,0.1414)$ & 0.2795 & 0.0396   \\
$(0.2828,0,0.2828)$ & 0.2806 & 0.0397   \\
$(0.4243,0,0.4243)$ & 0.2811 & 0.0398  \\
$(0.495,0,0.495)$ & 0.2815 & 0.0399   \\
$(0.5657,0,0.5657)$ & 0.2820 & 0.0400  \\
$(0.6364,0,0.6364)$ & 0.2828 & 0.0401  \\  \hline
$(0.1,0,0)$ & 0.2800 & 0.0397  \\
$(0.2,0,0)$ & 0.2801 & 0.0397  \\
$(0.4,0,0)$ & 0.2802 & 0.0397   \\
$(0.6,0,0)$ & 0.2805 & 0.0397  \\
$(0.7,0,0)$ & 0.2812 & 0.0398  \\
$(0.8,0,0)$ & 0.2817 & 0.0399   \\
$(0.9,0,0)$ & 0.2820 & 0.0399  \\
\end{longtable}
\end{center}

\begin{figure}[H]
\begin{subfigure}[b]{0.45\textwidth}
 \includegraphics[width=\textwidth,height=7cm]{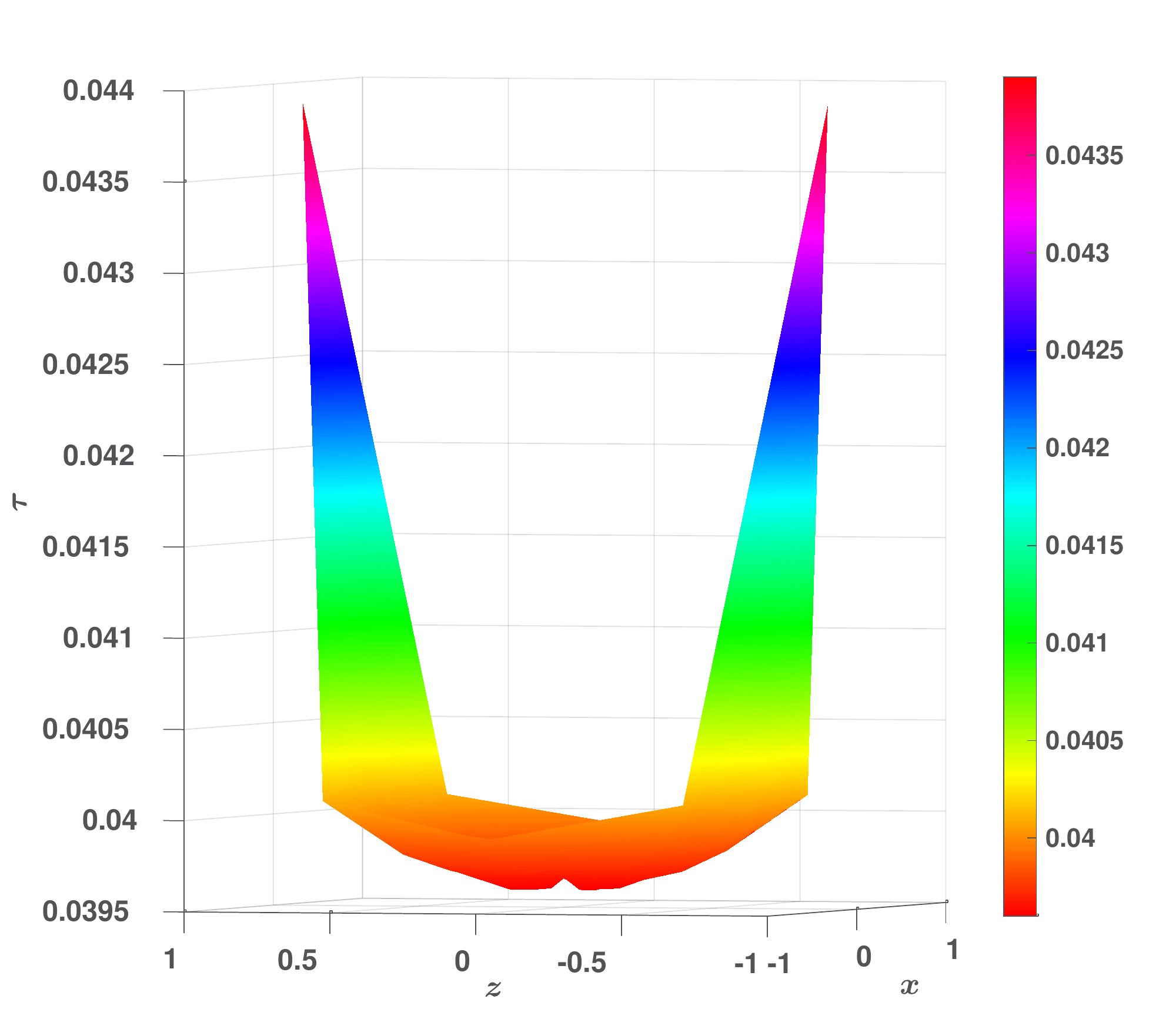}
 \caption{\label{fig:bdd_2_del_2_surface}}
\end{subfigure}
\hfill
\begin{subfigure}[b]{0.45\textwidth}
\includegraphics[width=\textwidth,height=7cm]{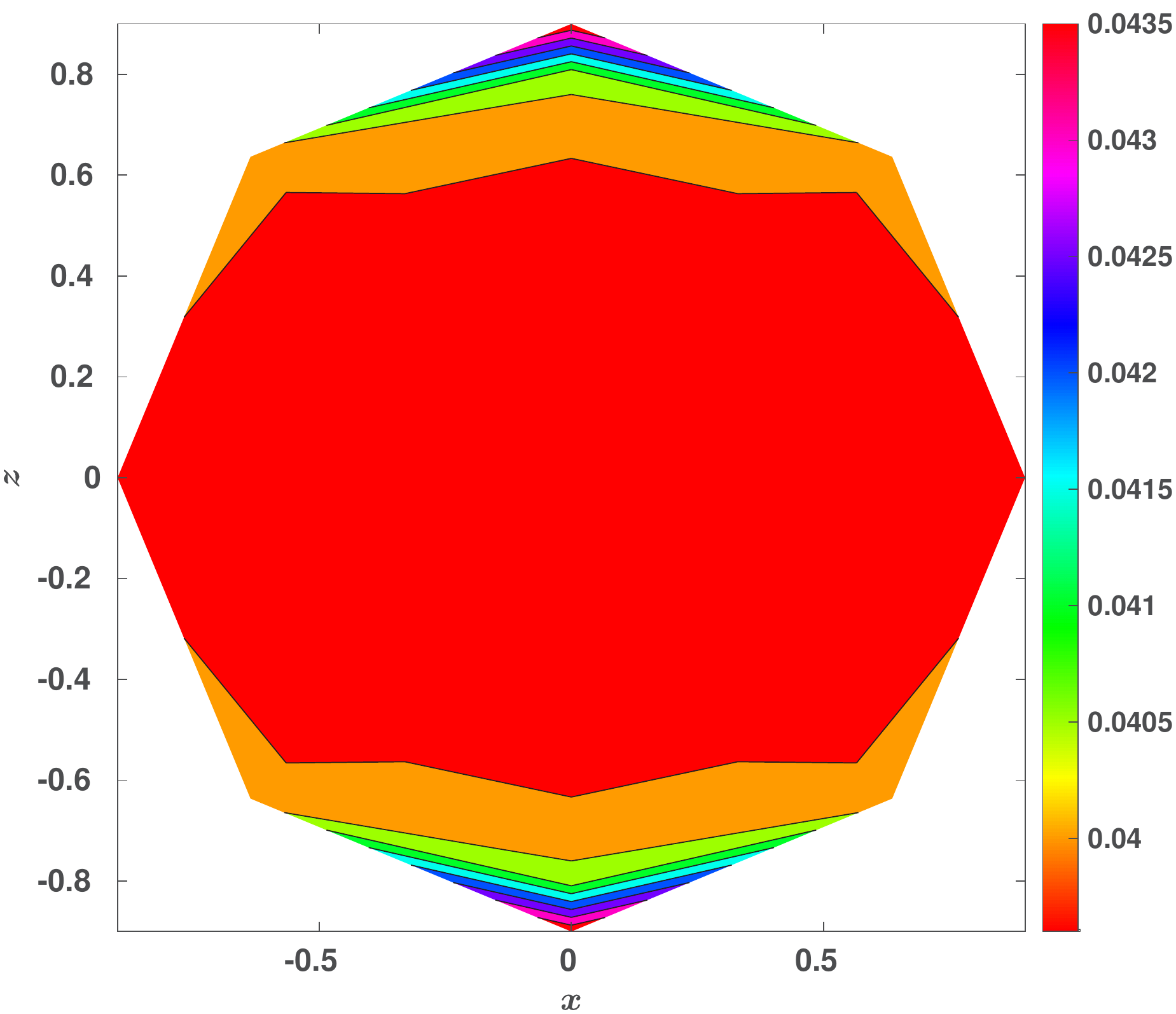}
\caption{\label{fig:bdd_2_del_2_contour}}
\end{subfigure}
\hfill
\caption{ (\subref{fig:bdd_2_del_2_surface}) Surface plot perspective view and (\subref{fig:bdd_2_del_2_contour}) contour plot of the relative boundary time $\tau(\phi, r)$, a Brownian particle spends in $\delta_{\text{region}}$ when $\delta=0.01$ in the unit sphere with the two-trap configuration. }
\label{fig:bdd_2_del_2}
\end{figure}

For the two-trap configuration, the boundary time $\tau$, when expressed as a function of launching coordinates, $\tau (\phi, r)$ is constant $\sim 28.21 \%$  throughout the domain, inside $\delta_{\text{region}}$ for $\delta=0.1$, which encompassed the $\sim 27.1 \%$ of the total volume of the domain. The study reveals that the $\tau(\phi,r)$ is constant $\sim 4.002 \%$ for $\delta_{\text{region}}$ when $\delta=0.01$, which is $\sim 2.97 \%$ of total volume of the sphere with two absorbing windows at the north and the south pole.

\begin{center}
\begin{longtable}{|c|c|c|c|}
\hline
$\delta$ & {{Boundary layer volume fraction $|\Omega_\delta|/|\Omega|$}} &$\tau$, single-trap & $\tau$, two-trap    \\ \hline\hline
\endfirsthead

\hline
$\delta$ & {{Boundary layer volume fraction $|\Omega_\delta|/|\Omega|$}} &$\tau$, single-trap & $\tau$, two-trap    \\ \hline\hline
\endhead

\hline
\endfoot

\hline
\caption{  Average boundary time fraction values $\tau$ \eqref{eq:tau:delta} vs.~volume fraction of the boundary region in the case of isotropic diffusion of Brownian particles in the unit sphere with one and two polar traps of radius $\epsilon=0.01$, for boundary layer thicknesses $\delta = 0.1$ and $\delta = 0.01$.}
\label{table:boundary_dynamics_1_2trap}
\endlastfoot

0.1    & 27.10\%  & 28.29\% & 28.21\% \\
\hline

0.01   & 2.970\% & 4.014\% & 4.002\% \\
\end{longtable}
\end{center}

\section{Comparison of Brownian Dynamics Simulations for Isotropic vs.~Anisotropic
diffusion}\label{sec:var_vs_constant_diffisuion}

This section presents the study concerning Brownian dynamics simulations' effect when a particle undergoes anisotropic diffusion. We also exhibit the comparison of direct numerical simulations used for computation of MFPT and study of boundary dynamics for a unit sphere with a single boundary trap at the north pole for both the anisotropic and isotropic diffusion.

\begin{figure}[H]
\begin{subfigure}[b]{0.45\textwidth}
 \includegraphics[width=1.1\textwidth]{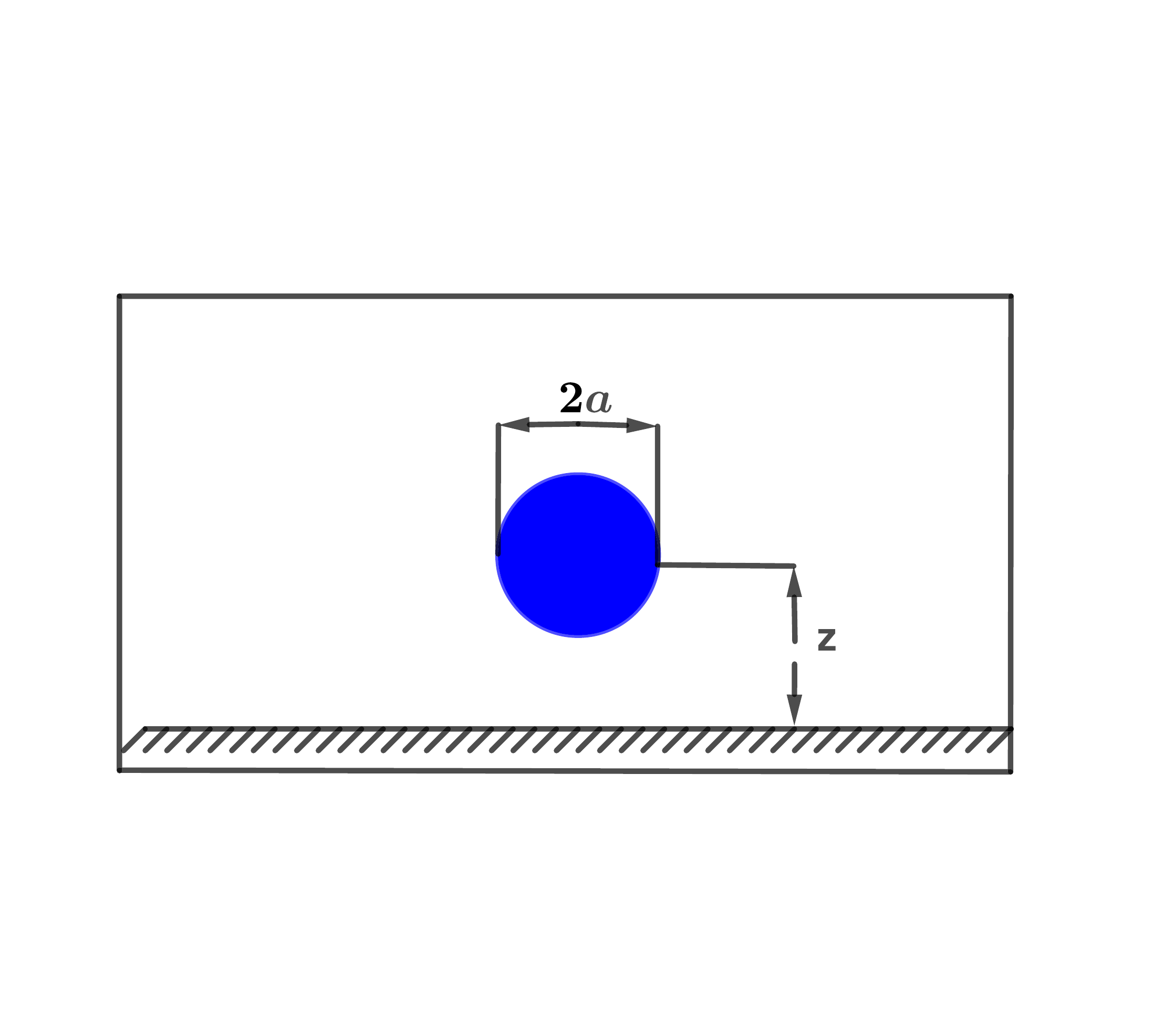}
\caption{}
\end{subfigure}
\hfill
\begin{subfigure}[b]{0.47\textwidth}
\includegraphics[width=\textwidth]{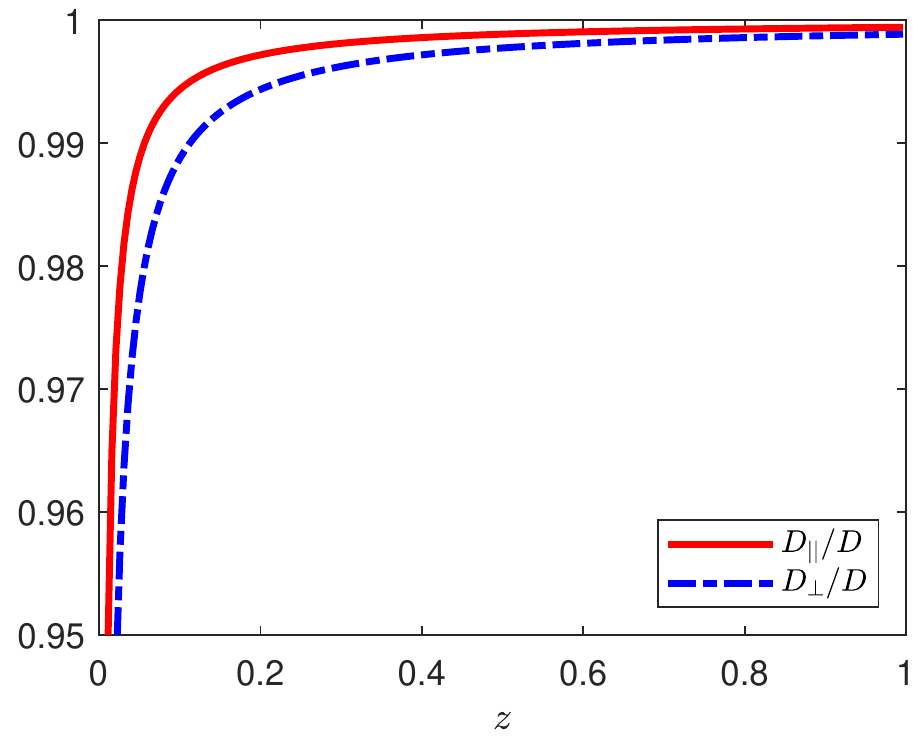}
 \caption{}
\end{subfigure}
\hfill
\caption{  (a) Brownian particle size and distance to the boundary. (b) Ratios $D_{\parallel}/D$ and $D_{\perp}/D$ for anisotropic diffusion coefficients as functions of $z$ when $a=0.001$.}
\label{fig:intro_var_dif_1trap}
\end{figure}

For this work, have launched the Brownian particles from the various position on the $Z$-axis, for $\sim 1.3 \times 10^4$ times. For the Brownian particle with anisotropic diffusion, we have used the theory and results by Lin et al. \cite{Lin}. They presented the theory and provided the experimental data, which shows that the diffusion of Brownian particles near the wall is different, i.e., the diffusion coefficient changes based on the distance to the wall to particle size ratio. They also showed that the parallel \eqref{eq:diff_parallel} and perpendicular \eqref{eq:diff_perp} motions to the wall have different diffusion coefficients.

In \cite{Lin,booklindif}, it is explicitly shown that for a Brownian particle with radius $a$ and distant $z$ from the boundary of a flat wall (See Fig.~\ref{fig:intro_var_dif_1trap}), the diffusion coefficients for parallel $D_{\parallel}$ and perpendicular $D_{\perp}$ to one flat wall are

\begin{align}
D_{\parallel} = \dfrac{k_B T}{6 \pi \eta a\lambda_{\parallel}a } = \lambda_{\parallel}^{-1} D_0
\end{align}

and

\begin{align}
D_{\perp} = \dfrac{k_B  T}{6 \pi \eta a\lambda_{\perp} a } = \lambda_{\perp}^{-1} D_0,
\end{align}

where

\begin{equation}\label{eq:D_O}
 D_0 = \dfrac{k_B T}{6 \pi \eta a},
\end{equation}

\begin{equation}\label{eq:diff_parallel}
\lambda_{\parallel}^{-1} \cong 1 - \dfrac{9}{16} \Big(\dfrac{a}{z} \Big) + \dfrac{1}{8} \Big(\dfrac{a}{z} \Big)^3 - \dfrac{45}{256} \Big(\dfrac{a}{z} \Big)^4 -\dfrac{1}{16} \Big(\dfrac{a}{z} \Big)^5 + \mathcal{O} \Big( \dfrac{a}{z} \Big)^6
\end{equation}

and

\begin{equation}\label{eq:diff_perp}
\lambda_{\perp}^{-1} \cong 1 - \dfrac{9}{8} \Big(\dfrac{a}{z} \Big) + \dfrac{1}{2} \Big(\dfrac{a}{z} \Big)^3 + \mathcal{O} \Big( \dfrac{a}{z} \Big)^4.
\end{equation}

\begin{figure}[htbp]
\begin{subfigure}[b]{.45\textwidth}
  \centering
  \subcaptionbox{\label{fig:dper_vs_cons_diff}}	{\includegraphics[width=\textwidth,height=6cm]{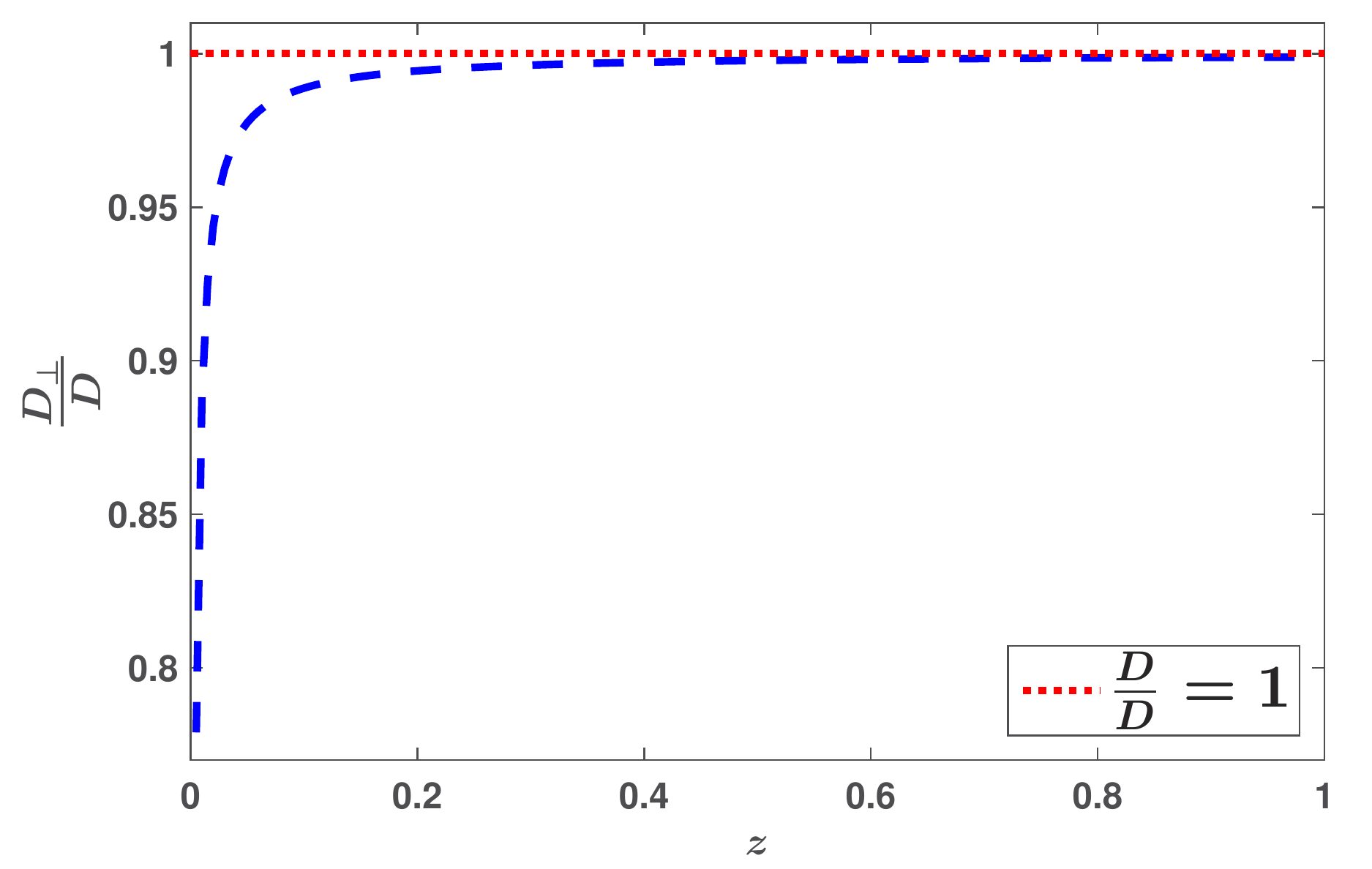}}
\end{subfigure}
\begin{subfigure}[b]{.45\textwidth}
\centering
\subcaptionbox{\label{fig:dpar_vs_cons_diff}}		{\includegraphics[width=\textwidth,height=6cm]{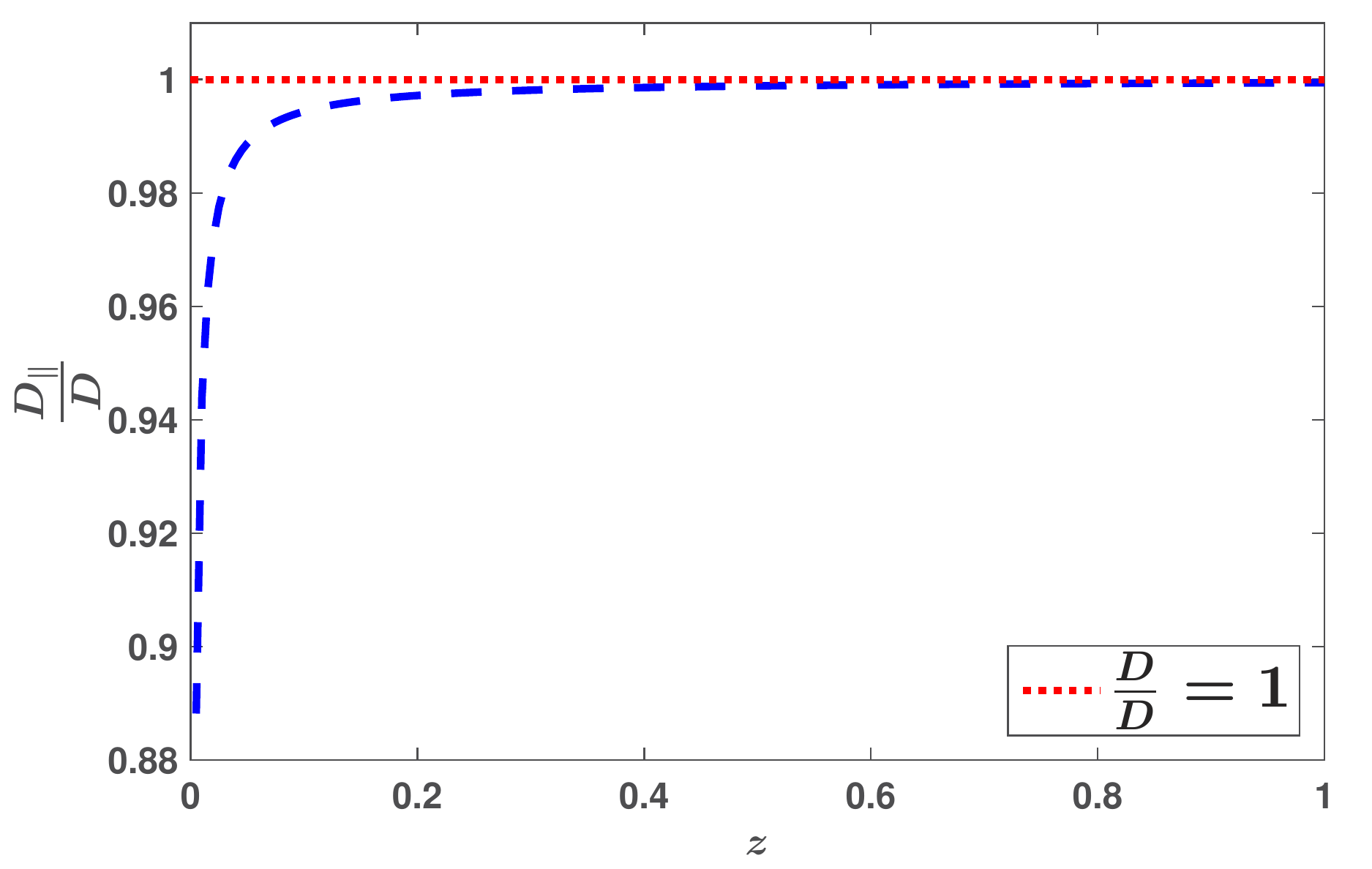}}
\end{subfigure}
\hfill
\caption{Comparison of (\subref{fig:dper_vs_cons_diff}) $D_{\perp}/D$ and (\subref{fig:dpar_vs_cons_diff}) $D_{\parallel}/D$ ( where $D_{\perp}$ \eqref{eq:diff_perp} and $D_{\parallel}$ \eqref{eq:diff_parallel} respectively represents the diffusion coefficients for perpendicular and parallel motion of a Brownian particles relative to the boundary) to $D/D$, where $D$ represents the constant of isotropic diffusion.}
\label{fig:aniso_vs_iso_diff}
\end{figure}

In the above formula for diffusion coefficient of the sphere $D_0$ \eqref{eq:D_O}, $k_B$ is the Boltzmann constant and the $T$ is the temperature of the system. For our work, the constant $D_0$ equals $D$, which is unit for our case.
With all these piece of information, we modified our working \verb|MATLAB| code to account for the Brownian dynamics simulations for the case of anisotropic diffusion. After drawing random displacement vectors, we can interpret the Brownian motion step as parallel and perpendicular to the wall due to the Stokes equation's linearity. Then, we scaled them accordingly by multiplying them by corresponding diffusivities formulas (See ~\eqref{eq:diff_parallel} \eqref{eq:diff_perp}).

The anisotropic diffusion kicks in when the distance to the boundary $z$ (See Fig.~\ref{fig:intro_var_dif_1trap}) is less than the layer width of the boundary, defined as the size of the particles times boundary factor. Here boundary factor is a constant, which is about 22.5, and we have used the particle size as $10^{-3}$. At distances about this (and more) from the boundary, both the perpendicular diffusivity $D_{\perp}$ and parallel diffusivity $D_{\parallel}$ are $> 0.95 D$. It can be assumed to equal to the constant of isotropic diffusivity $D$, which is the unit for our work. Hence, for this case, anisotropy and boundary effects can be ignored.

Fig.~\ref{fig:aniso_vs_iso_diff} displays the comparison of $\dfrac{D_{\parallel}}{D}$ and $\dfrac{D_{\perp}}{D}$, when expressed as a function of distance from the boundary, i.e., $z$ to the $\dfrac{D}{D}$, where $D$ is the isotropic diffusion constant.
\subsection{Averaged Brownian Escape Times for Anisotropic Diffusion}\label{sec:var_vs_constant_diffisuion_escape_time}

\begin{figure}[H]
  \centering
  \includegraphics[width=12cm,height=7cm]{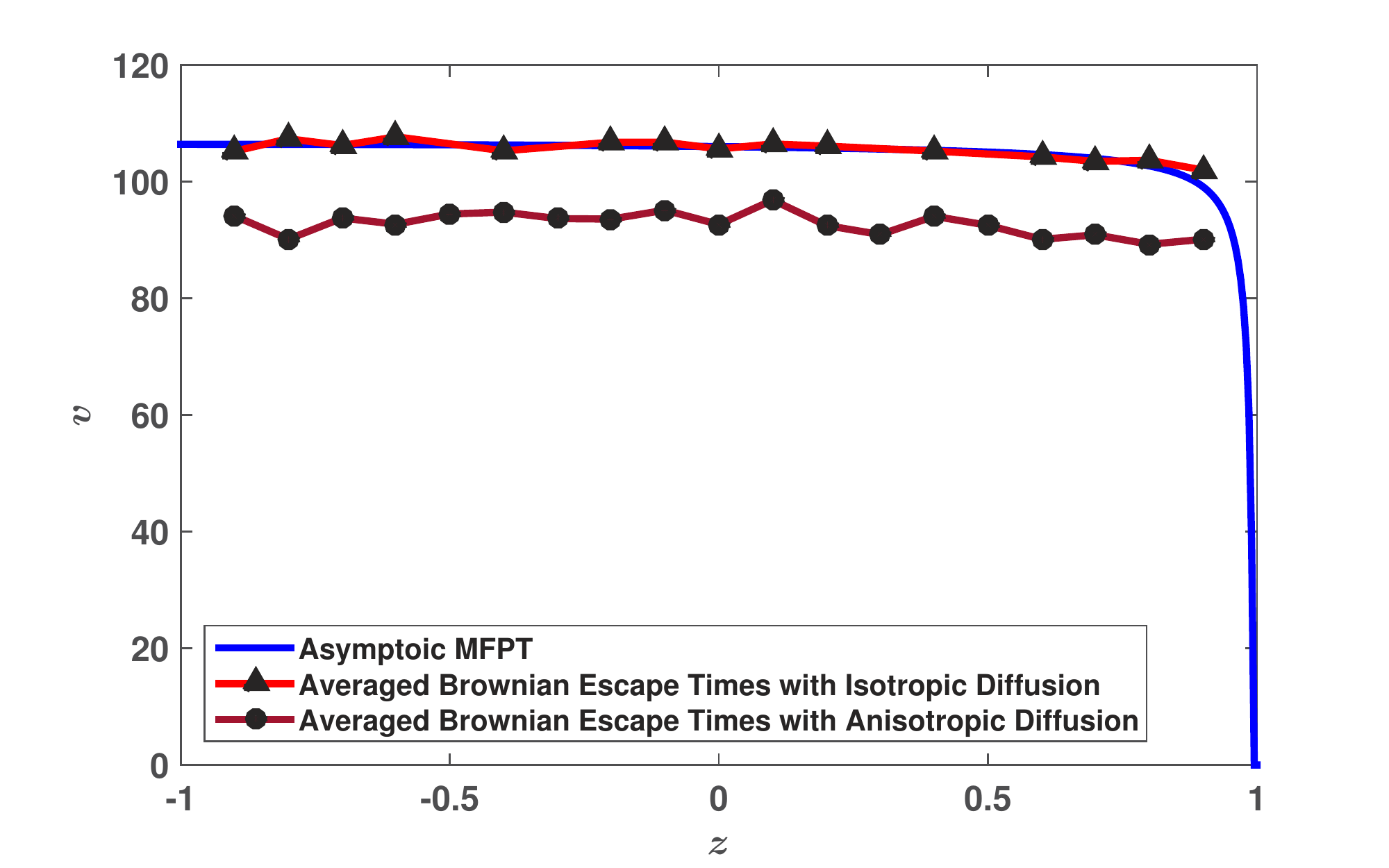}
\caption{Averaged Brownian escape times for anisotropic and isotropic diffusion vs MFPT, when the particles are launched from the $Z$-axis}
\label{fig:mfpt_z_var_dif}
\end{figure}

This section mulls over the averaged Brownian escape times for a Brownian particle undergoing the anisotropic diffusion. For this work, we have performed a direct numerical simulation using the modified \verb|MATLAB| code, which accounts for the boundary reflections, anisotropic diffusion, and escape conditions rigorously. For better analysis, we have launched the Brownian particles from the various position on the $Z$-axis, for $\sim 1.3 \times 10^4$ times, and recorded the averaged Brownian escape times for these launching coordinates (See Table.~\ref{table:mfpt_1trap_diffusion}). Fig.~\ref{fig:mfpt_z_var_dif} compares the averaged Brownian escape times for both the anisotropic and isotropic diffusion with the asymptotic MFPT.

\begin{center}
\begin{longtable}{|c|c|}
\caption{A comprehensive list illustrating the averaged Brownian escape times when a Brownian particle is launched in the unit sphere with the one trap configuration undergoing anisotropic diffusion.
}
\label{table:mfpt_1trap_diffusion}
\\
\hline \multicolumn{1}{|c|}{\textbf{Launching Coordinates}} & \multicolumn{1}{c|}{\textbf{Averaged Brownian Escape Times}}\\ \hline
\endfirsthead

\multicolumn{2}{c}%
{{\bfseries \tablename\ \thetable{} -- Continued}} \\
\hline \multicolumn{1}{|c|}{\textbf{Launching Coordinates}} & \multicolumn{1}{c|}{\textbf{Averaged Brownian Escape Times}} \\ \hline
\endhead

\hline \multicolumn{2}{|r|}{{Continued on next page}} \\ \hline
\endfoot

\hline
\endlastfoot
$(0,0,-0.9)$ & 94.1280 \\
$(0,0,-0.8)$ & 90.1409\\
$(0,0,-0.7)$ & 93.7804\\
$(0,0,-0.6)$ & 92.7116\\
$(0,0,-0.5)$ & 94.4702 \\
$(0,0,-0.4)$ & 94.7872\\
$(0,0,-0.3)$ & 93.6790\\
$(0,0,-0.2)$ & 93.5953 \\
$(0,0,-0.1)$ & 95.1294 \\ \hline
$(0,0,0.0)$ & 92.6663 \\
$(0,0,0.1)$ & 96.8617 \\
$(0,0,0.2)$ & 92.4898 \\
$(0,0,0.3)$ & 90.9424 \\
$(0,0,0.4)$ & 94.0706\\
$(0,0,0.5)$ & 92.5329\\
$(0,0,0.6)$ & 90.1264\\
$(0,0,0.7)$ & 90.8989\\
$(0,0,0.8)$ & 89.2767 \\
$(0,0,0.9)$ & 90.1289\\
\end{longtable}
\end{center}

\subsection{Boundary Times for Anisotropic Diffusion}\label{sec:var_vs_constant_diffisuion_boundary_time}

This section presents the study of boundary time $\tau$ when a Brownian particle is launched from inside a unit sphere with an absorbing window at the north pole undergoing the anisotropic diffusion. To study the Brownian particle's boundary dynamics, we have launched them from the various position on the $Z$-axis and calculated the boundary time $\tau$ for each of the launches.

\begin{figure}[H]
\begin{subfigure}[b]{.45\textwidth}
  \centering
  \subcaptionbox{\label{fig:del_1_var_diff}}	{\includegraphics[width=1.1\textwidth,height=6cm]{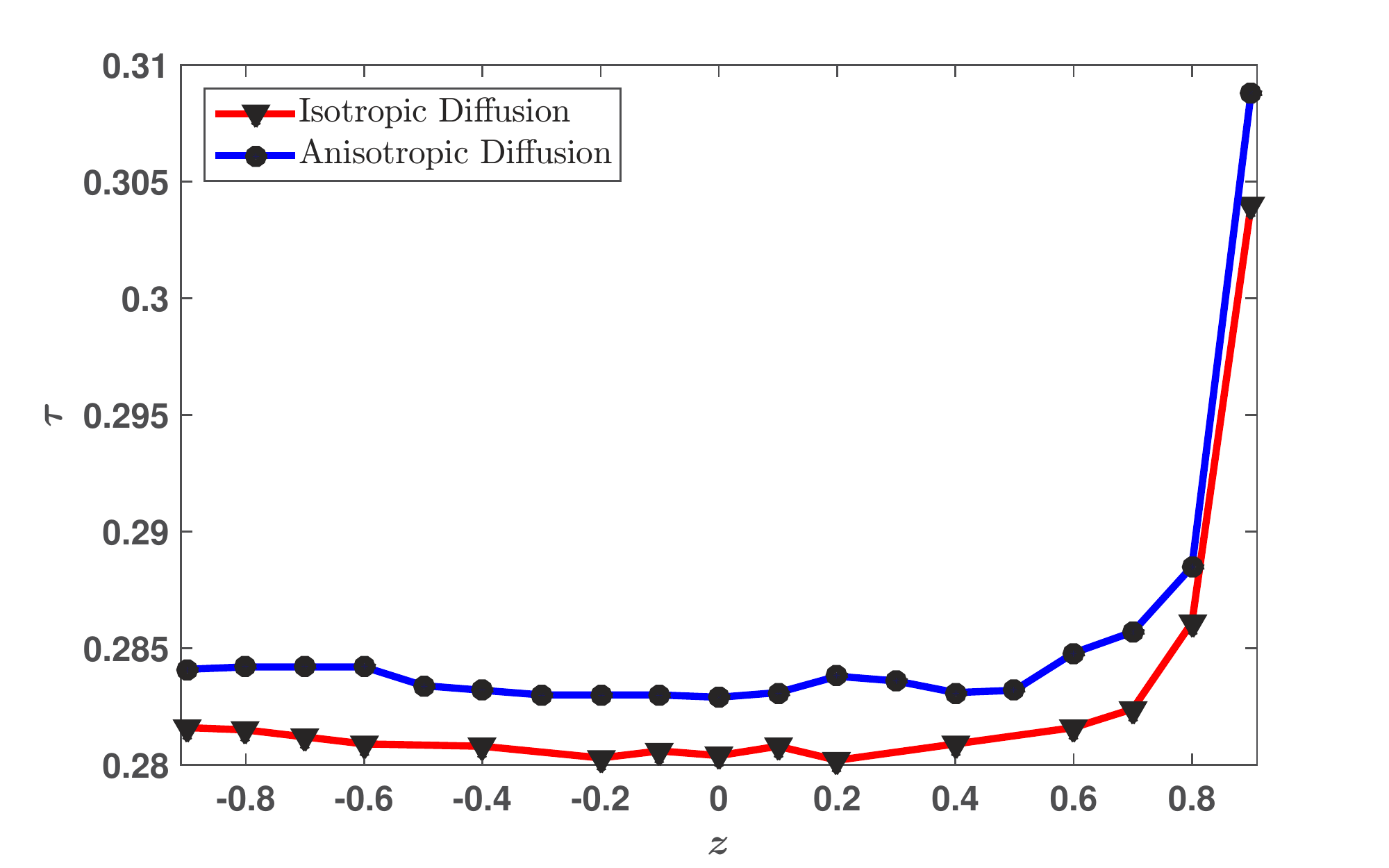}}
\end{subfigure}
\begin{subfigure}[b]{.45\textwidth}
\centering
\subcaptionbox{\label{fig:del_2_var_diff}}		{\includegraphics[width=1.1\textwidth,height=6cm]{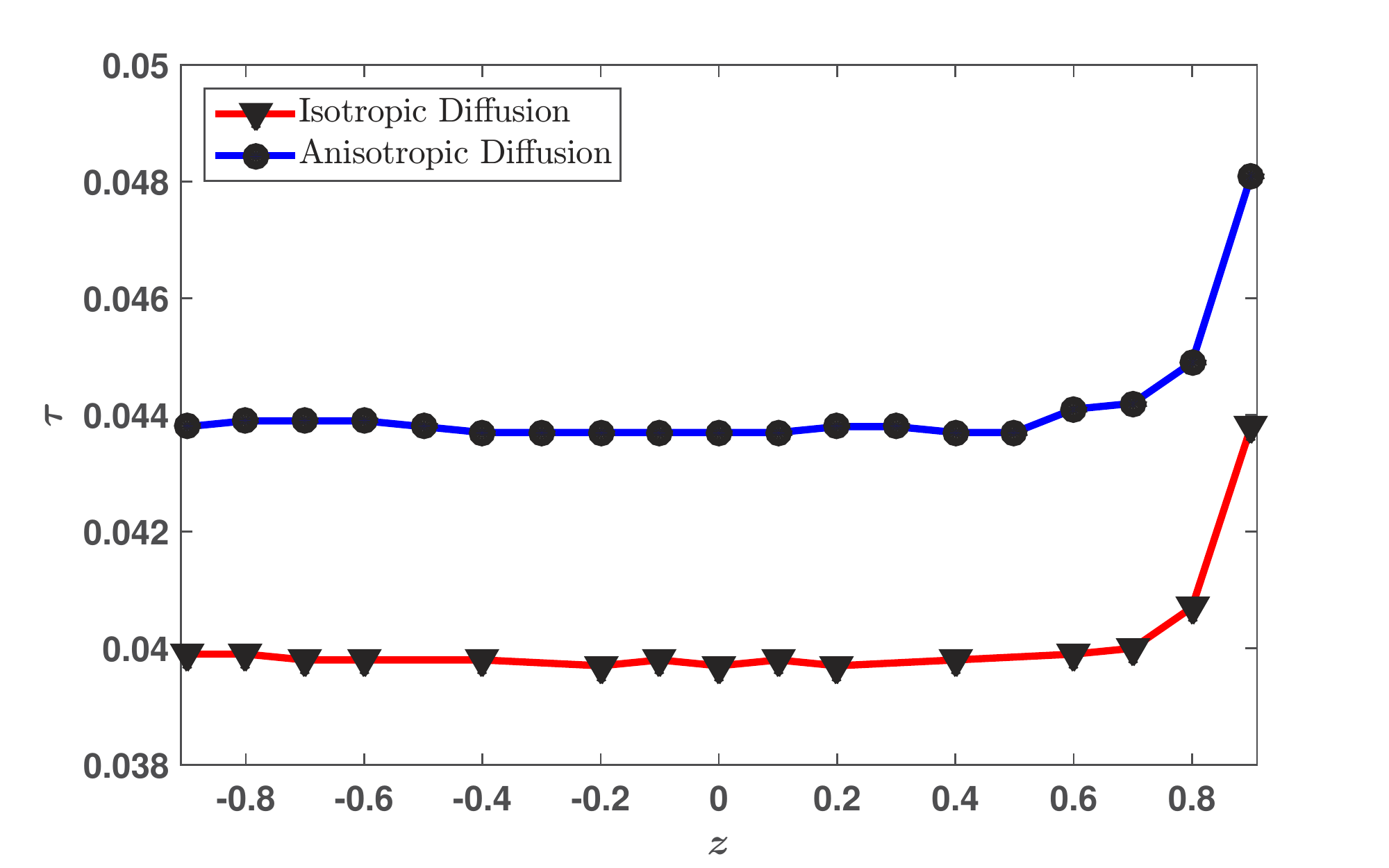}}
\end{subfigure}
\hfill
\caption{ Comparison of boundary time $\tau$ for isotropic vs anisotropic diffusion when a Brownian particle is inside $\delta_{region}$ for (\subref{fig:del_1_var_diff}) $\delta=0.1$, (\subref{fig:del_2_var_diff}) $\delta =0.01$.}
\label{fig:del_var_diff}
\end{figure}

\begin{center}
\begin{longtable}{|c|c|c|}
\caption{A comprehensive list illustrating the boundary time $\tau$, when a Brownian particle is launched in the unit sphere with the one-trap configuration undergoing the anisotropic diffusion.}
\label{table:boundary_dynamics_1trap_diffusion}
\\
\hline
\multirow{2}{*}{\textbf{Launching Coordinates}} &
\multicolumn{2}{c|}{\textbf{Boundary Time}}  \\
& $\delta=0.1$ & $\delta=0.01$\\ \hline

\endfirsthead

\multicolumn{2}{c}%
{{\bfseries \tablename\ \thetable{} -- Continued}} \\
\hline \multirow{2}{*}{\textbf{Launching Coordinates}} & \multicolumn{2}{c|}{\textbf{Boundary Time}} \\
& $\delta=0.1$ & $\delta=0.01$ \\ \hline
\endhead

\hline \multicolumn{3}{|r|}{{Continued on next page}} \\ \hline
\endfoot

\hline

\hline Average $\tau$ & 0.2853
 & 0.0441
  \\ \hline

  \endlastfoot

$(0,0,-0.9)$ & 0.2841 & 0.0438  \\
$(0,0,-0.8)$ & 0.2842 &  0.0439\\
$(0,0,-0.7)$ & 0.2842 & 0.0439 \\
$(0,0,-0.6)$ & 0.2842 & 0.0439 \\
$(0,0,-0.5)$ & 0.2834 & 0.0438 \\
$(0,0,-0.4)$ & 0.2832 & 0.0437 \\
$(0,0,-0.3)$ & 0.2830 & 0.0437\\
$(0,0,-0.2)$ & 0.2830 & 0.0437\\
$(0,0,-0.1)$ & 0.2830 &  0.0437\\ \hline
$(0,0,0.0)$ & 0.2829 & 0.0437\\
$(0,0,0.1)$ & 0.2831 & 0.0437\\
$(0,0,0.2)$ & 0.2838 & 0.0438 \\
$(0,0,0.3)$ & 0.2836 & 0.0438\\
$(0,0,0.4)$ & 0.2831 & 0.0437\\
$(0,0,0.5)$ & 0.2832 & 0.0437 \\
$(0,0,0.6)$ & 0.2848 &  0.0041\\
$(0,0,0.7)$ & 0.2857 & 0.0442 \\
$(0,0,0.8)$ & 0.2885 & 0.0449 \\
$(0,0,0.9)$ & 0.3088 & 0.0481 \\
\end{longtable}
\end{center}

Let $\tau_{aniostropic}$ denotes the boundary time for a Brownian particle undergoing anisotropic diffusion, and $\tau_{isotropic}$ denotes the boundary time for a Brownian particle undergoing isotropic diffusion. Define the relative percentage boundary time $\delta \tau$

\begin{equation}\label{eq:relative_boundary_dynamics}
\delta{\tau} = \dfrac{|\tau_{isotropic} - \tau_{anisotropic}|} {\tau_{isotropic}} \times 100\%
\end{equation}

Table.~\ref{table:rel_boundary_time} presents the comparison of boundary time for both the isotropic and anisotropic diffusion when a Brownian particle is inside the $\delta_{region}$. The relative percentage boundary time is $0.8484\%$ when the Brownian particles are inside $\delta_{region}$ for $\delta=0.1$, whereas it is $9.8655 \%$ when a Brownian particle is inside the inside $\delta_{region}$ for $\delta=0.01$. This behavior can also be justified as the constant of anisotropic diffusion, i.e., $D_{\parallel}$ and $D_{\perp}$ are smaller than isotropic diffusivities $D$ (See Fig.~\ref{fig:aniso_vs_iso_diff}).

\begin{center}
\begin{longtable}{|c|c|c|c|c|}
\hline
$\delta$ & {{Boundary layer volume fraction $|\Omega_\delta|/|\Omega|$}} &$\tau_{isotropic}$ & $\tau_{anisotropic}$& $\delta{\tau}$   \\ \hline\hline
\endfirsthead

\hline
$\delta$ & {{Boundary layer volume fraction $|\Omega_\delta|/|\Omega|$}} & $\tau_{isotropic}$ Isotropic & $\tau_{anisotropic}$ Anisotropic & $\delta{\tau}$  \\ \hline\hline
\endhead

\hline
\endfoot

\hline

\caption{A comprehensive list illustrating the relative percentage boundary time $\delta \tau$ between the anisotropic boundary time $\tau_{anisotropic}$ and isotropic boundary time $\tau_{isotropic}$, when a Brownian particle is launched in the unit sphere with the absorbing window at the north pole $(0,0,1)$.}
\label{table:rel_boundary_time}
\endlastfoot

0.1    & 27.10\%  & 28.29\% & 28.53\% & 0.8484\% \\
\hline

0.01   & 2.970\% & 4.014\% & 4.410\% & 9.8655\%\\
\end{longtable}
\end{center}

\section{Discussion and Conclusions}\label{sec:conclusion_disc}

The narrow escape problem is a first-passage problem concerned with randomly moving particles in a physical domain being trapped by absorbing surface traps or windows, such that the measure of these traps is small compared to the domain size. The mean first passage time (MFPT) $v(x)$ defined as the expectated value of time $\tau(x)$ required for the particles having Brownian trajectories $\{W(t)\}_{t \ge 0}$ to escape through one of the boundary traps when launched from $x=W(0)$. The initial general model for all random walk-type/diffusion problems is Brownian motion, possibly with variable diffusivity. The Poisson problem \eqref{eqn:NEP_Dir_Neu} approximates such diffusion problems in the Narrow Escape context, averaging trajectories over multiple runs. A model validating the MFPT as solution of a Poisson partial differential equation (PDE) problem has been proposed and examined in the recent literature (cf.~\cite{HolcmanD.2004ETaS, RednerSidney1951-2001Agtf}), where the approximate asymptotic expressions for MFPT and average MFPT concerning narrow escape kinetics for various two and three-dimensional domain have been studied. The validity of the quality of the approximation of Brownian motion by Poisson's problem is thoroughly studied in \cite{RednerSidney1951-2001Agtf,HeinzStefan1958-2011Mm}.

Recent literature \cite{HolcmanD.2004ETaS,SingerA.2006NEPI,CheviakovAlexei2010AAAO,Gomez} contains the various asymptotic results that have been inferred concerning the narrow escape problems use the method of matched asymptotic expansion. The problem for the sphere or another domain with a single trap has been solved approximately in \cite{Brow_leak}. For the two-dimensional spaces with a smooth boundary and a small surface trap, the asymptotic expansion for MFPT is obtained in \cite{HolcmanD.2004ETaS,SingerA.2006NEPI}. Furthermore, these asymptotic results have been extended for a two-dimensional domain having a cluster of absorbing windows in \cite{PillayS2010AAAO}.

For the case of the unit sphere with $N$ well-separated asymptotically small absorbing traps or windows at the surface, Cheviakov et al.  \cite{CheviakovAlexei2010AAAO} have obtained an three-term asymptotic expansion for MFPT applying the method of matched asymptotic expansions. The asymptotic results for some non-spherical domains have studied and results are presented in \cite{Gomez}.

The goal of this work is to perform a direct numerical simulation of multiple particles undergoing Brownian motion in a spherical 3D domain with boundary traps, to compute MFPT values by averaging of Brownian escape times, and to compare the results with those predicted by
approximate solutions of the MFPT PDE problem. The current study aims at understanding aspects of the narrow escape kinetics that could not be extracted from asymptotic MFPT results. First, we were interested in knowing how many single-particle simulations are actually required to
closely match the predicted asymptotic averaged MFPT values. Second, having multiple simulated trajectories of the Brownian particle at our disposal, it was of interest to study other statistics of narrow escape dynamics, in particular, the dynamics in the bulk and near the boundary, estimating the time spend by Brownian particles near the domain boundary.

As a result an efficient and flexible, fully parallelized \verb|MATLAB| code was developed and tested, that can be applied to study diffusion processes/average values as well as multiple other statistical characteristics for Brownian motion-based diffusion processes. The code is able to model and collect $3$-D trajectories of Brownian particles starting from a given point up to their escape through a given trap, taking into account boundary reflections in a rigorous way. A comprehensive study of results obtained from the simulations shows that the asymptotic results for the escape times are in good agreement with the asymptotic results even when the radius of the boundary trap is small.

A comprehensive study of results obtained from the simulations showed that averages of $\sim 10^{4}$ single-particle simulations are sufficient to closely match the asymptotic MFPT values for the unit sphere.
	
With respect to the unit sphere having a boundary trap at the north pole Table.~\ref{table:mfpt_1trap} presents a comprehensive list showing the comparison between the asymptotic MFPT formula \eqref{eq:mfpt_nep} and the averaged Brownian numerically simulated escape times \eqref{eq:average_brow_escape_times}. The calculation illustrates that the simulation shows a close approximation of the mean first passage time MFPT. The study exhibits that the relative percentage difference error $\delta v(x)$ between the asymptotic results mostly not exceed $1\%$ accuracy, even when the radius of the boundary trap is $10^{-2}$ (See Fig.~\ref{fig:accu_1}).

For the two-trap configuration, the good agreement between the asymptotic and simulation results is illustrated in the Table.~\ref{table:mfpt_2trap} for the unit sphere having two boundary traps located sequentially at the north pole and the south pole. The study reveals that the relative percentage difference error $\delta v(x)$ between the results computed from the simulations is within the $0.7\%$ when the Brownian particles are launched from the $Z$-axis, whereas $\delta v(x)$ is mostly within $1\%$ when the particle is launched from various positions specified by \eqref{eq:lauch_coords} even when the radius of the boundary traps are $10^{-2}$ (See Fig.~\ref{fig:accu_22}). The surface plots with both perspective and top view, and contour plot (See Fig.~\ref{fig:accu_22}) are added, which help us discern the obtained results.

The MFPT PDE problem given by \eqref{eq:NEP_pde} does not retain any information about $3$-D trajectories of the Brownian particle. Our working \verb|MATLAB| code is capable of modeling and accumulating the $3$-D trajectories of Brownian particles, and following the physical laws of reflection everywhere on the boundary except the traps. Afterwards, with the help of our code, we try to understand the dynamics of the Brownian particle near the boundary by answering how much dimensionless seconds a Brownian particle spends on average in a delta-sized region $\delta_{\text{region}}$ near the boundary. It is shown that the Brownian particles spend more in the boundary layer than predicted by the boundary layer relative volume, with the effect being more pronounced in a narrow layer near the spherical wall.

For the one-trap configuration, the boundary time $\tau$, when expressed as a function of launching coordinates, $\tau (\phi, r)$ is constant $\sim 28.29 \%$  throughout the domain, inside $\delta_{\text{region}}$ for $\delta=0.1$, which encompassed the $\sim 27.1 \%$ of the total volume of the domain. The study reveals that the $\tau(\phi,r)$ is constant $\sim 4.01 \%$ for $\delta_{\text{region}}$ when $\delta=0.01$, which is $\sim 2.97 \%$ of the total volume of the sphere with one absorbing window at the north pole.

For the two-trap configuration, the boundary time $\tau$, when expressed as a function of launching coordinates, $\tau (\phi, r)$ is constant $\sim 28.21 \%$  throughout the domain, inside $\delta_{\text{region}}$ for $\delta=0.1$, which encompassed the $\sim 27.1 \%$ of the total volume of the domain. The study reveals that the $\tau(\phi,r)$ is constant $\sim 4 \%$ for $\delta_{\text{region}}$ when $\delta=0.01$, which is $\sim 2.97 \%$ of total volume of the sphere with two absorbing windows at the north and the south pole.

In \S \ref{sec:var_vs_constant_diffisuion}, we extend our work and use our working \verb|MATLAB| code to present and compare the Brownian dynamics simulation for computing MFPT and boundary dynamics of a particle for both the anisotropic and isotropic diffusion.

It is found that accounting the modified near-wall discussion leads to smaller escape times than predicted by the PDE model \eqref{eqn:NEP_Dir_Neu}, and a larger time spent by particles near the boundary. The Table.~\ref{table:rel_boundary_time} presents the comparison of boundary time for both the isotropic and anisotropic diffusion when a Brownian particles in inside the $\delta_{region}$. The relative percentage boundary time is $0.8484\%$ when the Brownian particles is inside $\delta_{region}$ for $\delta=0.1$, whereas it is $9.8655 \%$ when a Brownian particle is inside the inside $\delta_{region}$ for $\delta=0.01$. This behavior can also be justified as the constant of anisotropic diffusion, i.e., $D_{\parallel}$ and $D_{\perp}$ are smaller than isotropic diffusivities $D$ (See Fig.~\ref{fig:aniso_vs_iso_diff}).

The ubiquitous nature of the problem concerning the narrow escape kinematics may warrant further study. The developed code can be used to study the dynamics of Brownian particles in any $3$-D domain. For instance, we can study the dynamics of a Brownian particle inside domains with long neck \cite{long_neck}, which will be helpful in understanding and estimating the MFPT concerning the domain with dendritic spine geometry. The code can also be used to study the nanoparticle diffusion within inverse opals and related man-made materials with cavities \cite{nanoparticle} through which a naonoparticle can escape. In \cite{Bressloff}, Bressloff et al. presented a detailed study of protein receptor trafficking within the membrane of a cylindrical dendrite. With some modification in our working \verb|MATLAB| code, we can numerically estimate MFPT in some good agreement with asymptotic results. Moreover, it will also be interesting to examine the Brownian particle's boundary dynamics in such complex domains.

In \cite{Schwarz}, Schwarz et al. presented various numerical methods for homogeneous and inhomogeneous intermittent random search strategies. We can make use of all these techniques to optimize our code which will further increase the computational efficiency.

The Brownian simulation adopted in this work may be further modified and possibly improved through the use of a more general Langevin's velocity-based Brownian motion model \cite{HeinzStefan1958-2011Mm}, defined as the stochastic difference equation
\begin{equation}\label{eq:langevin}
\begin{array}{ll}
\dfrac{x_{n+1} - x_{n}}{\Delta t} = v_{n}, \\[3ex]
\dfrac{v_{n+1} - v_{n}}{\Delta t} = -\dfrac{1}{\tau} \Big( v_{n} - \dfrac{\Delta x_{n}}{\Delta t} \Big),
\end{array}
\end{equation}
where $x_n$ represents position and $v_n$ describes velocity of a particle undergoing Brownian motion, $D$ is the constant of the diffusivity, $\tau$ is the characteristic time scale, ${\Delta t}$ is the time step, and the change in Wiener process, i.e., normally distributed random displacements $\Delta x_n$ is given by
\begin{align*}
    \Delta x_n=\sqrt{3D\tau}\, \gamma_{n-1}; \quad \gamma_{n} \sim \mathcal{N}(0,1).
\end{align*}
The above set of equations can be reduced to linear second-order difference equation given by
\begin{align*}
x_{n+1} &= x_n + \Big( 1- \dfrac{\Delta t}{\tau} \Big) \Big( x_n - x_{n-1} \Big) +\dfrac{\Delta t}{\tau} \Big( \Delta x_{n-1} \Big).
\end{align*}
On setting $\Delta t = \tau$, the Langevin's velocity-based model given by the above second-order difference equation is reduced to Brownian motion modeled by continuous Wiener process with independent random time increments given by \eqref{eq:weiner_process}. This also implies that the Brownian motion modeled by the continuous Wiener process describes a coarse Brownian motion model for which the time step $\Delta t$ equals the typical relaxation time $\tau$.

In the future work it would be of interest to modify the continuum PDE-based MFPT model \eqref{eqn:NEP_Dir_Neu} to include variable and anisotropic diffusion, which could be redefined in the tensor form, with the Poisson equation $D \Delta v = -1$ replaced by $\div(\mathbf{D}\cdot\grad v)=-1$, develop numerical or closed-form exact or approximate solutions, and compare with the results of Section \ref{sec:var_vs_constant_diffisuion} of this contribution. It is also of interest to perform direct Brownian simulations more complex physical settings, including randomly moving receptors on the surface of the domain \cite{lawley2019receptor}. Thus, we can use this working \verb|MATLAB| code to study Brownian dynamics for both the anisotropic and isotropic diffusion for simple and complex domains.

\section*{Acknowledgements}

The authors are grateful to Jason Gilbert for discussions and initial versions of the Brownian code. A.~C.~thanks NSERC of
Canada for research support through a Discovery grant RGPIN-2019-05570. V.S. is grateful to the University of Saskatchewan and the Indian Institute of Technology Gandhinagar for research opportunities and financial support.

{\footnotesize
\bibliography{biblo}
\bibliographystyle{ieeetr}
}

\end{document}